\documentclass{emulateapj}

\usepackage{epsfig}

\usepackage{textcomp}

\usepackage{float}

\usepackage{amssymb}

\newcommand{\lamsq} {$\lambda^2$}

\newcommand{\Deltalamsq} {$\Delta\lambda^2$}

\newcommand{\deltalamsq} {$\delta\lambda^2$}

\newcommand{\lamsqmin} {$\lambda^2_{min}$}

\newcommand{\lamsqmax} {$\lambda^2_{max}$}

\newcommand{\radmsq} {rad/m$^2$}

\begin{document}

\title{Integrated Polarization of Sources at $\lambda\sim$1\lowercase{m} and New Rotation Measure Ambiguities}

\author{Damon Farnsworth, Lawrence Rudnick}
\affil{Department of Astronomy, University of Minnesota, 116 Church St. S.E., Minneapolis, MN 55455, USA}
\author{Shea Brown}
\affil{CSIRO, Australia Telescope National Facility, PO Box 76, Epping NSW 1710, Australia}

\begin{abstract}
We present an analysis of the polarization of compact radio sources from six pointings of the Westerbork Synthesis Radio Telescope (WSRT) at 350 MHz with 35$\%$ coverage in \lamsq.  After correcting for the off-axis instrumental polarization with a simple analytical model, only a small number of 585 strong sources have significant polarizations at these wavelengths.  The median depolarization ratio from 1.4~GHz for the strongest sources is $<$0.2, reinforcing the likelihood that radio galaxies are found in magnetized environments, even outside of rich clusters.  Seven sources with significant 350~MHz polarization were selected for a more in-depth Faraday structure analysis.
We fit the observed values $Q/I$ and $U/I$ as a function of \lamsq~ using both a depolarizing screen and two component models.  We also performed RM Synthesis/Clean and standard fitting of polarization angle vs. \lamsq.  We find that a single rotation measure (RM), as found using polarization angle fitting or simple screen models, commonly provides a poor fit when the solutions are translated back into $Q$, $U$ space.
Thus, although a single ``characteristic'' rotation measure may be found using these techniques, the Faraday structure of the source may not be adequately represented.
We also demonstrate that RM Synthesis may yield an erroneous Faraday structure in the presence of multiple, interfering RM components, even when cleaning of the Faraday spectrum is performed.
We briefly explore the conditions under which rotation measures and Faraday structure results can be reliable.
Many measurements in the literature do not meet these criteria; we discuss how these influence the resulting scientific conclusions and offer a prescription for obtaining reliable RMs.
\end{abstract}


\section{Introduction}
\label{sec:introduction}
By characterizing the Faraday structure in radio synchrotron sources, properties of the magneto-ionic medium can be probed, such as magnetic field strength and orientation, as well as distribution of the relativistic and thermal electron populations.
Radio arrays such as the Westerbork Synthesis Radio Telescope (WSRT), the Expanded Very Large Array (EVLA), the Low Frequency Array (LOFAR), the Australia Telescope Compact Array (ATCA), the Allen Telescope Array (ATA), and the planned Australian Square Kilometer Array Pathfinder (ASKAP) are well suited for Faraday structure studies due to their enhanced \lamsq~ sampling capabilities, e.g., wide relative bandwidth (\Deltalamsq/\lamsqmin), and high spectral resolution (\deltalamsq).

We represent the complex linear polarization by 
\begin{equation}
  \mathbf{P} = Ipe^{2i\chi} = Q + iU \equiv I(q + iu)
\end{equation}
where $p$ and $\chi$ are the degree and angle of polarization, given by 
\begin{equation}
  p = \frac{P}{I} = \sqrt{q^2 + u^2}
\end{equation}
\begin{equation}
  \chi = \frac{1}{2} \arctan \frac{U}{Q}
\end{equation}
and $I$, $Q$, $U$ are the Stokes parameters for the total and orthogonal components of the linearly polarized intensities.
We use $q$, $u$ to represent the fractional values $Q/I$, $U/I$.

Traditionally, most polarization studies have determined rotation measures by fitting
\begin{equation}
 \chi(\lambda^2) = \chi_0 + \lambda^2 \rm RM,
 \label{eqn:linearChiEqn}
\end{equation}
where RM is the Faraday rotation measure, with little or no attention paid to the behavior of the fractional polarization.
A common practice has been to restrict RM fitting to regions of \lamsq~ space where $p$(\lamsq) is constant or decreases monotonically (e.g., \citealt{simard81}), which would occur for a foreground rotating or depolarizing screen.  This is sometimes done even when data showing a rise in $p$(\lamsq) at shorter wavelengths exists, ignoring evidence that multiple RM components may be present.
Others restrict their fitting to $\lambda < \lambda_{1/2}$ (defined by $p(\lambda_{1/2})/p(0)=0.5$), beyond which \cite{burn66} suggests that significant non-linear behavior in $\chi$(\lamsq) is expected (e.g., \citealt{haves75}).
In other cases, significant non-linear behavior in $\chi$(\lamsq) is observed (e.g., \citealt{morris64}, \citealt{roy05}), but no modeling of this anomalous behavior is made and the RM from the poor linear fit to $\chi$(\lamsq) is reported.
Others require that $p$ is above some threshold and/or that a minimum signal-to-noise value is present in the observations but do not report the behavior of $p$(\lamsq), which may hold information regarding the underlying Faraday structure (e.g., \citealt{clarke01}, \citealt{brown07}).

The only situation where d$\chi$/d\lamsq~ and $p(\lambda^2)$ are constant, and therefore unimportant in determining the Faraday structure, is when there is a single uniform Faraday screen completely in the foreground.
In all other cases, including all or most physically realistic ones, more sophisticated modeling is required.
For example, \cite{fletcher04} consider both polarization degree and angle in their study of the magnetic field of M31.  In addition, \cite{rossetti08} and \cite{fanti04} employed simple models of depolarization and $\chi$ rotation to examine Compact Steep-Spectrum (CSS) sources at $\lambda \leq 21$~cm.
We will briefly summarize some of the classic models where $p(\lambda^2) \neq$ constant.  For a detailed discussion on depolarization effects, we refer the reader to \cite{sokoloff98}.

If the thermal electrons are spatially coincident with the relativistic, synchrotron emitting electrons, e.g., then
\begin{equation}\label{burn.slab.eqn}
p(\lambda^2) \propto \frac{\sin (\lambda^2 \rm F_c)}{\lambda^2 \rm F_c}.
\end{equation}
as in the uniform slab model of \cite{gardner66}, where F$_{\rm c}$ is the Faraday depth through the slab; F$_{\rm c}$ can be thought of as an ``internal'' RM.
\cite{cioffi80} showed that the observed depolarization and $\chi$ rotation can have considerable differences depending on the geometry assumed, even for simple cases such as cylinders and spheres.

For a foreground screen consisting of many unresolved components with a random distribution of RMs, \cite{burn66} modeled the observed fractional polarization as
\begin{equation}\label{burn.screen.eqn}
p(\lambda^2) \propto \exp (-2 \sigma_{RM}^2 \lambda^4)
\end{equation}
where $\sigma_{RM}^2$ is the variance of a Gaussian dispersion in RM across this so-called ``mottled'' screen.
Modifications to this model have been proposed, e.g., by \cite{rossetti08} who include the effect of filling factors.

Two interfering foreground RM components will also produce non-\lamsq~ behavior in angle, and changes in fractional polarization that can rise or fall with increasing wavelength. \cite{goldstein84} describe the observed polarized flux from two such components as:
\begin{equation}\label{goldstein.2comp.eqn}
 P_{obs} = P_{1} [ 1 + k^2 + 2 k \cos(\chi_1 - \chi_2) ]^{1/2}
\end{equation}
where $k = P_2 / P_1 \leq 1$ is the ratio of the the polarized fluxes, $\chi_1$ and $\chi_2$ are the polarization angles at the observation frequency.

With adequately sampled data in \lamsq \space space, all of the above cases can, in principle, be distinguished.
In practice, however, \lamsq \space sampling is inadequate to map out the Faraday structure, and even the large fractional bandwidths of the WSRT or the EVLA can be insufficient.
As we will illustrate below, determination of the Faraday structure requires observations which detect the \textit{variations} in both $p(\lambda^2)$ and  d$\chi$/d\lamsq.
In particular, the result d$\chi$/d\lamsq \space $\approx$ constant can occur over a substantial range in \lamsq~ even with underlying Faraday structure.
Whether or not failure to diagnose the presence of underlying Faraday structure is acceptable depends on the particular scientific goals, as we discuss further below.

In Section \ref{sec:observations} we present our WSRT observations at 350 MHz and the determination and removal of the off-axis instrumental polarization.
We present the results of our polarization and Faraday structure analyses in Section \ref{sec:results}.  There we characterize the depolarization of our sample of 585 compact sources and give a brief overview of the polarization diagnostics and Faraday structure modeling employed.  We then detail the modeling results on seven sources with significant 350~MHz polarization and the discrepancies between fitting $q$(\lamsq) and $u$(\lamsq) and other other techniques.
In Section \ref{sec:RMexperiments} we use the results of some simple experiments to demonstrate some of the inadequacies of common RM determination methods such as $\chi$(\lamsq) fitting and RM Synthesis.  We also offer some recommendations for reliable RM determinations.
A discussion of our findings, including the science implications of RM ambiguities, is presented in Section \ref{sec:discussion}.

\begin{table*}
	\caption{Summary Of WSRT 350 MHz Observations}
	\centering
		\begin{tabular}{ l c c c c c c c }
		\hline\hline
		Field & RA      & DEC     & \textit{b}$^1$ & Array         & Exposure & Common & Calibrators \\
		      & (J2000) & (J2000) & (\textdegree) & Configuration & (Hours)  & Beam ($\arcsec$) \\
		\hline
    Aries-Pisces & 01:09:14.30 & +13:09:58.0 & -49 & Mini-short & 12 & 325 x 70 & 3C147, 3C295 \\
    Coma SW & 12:54:08.00 & +26:42:00.0 & +89 & Special$^2$ & 12 & 70 x 70 & 3C147, 3C295 \\
    Coma NW & 12:54:08.00 & +27:58:00.0 & +88 & Special$^2$ & 12 & 70 x 70 & 3C147, 3C295 \\
    Coma NE & 12:59:52.00 & +27:58:00.0 & +88 & Special$^2$ & 12 & 70 x 70 & 3C147, 3C295 \\
    A & 14:53:00.00 & +40:25:00.0 & +62 & Maxi-short & 4.2 & 125 x 70 & 3C147, 3C48 \\
    B & 16:20:00.00 & +60:12:00.0 & +42 & Maxi-short & 12  & 105 x 70 & 3C48, 3C295 \\
		\hline
		\end{tabular}

    $^1$Approximate Galactic latitude at field center.  \\
		$^2$Special array configuration is 36m+54m+72m+90m.
	\label{tab:SummaryOfWSRT350MHzObservations}
\end{table*}

\begin{table*}
  \caption{Number of 78 kHz Channels Used For IF Band Averaging}
  \centering
    \begin{tabular}{ l  c  c  c  c  c  c  c  c  c }
    \hline\hline
    Field & Stokes  & IF1 & IF2 & IF3 & IF4 & IF5 & IF6 & IF7 & IF8 \\
          &         & (376.4)$^1$ & (367.7) & (358.9) & (350.2) & (341.4) & (332.7) & (323.9) & (315.2) \\
    \hline
     & I & 100 & 97 & 68 & 79 & 100 & 99 & 64 & 32 \\
    Aries-Pisces & Q & 100 & 99 & 76 & 79 & 100 & 99 & 64 & 32 \\
     & U & 69 & 96 & 58 & 24 & 85 & 28 & 64 & 32 \\
    \hline
     & I & 101 & 101 & 59 & 62 & 101 & 98 & 96 & 92 \\
    Coma SW & Q & 101 & 100 & 58 & 69 & 100 & 97 & 97 & 97 \\
     & U & 86 & 52 & [7]$^2$ & 40 & 96 & 38 & 39 & 59 \\
    \hline
     & I & 101 & 101 & 70 & 96 & 101 & 99 & 99 & 93 \\
    Coma NW & Q & 101 & 100 & 73 & 79 & 99 & 97 & 98 & 96 \\
     & U & 91 & 84 & 21 & 75 & 100 & 27 & 78 & 32 \\
    \hline
     & I & 101 & 101 & 56 & 29 & 101 & 99 & 91 & 87 \\
    Coma NE & Q & 101 & 100 & 64 & 62 & 100 & 98 & 96 & 95 \\
     & U & 98 & 85 & [5]$^2$ & 16 & 99 & 54 & 68 & 57 \\
    \hline
     & I & 92 & 101 & 71 & 96 & 96 & 99 & 101 & 88 \\
    A & Q & 95 & 101 & 76 & 96 & 96 & 99 & 101 & 88 \\
     & U & 30 & [1]$^2$ & 26 & 78 & 65 & [2]$^2$ & 17 & 88 \\
    \hline
     & I & 86 & 94 & 42 & 41 & 100 & 92 & 90 & 70 \\
    B & Q & 86 & 97 & 64 & 76 & 101 & 97 & 101 & 93 \\
     & U & 52 & 98 & 76 & 88 & 101 & 97 & 101 & 69 \\
    \hline
    \hline
    \end{tabular}

    $^1$IF band central frequencies are given below IF number in MHz. \\
    $^2$Bracketed values identify IFs where $U$ averaging was not performed due to too few channels.
  \label{tab:NumberOfChannelsUsedForBandAveraging}
\end{table*}

\section{Westerbork 350 MHz Observations and Instrumental Polarization}
\label{sec:observations}

\subsection{Observations and Data Reduction}
We observed six fields with the WSRT in 2008 and 2009, originally selected for possible large-scale diffuse polarization found in the NRAO Very Large Array Sky Survey (NVSS, \citealt{condon98}) through a reprocessing by \cite{rudnick09}.  To minimize the contribution of polarized Galactic foreground emission, we have selected fields with $|b|$$\gtrsim$42\textdegree.
Observations were made in spectral line mode with a central frequency of 345 MHz, 70 MHz bandwidth, and 1024 channels over 8 intermediate frequency (IF) sub-bands, yielding full Stokes parameters.
Even though the central frequency is 345 MHz, we will continue to refer to this band as the 350 MHz band to comply with the established convention.
Various array configurations were used and are shown in Table \ref{tab:SummaryOfWSRT350MHzObservations}.
The nominal synthesized beam size varies with array configuration and observing frequency, but is approximately 70$''$ in RA for our observations.  Due to the East-West array configuration, the beam becomes elongated in the North-South direction by a factor of $\csc(\delta)$.

We will now summarize the key elements of the data reduction and calibration process; for a complete description we refer the reader to \cite{brown09}.
All reduction was done using standard techniques in AIPS, correctly accounting for the WSRT linearly polarized feeds, and including several iterations of amplitude and phase self calibration for total intensity.
Flux calibrators were observed, and are listed in Table \ref{tab:SummaryOfWSRT350MHzObservations}.
The AIPS procedure LPCAL was used to correct for polarization leakage between the $X$ and $Y$ orthogonal linear polarization receivers.
Additionally, calculation of a time-independent phase correction between linear polarizations $X$ and $Y$ was attempted for each channel using a polarized calibrator observed during the run.
The polarized calibrator 3C345 was used for all fields except Field B, for which DA240 was used instead.
Unfortunately, a solution was not found for every channel, rendering those channels without a solution useless for Stokes $U$ measurement.

Cleaning and imaging were also done in AIPS, where 4\textdegree x4\textdegree \space images in Stokes $I$, $Q$, and $U$ were created for each channel.  The community is just beginning to experiment with the much simpler problem of multifrequency synthesis/cleaning in total intensity, where one or two spectral parameters  can be used to characterize the frequency dependence, and the biases there have not yet been characterized.  $Q$ and $U$ have much more complex behavior as a function of frequency and will require extensive experimentation in the future.
Therefore, each channel and Stokes quantity was cleaned separately in AIPS with IMAGR using a loop gain of 0.1 and 15,000 clean components per field.
Images of Stokes $V$ (circular polarization) were made to verify that no leakage into $V$ was present, under the assumption that it is negligible for typical astrophysical sources.
Typical channel sky RMS values of $\sim$3-5 mJy/beam were obtained in the cleaned $I$ images, and $\sim$1-3 mJy/beam for the $Q, U$ images.
All images for a field with sky RMS $\leq5$ mJy/beam (uncorrected for primary beam attenuation) were convolved to a common beamsize, allowing channel averaging to be performed as described below.

Average images of Stokes $I,$ $Q,$ and $U$ for each of the eight IFs were constructed from the individual channel images, along with a total intensity map averaged over all eight IFs.
The number of channels used for each band average image is listed in Table \ref{tab:NumberOfChannelsUsedForBandAveraging}.  Channels with imaging problems, such as strong artifacts due to radio frequency interference (RFI), were excluded.
In addition, $U$ imaging was not performed on channels where no $X$-$Y$ phase correction was found.
We supplemented these data using images from the NVSS to provide measurements of Stokes $I$, $Q$, $U$ at 1.4 GHz.
The NVSS images were convolved to the corresponding WSRT field's beamsize.
See Table \ref{tab:SummaryOfWSRT350MHzObservations} for an overview of the field properties, including common beam convolution sizes.
The Coma fields were imaged using a (somewhat smaller than nominal) restoring beam of 70$''$x70$''$ as part of another study \citep{brown11}.

\begin{figure*}
  \centering
  \epsfig{file=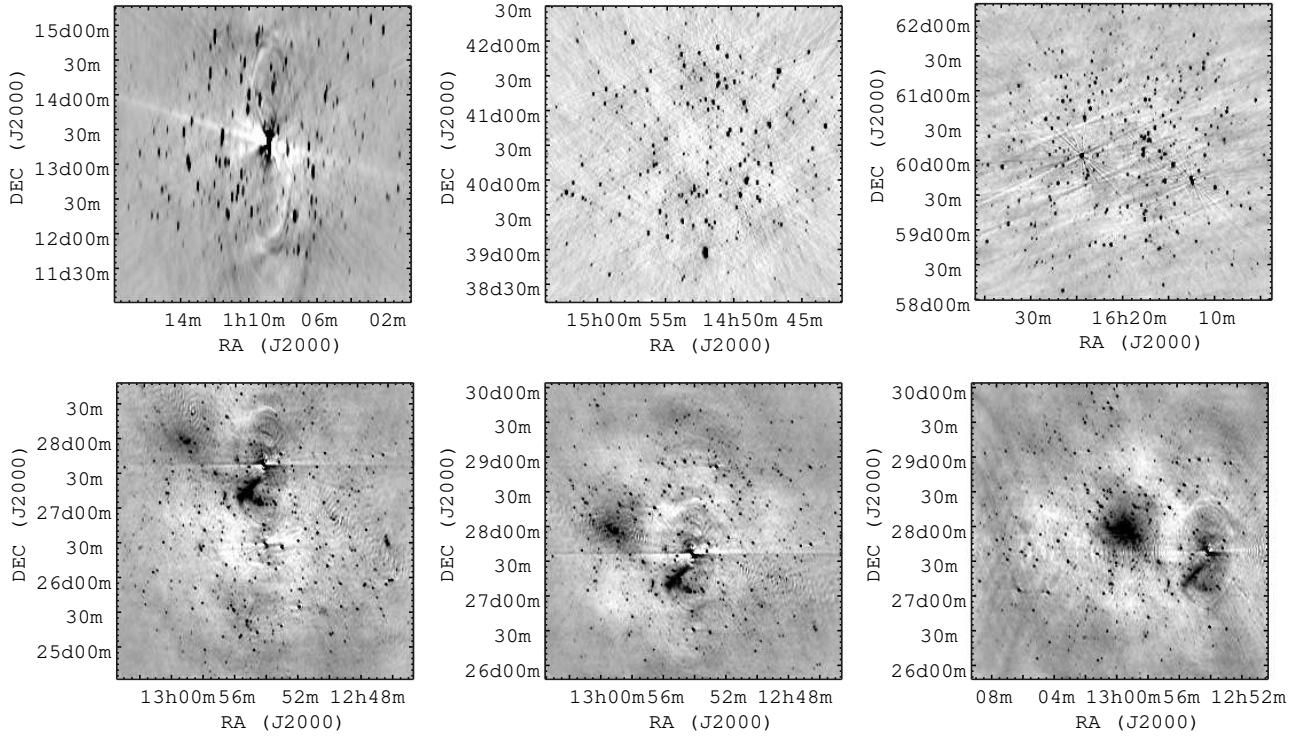,angle=0,width=0.95\textwidth}
  \caption{Total intensity images of the six fields observed at 350 MHz with the WSRT.  Top row, from left: Aries-Pisces, Field A, Field B.  Bottom row, from left: Coma SW, Coma NW, Coma NE (See Table \ref{tab:SummaryOfWSRT350MHzObservations}).  Images are 4x4 degrees.  Diffuse emission from the Coma halo and relic are visible in the Coma images.  Also visible are residual imaging artifacts near the strongest sources, common for the WSRT.}
  \label{fig:wsrt.Iavg}
\end{figure*}

\begin{figure*}
  \centering
  \epsfig{file=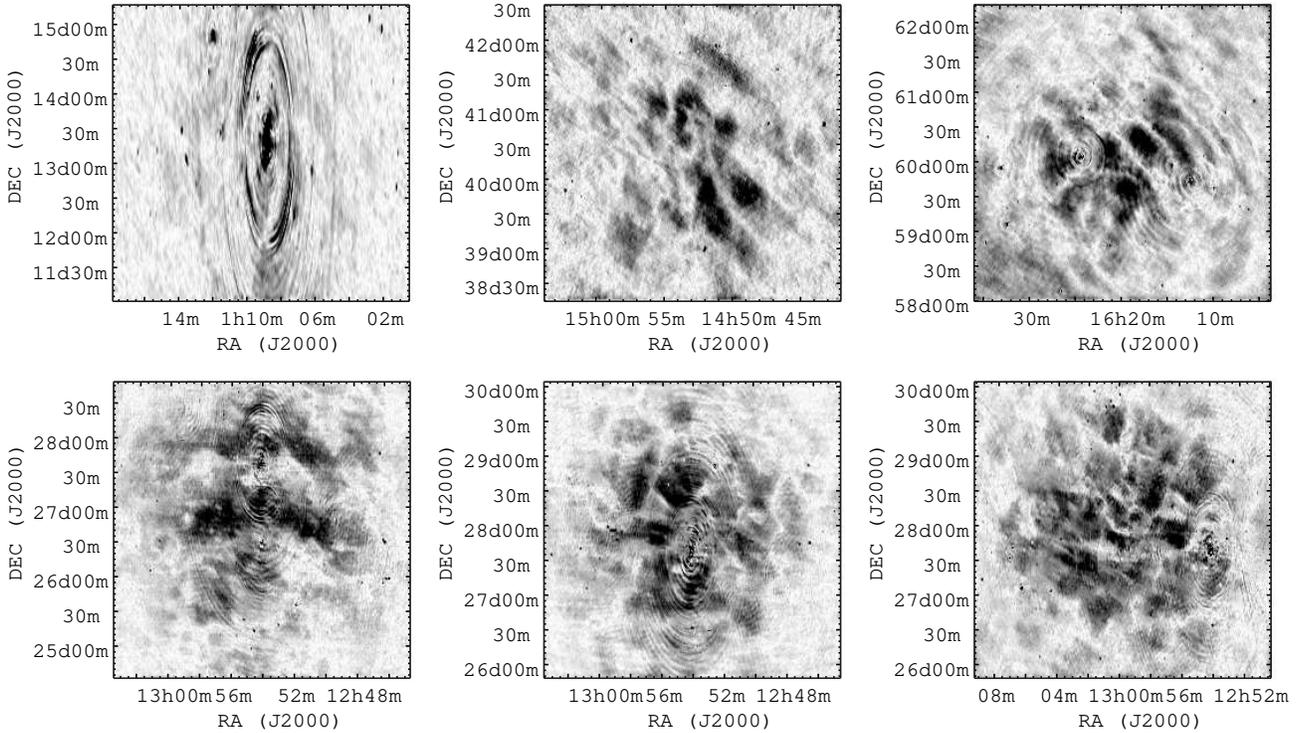,angle=0,width=0.95\textwidth}
  \caption{Linear polarization images at RM=0 of the six fields, taken from the results of RM Synthesis, observed at 350 MHz with the WSRT.  Top row, from left: Aries-Pisces, Field A, Field B.  Bottom row, from left: Coma SW, Coma NW, Coma NE (See Table \ref{tab:SummaryOfWSRT350MHzObservations}).  Images are 4x4 degrees.  Note the diffuse Galactic emission which pervades each field.  Also visible are residual imaging artifacts near the strongest sources, common for the WSRT.}
  \label{fig:wsrt.rm0}
\end{figure*}

Total intensity images of the six fields are shown in Figure \ref{fig:wsrt.Iavg}.  Images of the linear polarization at RM=0, taken from the results of RM Synthesis (see Section \ref{sec:sourceSelection}), are shown in Figure \ref{fig:wsrt.rm0}.  Note that the polarization maps at RM=0 are pervaded by diffuse Galactic emission (e.g., \citealt{brentjens05}, \citealt{schnitzeler09}, \citealt{wolleben10}, \citealt{bernardi10}).

\subsection{Instrumental Polarization of WSRT at 350 MHz}
\label{sec:instrumentalPolarization}
To identify sources with either real or instrumental polarization, we first selected sources in each field with $I/\sigma_{I}$ $\geq30$ in the all-IF Stokes $I$ image, yielding 585 total sources for the six fields. We then extracted Stokes $I$, $Q$, $U$ from each of the WSRT individual IF and NVSS images, at the peak location in the all-IF Stokes $I$ image.
A background subtraction was performed for each measurement using a rectangular region about the source, of inner dimension 1$\times$ the synthesized beam dimensions and outer dimension 2$\times$ the synthesized beam dimensions. The RMS deviation within each annulus was adopted as the statistical error in each measurement.

For the purpose of illustrating the instrumental polarization, we apply the simplest bias correction to the polarization amplitude:
\begin{equation}
\label{eqn:pcorr}
P_{corr} = \sqrt{P_{meas}^2 - \sigma_P^2},
\end{equation}
which is an approximation to the ``most probable estimator'' of \cite{wardle74}, good for $P_{corr}/\sigma_{P} > 0.5$.  This most probable estimator is the best available for $P_{corr}/\sigma_{P} > 0.7$ (\citealt{simmons85}), and we only report results well above this limit.
We use a propagated error calculation for $\sigma_P$ based on the observed errors in $Q$ and $U$.

\begin{figure}
  \centering
  \epsfig{file=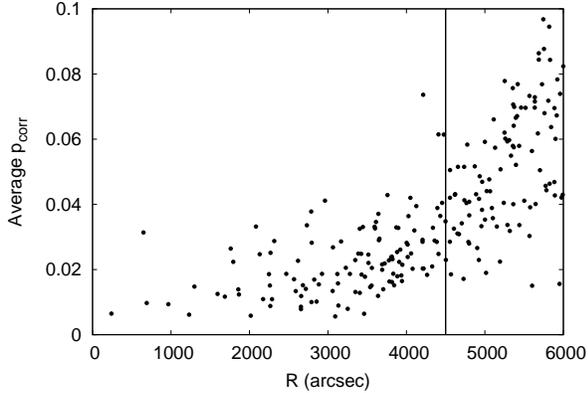,angle=0,width=0.45\textwidth}
  \caption{Plot of average $p_{corr}$ at 350~MHz for sources with at least six of the eight IF measurements satisfying $P_{meas}/\sigma_{P} \geq 1.2$ before instrumental correction has been applied.  The instrumental polarization increases with off-axis radius.  The vertical line at 4500$''$ corresponds to the radial limit of our instrumental polarization model fitting.}
  \label{fig:mcorr.1.2sig.noInstPolCorr.6samples}
\end{figure}

\begin{figure*}
  \centering
  \epsfig{file=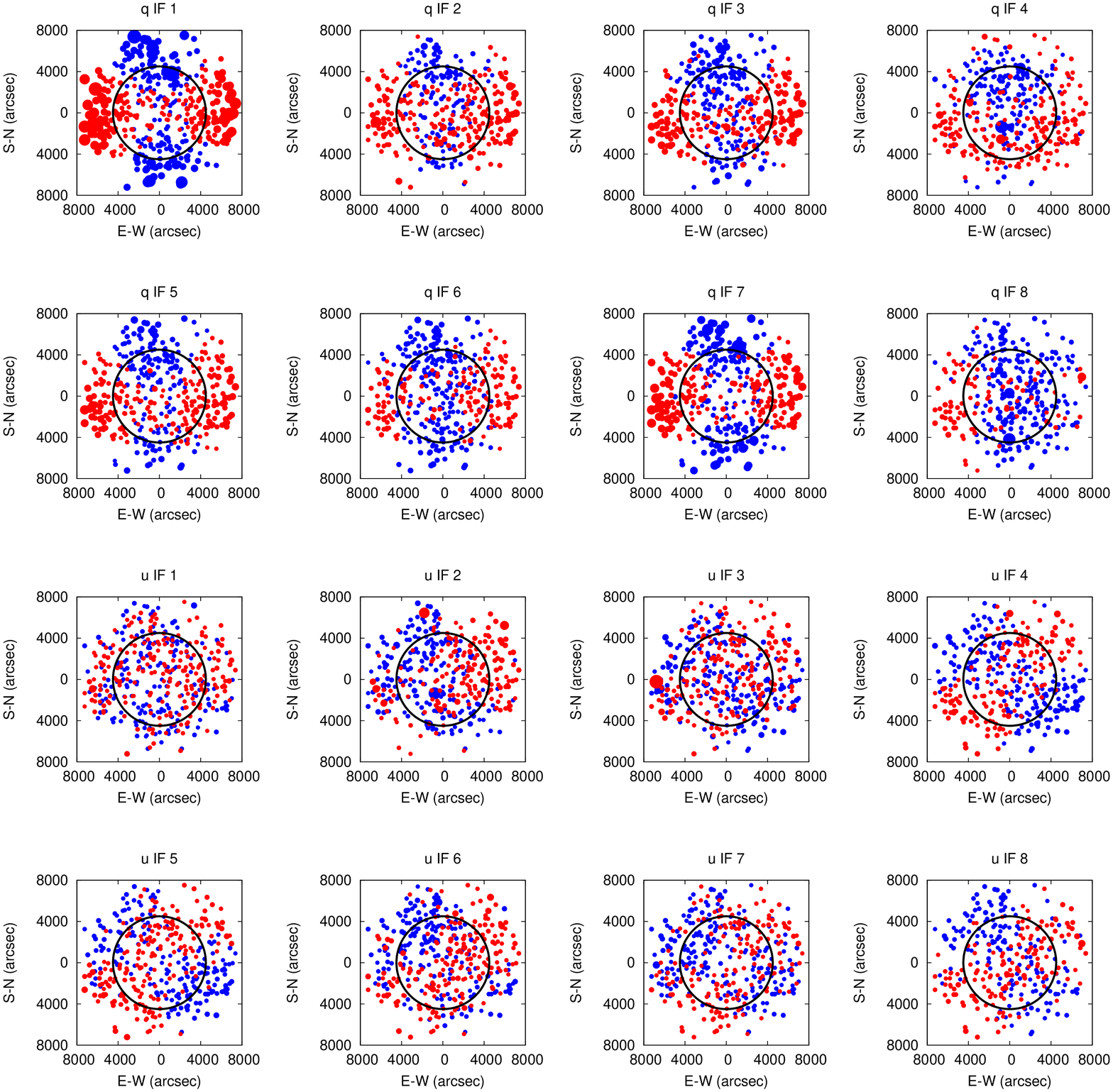,angle=0,width=0.95\textwidth} \\
  \caption{Plot of $q$ and $u$ measurements for sources with $P_{obs}/\sigma_{P} \geq 1.2$ from all six fields as projected on the sky, illustrating the radial and azimuthal behavior of off-axis instrumental polarization.  Blue points are negative, red points are non-negative; the point size is related to the magnitude of the measurement.  The quadrupole pattern described in the text is apparent.  The black circle, of radius 4500$''$, surrounds the region used for model fitting of instrumental polarization.}
  \label{fig:if1-8.qupolar.1.2x}
\end{figure*}

Figure \ref{fig:mcorr.1.2sig.noInstPolCorr.6samples} shows $p_{corr} \equiv P_{corr}/I$, averaged over multiple IFs, vs. off-axis radius for the WSRT data set, illustrating the instrumental enhancement of fractional polarization with radius as mentioned previously by \cite{debruyn05} and investigated at 1.4 GHz by \cite{popping08}.  In this work, we determined the instrumental polarization behavior in both $Q$ and $U$ for each IF in order to perform a first order correction.  For each IF, we first selected from the 585 initial sources those satisfying $P_{meas}/\sigma_{P} \geq 2$ and plot $q$ and $u$ as a function of their locations relative to the pointing center (Figure \ref{fig:if1-8.qupolar.1.2x}).
In several IFs there is a clear quadrupole pattern, in general possessing a greater magnitude in $q$ than $u$.
The observed $q$ quadrupole pattern is oriented coincident with the orientation of the $X$ and $Y$ linear dipole feeds on the WSRT, which face the sky perpendicular to each other and form Stokes $Q$ by the linear combination of $XX^*$ and $YY^*$. The observed $u$ quadrupole pattern, which is formed from a linear combination of $XY^*$ and $YX^*$, is offset 45$^\circ$ on the sky with respect to the $q$ pattern, as one would expect.

To quantitatively model the instrumental polarization for each IF, we made the following cuts to the data.
Outside 4500$''$ the instrumental polarization rises sharply and we do not attempt any correction beyond that limit, cutting the total number of sources from 585 to 335\footnote{This includes 36 duplicated sources observed at different off-axis positions due to the multiple pointings for the Coma field.}.
We then required $P_{meas}/\sigma_P \geq 2$, yielding roughly 100 sources per IF.
For each IF we fit a double cosine function to each set of $q$ and $u$ of the form
\begin{equation}
f(r,PA) = A e^{Br} cos{(2 PA + C)},
\end{equation}
which includes the distance from the pointing center, $r$, and position angle, $PA$, of the source.
This yielded 16 total sets of parameters, which are given in Table \ref{tab:instrModelFitParms}.
We then produced corrected $Q$, $U$ observations for each source by subtracting the modeled instrumental contribution.

\begin{table}
	\caption{Model Fit Parameters for Instrumental Polarization}
	\centering
		\begin{tabular*}{0.45\textwidth}{ @{\extracolsep{\fill}} c  c  c  c  c }
		\hline\hline
		IF & Fractional & A    & B                  & C     \\
		   & Stokes     & (\%) & $(\times 10^{-3}/\arcsec)$ & $(^{\circ})$ \\
		\hline
		1 & $Q$ & 0.08 & 1.1 & 83 \\
		  & $U$ & $8 \times 10^{-8}$ & 3.7 & 81 \\
		\hline
		2 & $Q$ & 0.03 & 1.1 & 99 \\
		  & $U$ & 0.007 & 1.3 & 48 \\
		\hline
		3 & $Q$ & 0.02 & 1.2 & 92 \\
		  & $U$ & 0.03 & 0.75 & 43 \\
		\hline
		4 & $Q$ & 0.10 & 0.64 & 109 \\
		  & $U$ & 0.30 & 0.26 & 47 \\
		\hline
		5 & $Q$ & 0.34 & 0.51 & 90 \\
		  & $U$ & 0.004 & 1.3 & 33 \\
		\hline
		6 & $Q$ & 0.29 & 0.4 & 91 \\
		  & $U$ & 0.008 & 0.7 & 4  \\
		\hline
		7 & $Q$ & 0.15 & 0.78 & 85 \\
		  & $U$ & 0.13 & 0.13 & 16 \\
		\hline
		8 & $Q$ & 0.0005 & 1.9 & 108\\
		  & $U$ & 0.06 & 0.65 & 62 \\
		\hline
		\hline
		\end{tabular*}
	\label{tab:instrModelFitParms}
\end{table}

The instrumental polarization is weak near the pointing axis, generally much less than $1\%$, but grows to as much as 6\% in $q$ near $R_{pb}$ (half power radius of the primary beam) for the odd numbered IFs.
In $u$ the instrumental contribution is $<$3\% at $R_{pb}$ for all IFs.

\begin{figure*}
  \centering
  \epsfig{file=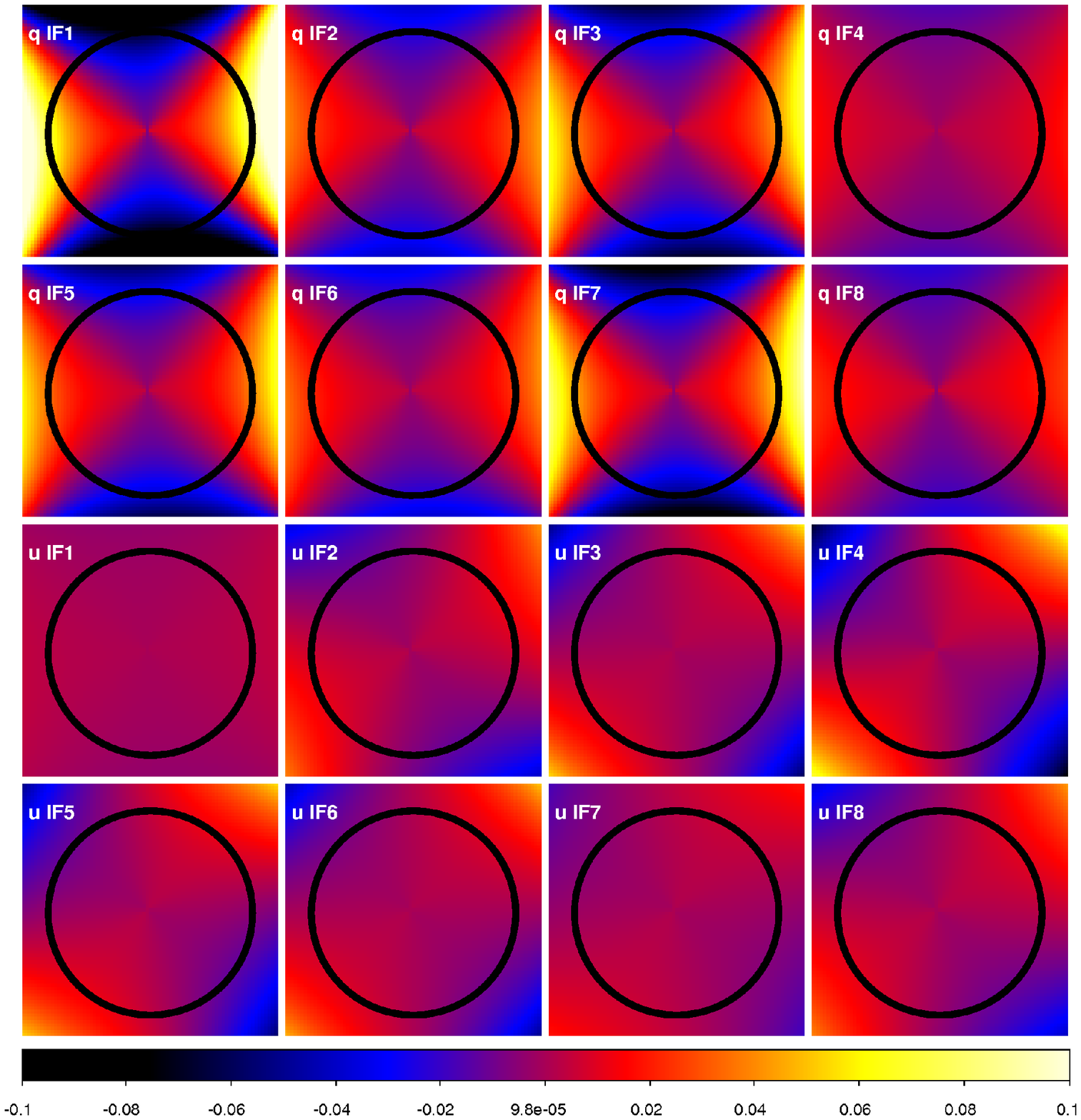,angle=0,width=0.95\textwidth} \\
  \caption{Model fits to the instrumental polarization of WSRT at 350 MHz as projected onto the sky (North is up, East is left).  Top row:  $q$ for IF1 - IF4.  Second row:  $q$ for IF5 - IF8.  Third row:  $u$ for IF1 - IF4.  Bottom Row:  $u$ for IF5 - IF8.  The frequency dependence can be seen in $q$ by noting the increased amplitude in the odd numbered IFs.  The black circle is of radius 4500$''$.}
  \label{fig:qumodel.idl}
\end{figure*}

By examining Figure \ref{fig:qumodel.idl}, one can see evidence of the 17 MHz modulation, as found by \cite{popping08}, in the $q$ models for IFs 1, 3, 5, and 7, which are separated by $\approx$17 MHz.
In these IFs, the instrumental polarization is stronger by factor of roughly 2-3 at $R_{pb}$ over the neighboring even numbered IFs.
This effect is much less pronounced in $u$, as seen in Figure \ref{fig:qumodel.idl}.

\begin{figure}
  \centering
   \epsfig{file=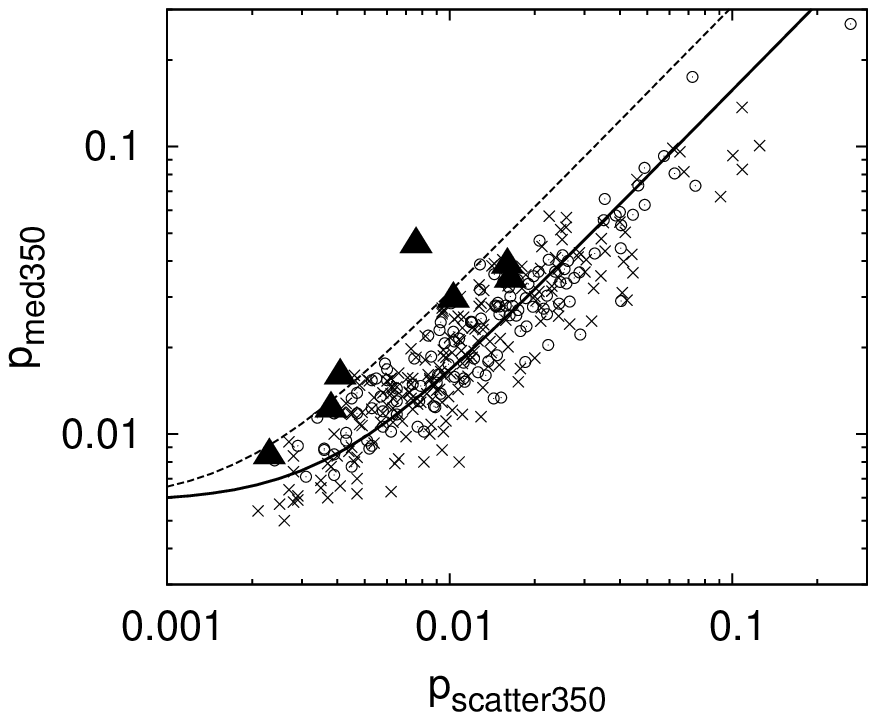,angle=0,width=0.45\textwidth}
  \caption{Plot of the median of $p$ in the eight WSRT IFs vs. the RMS scatter of $p$ among the IFs, to determine the polarization bias remaining after correction for instrumental polarization.  Circles represent sources exhibiting no polarization in the NVSS.  Xs represent sources exhibiting moderate or strong polarization in the NVSS.  Solid triangles show sources chosen for modeling as described in the text.  The solid line is the best fit for all bands for each of the 335 sources used to model the instrumental polarization, and the dashed line shows the defined upper limit discussed in the text.}
  \label{fig:detect90}
\end{figure}

After correction for instrumental polarization, there is still a significant polarized flux bias from a variety of factors which differ from one IF to another, including the noise bias (including random noise and residual sidelobe structures) and non-quadrupole components to the instrumental polarization as a function of IF and two-dimensional location within the primary beam.  These are not well modeled by Equation \ref{eqn:pcorr}, so in order to make a practical model for the polarized flux bias we took an empirical approach and measured the median ($p_{med350}$) and RMS scatter ($p_{scatter350}$) among the 350 MHz IFs of the polarized fraction for each of 335 sources with $r < 4500''$.  We expect that residual instrumental polarizations, sidelobe structures, and noise will all vary from IF to IF, and that $p_{scatter350}$ will therefore provide an estimate of all of these contributions.  On the other hand, $p_{med350}$ provides an estimate of the true polarized flux, along with a bias related to $p_{scatter350}$.  These are plotted vs. each other in Figure \ref{fig:detect90}.  Different symbols represent different levels of NVSS polarized flux for the same sources.

There was no significant difference in the distribution as a whole between sources with no NVSS polarization and sources with moderate or strong NVSS polarization.  Therefore, the bulk of $p_{med350}$ values are likely due to the instrumental contributions described above, as opposed to intrinsic polarizations.  We fit the distribution and found 
\begin{equation}
\label{eqn:polbias}
 p_{med350} = \sqrt{(1.57 \times p_{scatter350})^2 + (0.006)^2}
\end{equation}
We then adopted this calculated value as the effective polarization bias to be subtracted in quadrature from each of the measurements when doing statistical analyses.  If a source had an intrinsic polarization equal to 1.5$\times p_{scatter350}$ which would add in quadrature to the calculated value of $p_{med350}$, then the source would be found on average at the dotted line in Figure \ref{fig:detect90}.  Only three sources out of 335 exceed this value (and at least two do have well-behaved polarization behaviors), so we adopt this as our upper limit for the purposes of calculating depolarization ratios.

We note that changes to the empirical fit in Equation \ref{eqn:polbias} will have a small effect on the statistical analyses in which it is employed.  For example, if the fit value of $p_{med350}$ is overestimated the above procedure may eliminate some sources that have significant real structure in $p$(\lamsq).  However, the number of such sources is small, as discussed further below, so we ignore that issue in order to examine the depolarization properties of the sample as a whole in the following section.  Since the residual bias correction from Equation \ref{eqn:polbias} is not applied in the individual source modeling described in Section \ref{sec:modelingTechniques}, it has no effect on the outcome of those analyses.

\section{Results}
\label{sec:results}

\subsection{Polarization Properties of the Overall Sample}
\label{sec:samplePolarization}
\begin{figure}
  \centering
   \epsfig{file=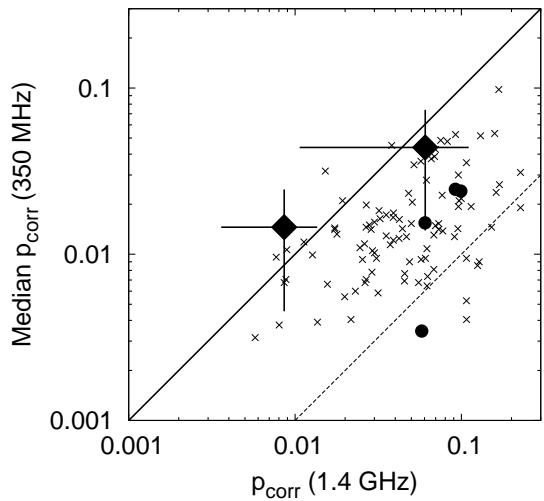,angle=0,width=0.45\textwidth}
  \caption{Plot of bias corrected median $p_{350}$ vs. bias corrected $p_{1.4}$ for 102 sources with significant polarization at 1.4 GHz, defined as $P/\sigma_P>2$.  Xs represent upper limits at 350 MHz for the median fractional polarization. Solid symbols represent the sources we modeled, except for NVSS J162740+603900 which did not have a significant detection in polarization at 1.4 GHz.  Circles represent sources whose median 350 MHz values are formally upper limits, although they were clearly detected in some IF bands.  Diamonds are significant detections at both bands, shown with their errors.}
  \label{fig:p90vp20}
\end{figure}

Starting with the sample of 335 sources discussed above, we determined their polarized fluxes in the 1.4 GHz NVSS survey.  We first convolved the NVSS $I$, $Q$ and $U$ images to the same beamsize as used in each corresponding WSRT field, then measured the $I$, $Q$ and $U$ fluxes at the locations of each total intensity peak in $I$ at 350 MHz.  Background subtraction and error estimation were performed using the same rectangular region about the source as described previously for our WSRT measurements.  We then calculated the polarized flux (and fractional polarization) after correcting for the noise bias, according to Equation \ref{eqn:pcorr}.  After correction, we found that 102 of the 335 sources had significant polarizations at 1.4 GHz ($P_{corr1.4}$/$\sigma_{P1.4}>$ 2), and for each we calculated the upper limit to their polarized fractions at 350 MHz. These are plotted in Figure \ref{fig:p90vp20}.

The upper limits on the 350 MHz polarized fractions are largely independent of the fractional polarizations at 1.4 GHz.  There is a rough upper limit to the distribution visible in Figure \ref{fig:p90vp20} likely due to the fact that at low fluxes, only large values of $p_{1.4}$ can be detected, and the upper limits on $p_{350}$ will therefore also be high. Lines of unity slope on this diagram indicate specific depolarization ratios ($p_{350}$/$p_{1.4}$).  Upper limits to the depolarization ratios vary from $<$0.03 to $<$2, with a median of $<$0.3 .

\begin{figure}
  \centering
   \epsfig{file=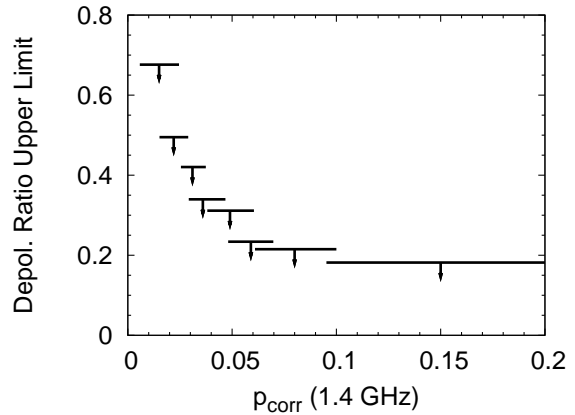,angle=0,width=0.45\textwidth}
  \caption{Plot of the upper limits to depolarization ratio from 1.4 GHz to 350 MHz vs. $p_{1.4}$ for the set of 335 sources described in Section \ref{sec:samplePolarization}.  Depolarization ratio is defined as $p_{350}/p_{1.4}$.}
  \label{fig:depolUL}
\end{figure}

In Figure \ref{fig:depolUL} we plot the median upper limit to the depolarization ratio as a function of $p_{1.4}$.  The decreasing values indicate the observational bias that we can only measure low depolarization upper limits for the highest values of $p_{1.4}$.  The median upper limit for the 20 highest $p_{1.4}$ sources is $\sim$0.2.
A more conservative requirement of $P_{corr1.4}$/$\sigma_{P1.4}>$ 3 did not change the overall distribution of the depolarization ratios.

\begin{table*}
  \caption{RM Synthesis Capabilities of WSRT}
    \centering
    \begin{tabular}{ c  c  c  c  c  c  c  c }
    \hline\hline
    Frequency & $\langle\lambda^2\rangle$ & \Deltalamsq & \lamsqmin & \deltalamsq & $\delta \phi$ & $\textnormal{max-scale}$ & $|\phi_{max}|$ \\
    (MHz) & (m$^2$) & (m$^2$) & (m$^2$) & (m$^2$) & (rad/m$^2$) & (rad/m$^2$) & (rad/m$^2$)\\
    \hline
    310-380   & 0.76 & 0.31 & 0.62 & $3.1 \times 10^{-4}$  & 12 & 5.0  & 5700  \\
    \hline
    \end{tabular}
  \label{tab:SummaryOfWSRTparms}
\end{table*}

\begin{table*}
  \caption{Table of Modeled Sources}
    \centering
    \begin{tabular}{ l  c  c  c  c }
    \hline\hline
    Source & RA      & DEC     & Off-axis           & Pos. Angle \\
           & (J2000) & (J2000) & Radius ($\arcsec$) & $(^\circ)$ \\
    \hline
    NVSS J010616+125116$^1$    & 01:06:16.8 & +12:53:22 & 2786 & 250 \\
    3C33S$^2$                  & 01:08:50.7 & +13:18:43 & 649  & 326 \\
    NVSS J011136+132437        & 01:11:36.2 & +13:25:41 & 2268 & 65  \\
    NVSS J011204+124118        & 01:12:04.5 & +12:42:39 & 2962 & 123 \\
    NVSS J125630+270108        & 12:56:30.5 & +27:01:10 & 3816 & 150 \\
    NVSS J162408+605400        & 16:24:08.8 & +60:54:04 & 3134 & 35  \\
    NVSS J162740+603900        & 16:27:41.0 & +60:39:05 & 3783 & 63  \\
    \hline
    \end{tabular}

  $^1$Resolved as double source in unconvolved NVSS image (with NVSS J010615+125210) \\
  $^2$NVSS J010850+131831
  \label{tab:interestingSources}
\end{table*}

\subsection{Model Fitting of Individual Sources}
\label{sec:modelFitting}

\subsubsection{Source Selection}
\label{sec:sourceSelection}
From the set of 335 sources (with $r<4500''$ and $I/\sigma_I>30$ at 350 MHz, and disregarding $p_{1.4}$), we identified a subset based on their Faraday Dispersion Function (FDF) using the RM Synthesis technique \citep{brentjens05}.  This allows for the best signal to noise averaging of all the data, since  $Q$(\lamsq) and $U$(\lamsq) can be summed as vectors after correcting for each assumed RM.

The observed FDF, $\tilde{F}(\phi)$, is constructed (using the formalism of \citealt{brentjens05}) thusly:
\begin{eqnarray}
 \tilde{F}(\phi) = F(\phi) * R(\phi) & = & K \sum\limits_{i}^{N} w_i P_i e^{-2i\phi(\lambda^2-\lambda^2_0)} \\
 R(\phi) & = & K \sum\limits_{i}^{N} w_i e^{-2i\phi(\lambda^2-\lambda^2_0)} \\
 K       & = & \left( \sum\limits_{i}^{N} w_i \right)^{-1}
\end{eqnarray}
at an arbitrary Faraday depth, $\phi$, which replaces the usual rotation measure; in practice one chooses a range of Faraday depths to reconstruct a Faraday spectrum.  The quantities $P_i$ and $w_i$ are the observed vector polarization and applied weight, respectively, at locations of sampled $\lambda^2$.  The quantity $\lambda^2_0$ is the mean $\lambda^2$ of the set of observations, and the reconstructed FDF is represented at $\lambda^2 = \lambda^2_0$.  Note that the actual $F(\phi)$ is obtained by deconvolving the Rotation Measure Spread Function (RMSF, $R(\phi)$), which is the normalized response function in Faraday space, from the observed $\tilde{F}(\phi)$.  We briefly discuss the deconvolution procedure, RM Clean, in Section \ref{sec:modelingTechniques}.  Unless otherwise noted, all FDFs and RMSFs in this study were constructed using uniform weighting.
For this paper we use an over-tilde to represent transformed polarization quantities unless otherwise noted, e.g., $\tilde{P}$ represents the magnitude of the FDF, $\tilde{Q}$ represents the real part of the FDF, and so on.

We used all channels where sky noise in Stokes $Q$ and $U$ were $\leq5$ mJy/beam (uncorrected for primary beam attenuation), with the number of channels listed in Table \ref{tab:NumberOfChannelsUsedForBandAveraging}.
A typical RMSF for the WSRT 350 MHz band is shown in Figure \ref{fig:1000.rmsf}; this RMSF was constructed using roughly 400 channels across the full band.
The main lobe of each RMSF had a characteristic FWHM $\sim$12 rad/m$^2$.  Nominal RM Synthesis capabilities of the WSRT 350 MHz band are given in Table \ref{tab:SummaryOfWSRTparms}.  We note that no instrumental polarization correction has been applied to the data used to construct these FDFs, since the corrections were determined only when the channels were averaged within each IF band.

A coarse initial search over Faraday depths between $\pm1000$ rad/m$^2$ was performed at a resolution of 10 rad/m$^2$.
Once we had determined that no significant power existed outside a Faraday depth of $\pm200$ rad/m$^2$, we performed a finer search between $\pm200$ rad/m$^2$ with a resolution of 1 rad/m$^2$.

To make an initial cut to the set of sources, the location ($\phi_{max}$) and amplitude ($\tilde{P}_{max}$) of the peak in $\tilde{P}(\phi)$ were determined for each source, and those with a peak amplitude of $\tilde{P}_{max} \geq 3$ mJy/beam/RMSF (uncorrected for primary beam attenuation) were selected.
In all, 116 of the original 335 sources passed this criterion, with a minimum signal to noise in $\tilde{P}_{max}$ of 3.4.  This removed many sources from the sample whose observed polarization may be enhanced artificially, e.g., by noise, which places power at all Faraday depths in the FDF.
We note that many of these remaining detections are due to instrumental polarization which is not corrected in the all-channel FDF.

\begin{figure*}
  \centering
  \begin{tabular}{|c|c|}
  \hline
  \epsfig{file=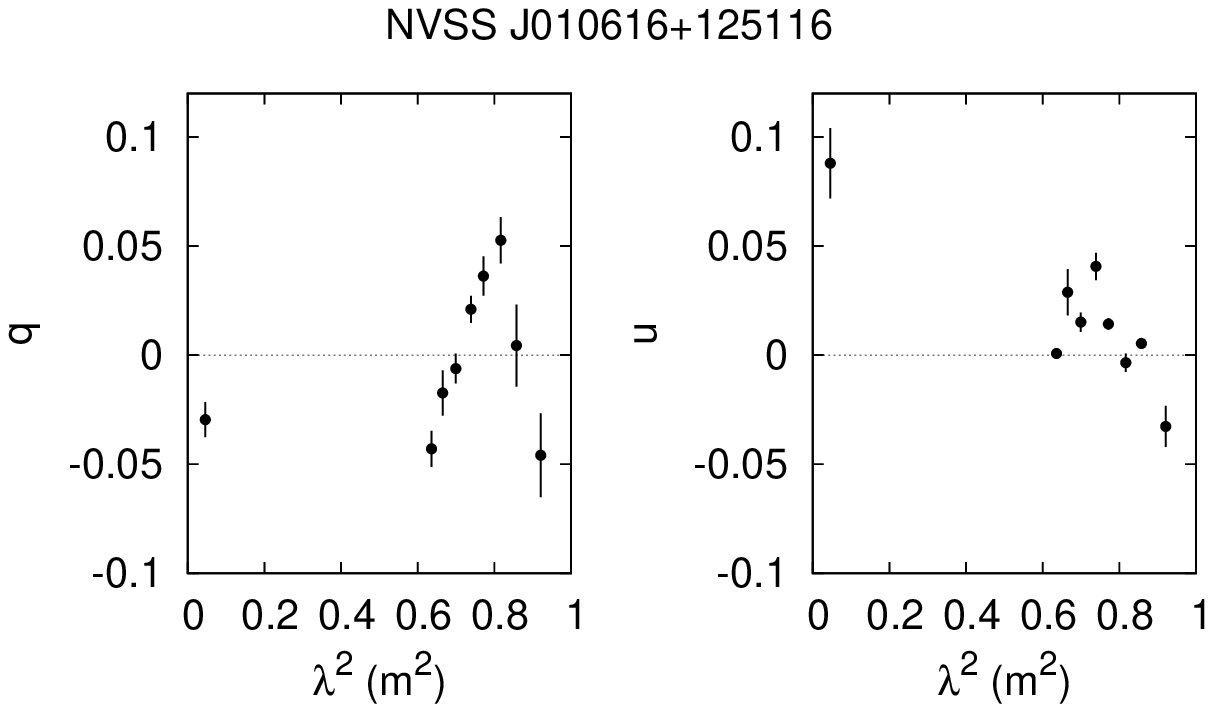,width=0.45\textwidth} &
  \epsfig{file=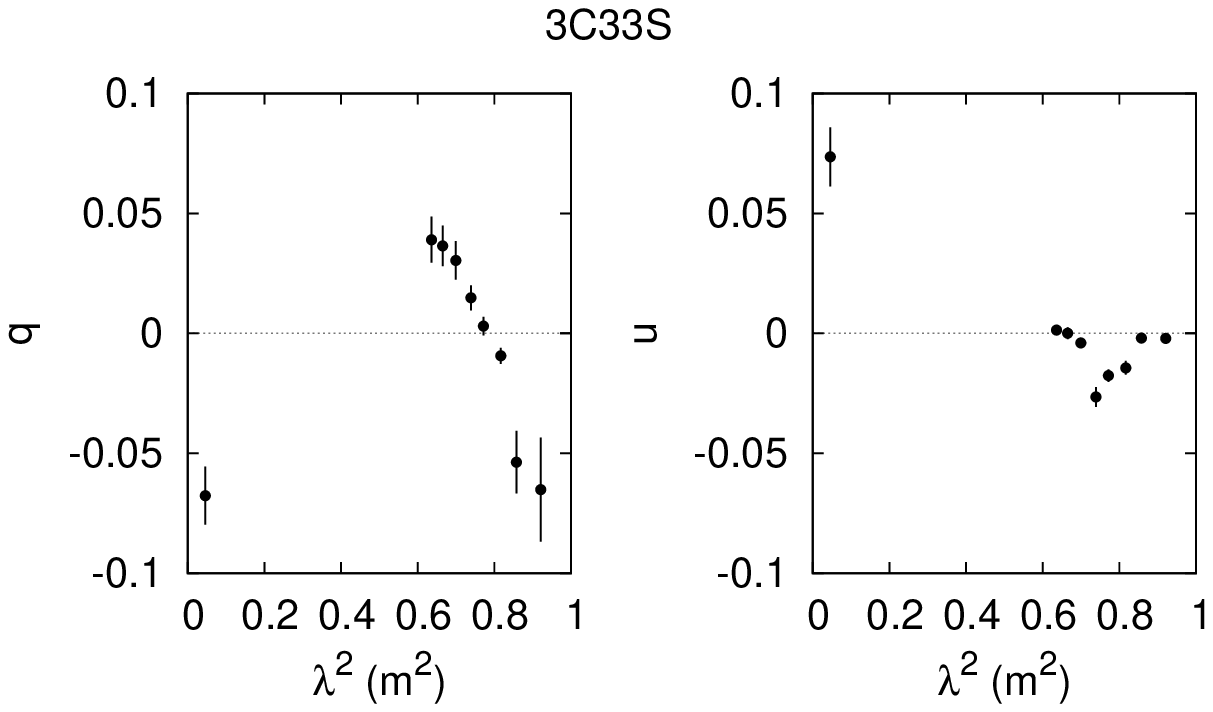,width=0.45\textwidth} \\
  \hline
  \epsfig{file=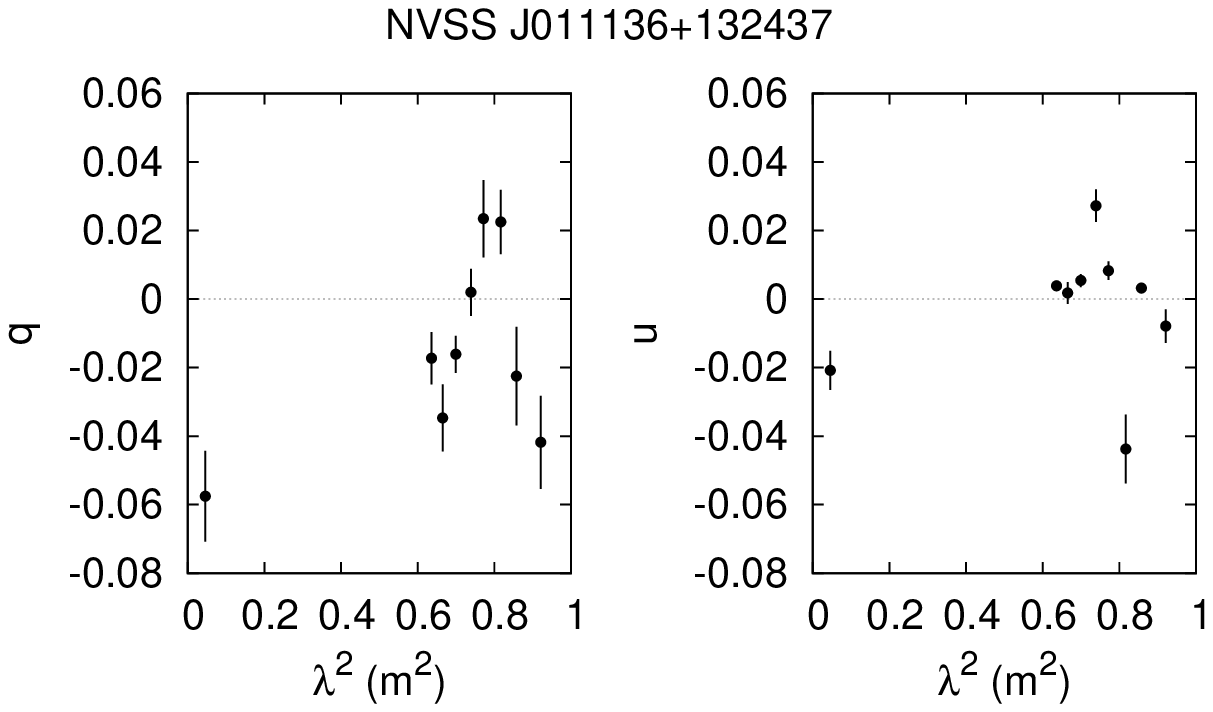,width=0.45\textwidth} &
  \epsfig{file=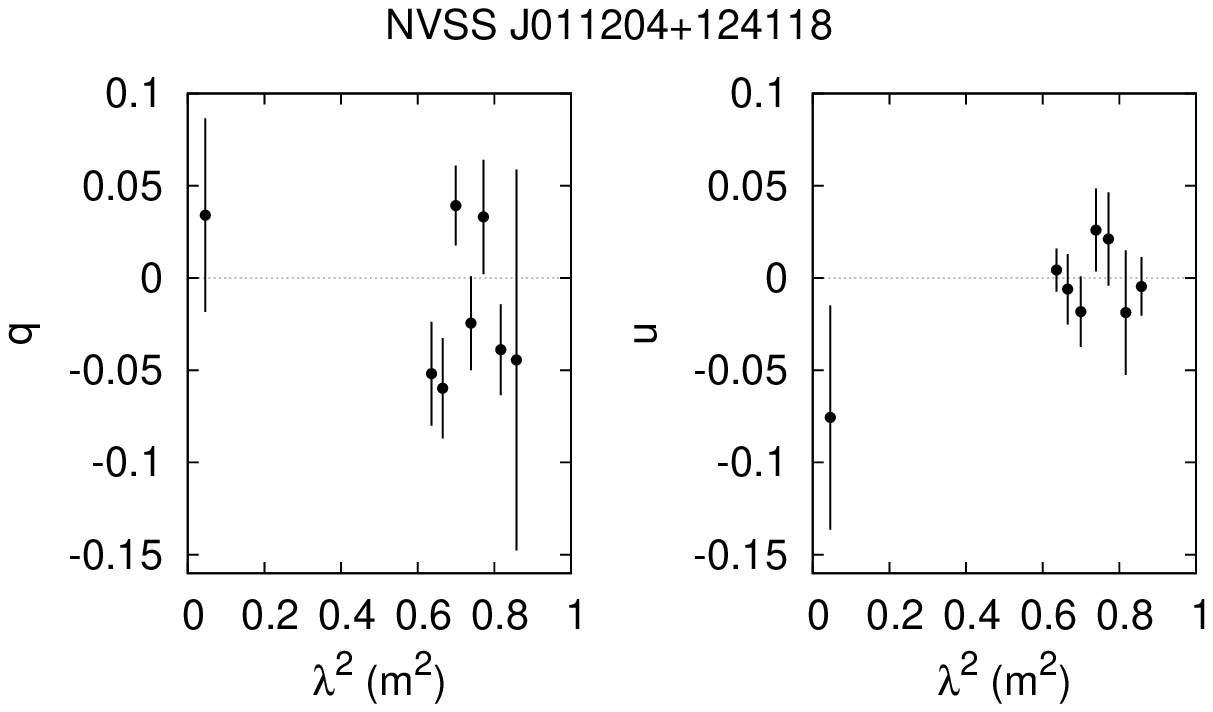,width=0.45\textwidth} \\
  \hline
  \epsfig{file=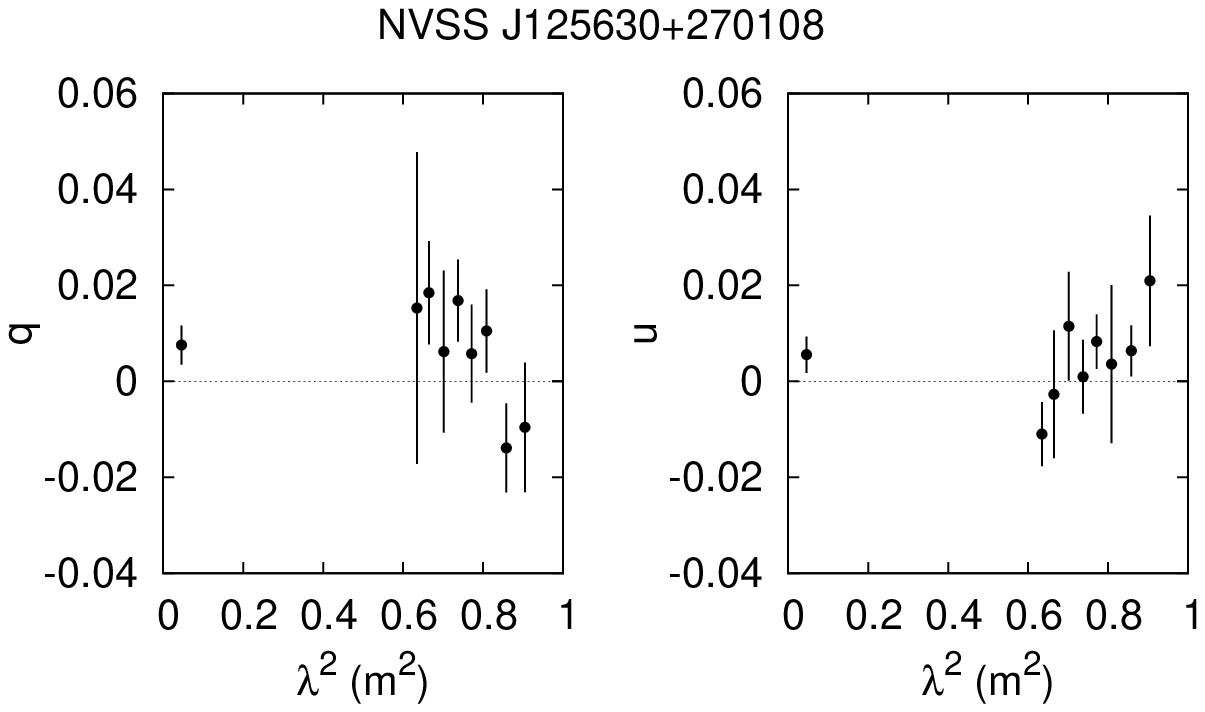,width=0.45\textwidth} &
  \epsfig{file=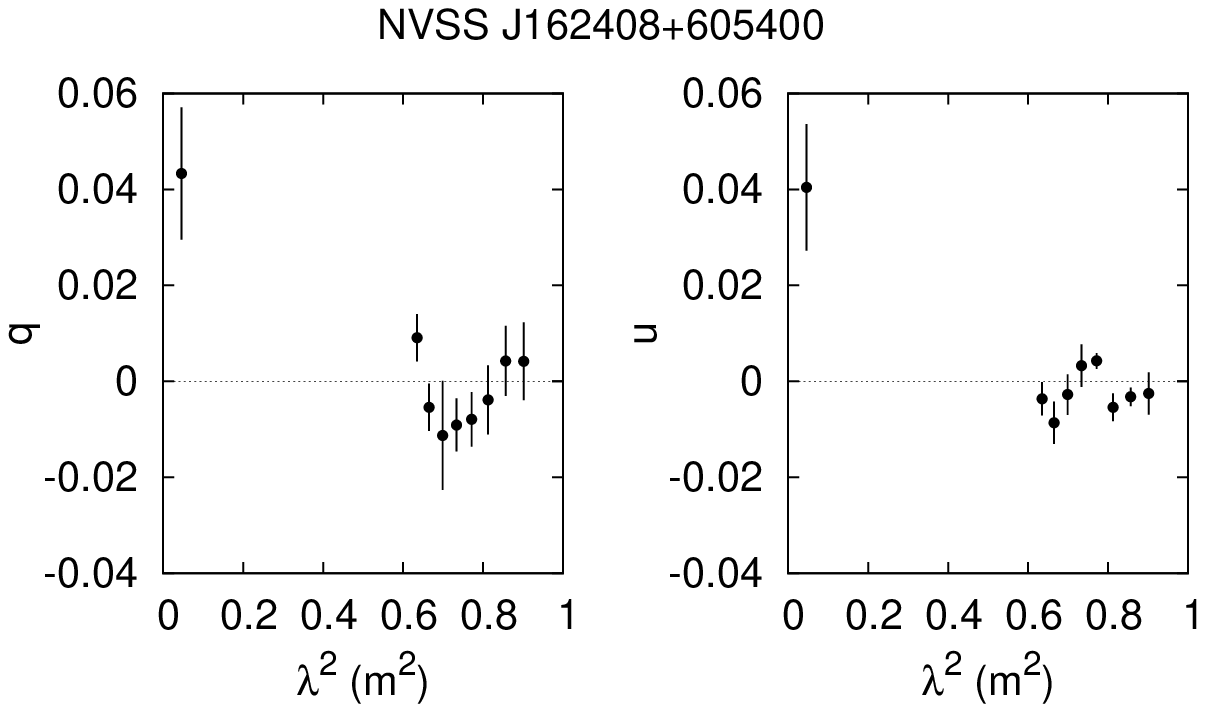,width=0.45\textwidth} \\
  \hline
  \epsfig{file=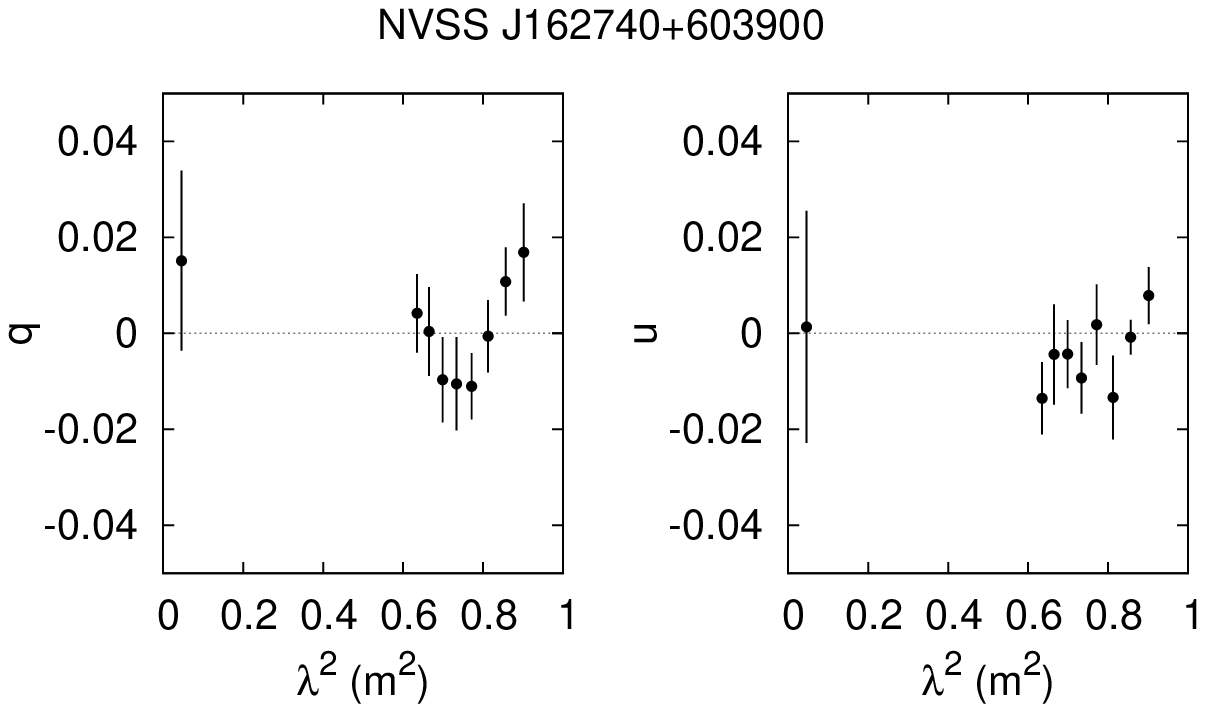,width=0.45\textwidth} & \\
  \hline
  \end{tabular}
  \caption{Observed NVSS + WSRT IF band averaged $q$(\lamsq) and $u$(\lamsq) for the seven sources modeled.  Background subtraction and removal of WSRT instrumental polarization has been performed for each source, as described in the text.}
  \label{fig:qandu}
\end{figure*}

For each of these 116 sources, we then examined the IF averaged $Q$ and $U$ measurements, corrected for instrumental polarization. Because sources could have different fractional polarizations for different IFs, we did not demand that they have strong signals in all IFs. Sources with at least four of the eight IFs satisfying $P/\sigma_P \geq 2$, and $U/\sigma_U \geq 4$ were then selected from the list of 116.
We used only the $U$ data for this cut because of the greater uncertainty in the instrumental correction for $Q$ and the presence of occasional spuriously high $Q$ values.
Sources which exhibited a regular modulation in $p(\lambda^2)$ corresponding to to the $\sim$17 MHz modulation found by \cite{popping08} were excluded.
All such sources were found beyond $R\sim4000''$, evidence of residual instrumental polarization not fully accounted for by our model.
Only three sources met all of these criteria.  To those, we added four additional sources for modeling based on their high ratios of $p_{med350}$/$p_{scatter350}$, putting them at or above the upper limit line shown in Figure \ref{fig:detect90}. These seven sources selected for modeling are listed in Table \ref{tab:interestingSources} along with selected properties from the literature.  Plots of $q$(\lamsq) and $u$(\lamsq) are shown in Figure \ref{fig:qandu}.

\subsubsection{A Note on Bandwidth Depolarization}
The NVSS data at 1.4 GHz were constructed from two 42 MHz wide bands, centered at 1364.9 MHz and 1435.1 MHz.  Bandwidth depolarization for sources with $|$RM$|\lesssim50$ \radmsq~ would yield $(p_0-p)/p_0\lesssim2\%$ in the NVSS \citep{condon98}.  Any source with $|$RM$|$ high enough to suffer significant bandwidth depolarization in the NVSS would be severely depolarized in the IF averaged 350 MHz data, and would not have been selected for further investigation.

\subsubsection{Modeling Techniques}
\label{sec:modelingTechniques}
As can be seen in Figure \ref{fig:qandu} the sources found to have sufficient polarization for modeling all showed structure in $q$(\lamsq) and $u$(\lamsq) inconsistent with a simple Faraday screen.  A simple screen would result in sine and cosine waves in $q$ and $u$ with matched frequencies and amplitudes.  In order to measure the Faraday structure of these sources, we therefore explored a variety of techniques.  In particular, we used:  A. Linear fit to $\chi(\lambda^2)$;  B. RM Synthesis/Clean;  C. Model fitting to $q$, $u$ vs. \lamsq \space using a two component foreground screen;  and D. Model fitting using a single foreground screen with a mean RM and a separate depolarizing function. We have omitted an internal Faraday dispersion model for the following reason:  internal depolarization in the Milky Way and nearby galaxies arises because the synchrotron and thermal plasmas are well mixed \citep{sokoloff98}.  This is not true for extragalactic sources, where the depolarization almost always arises with Faraday variations across the beam (e.g., \citealt{tribble91} and references therein). We now briefly discuss each of the models employed followed by the results.

\emph{A. $\chi(\lambda^2)$.}  We determined the RM for each source using the most common method, minimizing the sum of the weighted residuals (i.e. chi-squared statistic, $\chi^2$) from fitting Equation \ref{eqn:linearChiEqn} to the observed polarization angles $\chi(\lambda^2)$. There were often a number of different solutions with comparable values of $\chi^2_{min}$ based on our choices for the n$\pi$ ambiguities.  We therefore made these choices to most closely match the results for the RM for the foreground depolarizing screen model described below.  We calculated the errors in RM by the standard propagation of errors from the residuals to the fit, not from the errors in the original data points.  Note that the reduced $\chi^2$ values, $\chi^2_{\nu} \equiv \chi^2 / dof$ (where $dof$ = degrees of freedom), as listed in Tables \ref{tab:1037tab} - \ref{tab:3063tab} are generally quite high, suggesting that these are not good fits, despite the apparently small derived errors in RM.

\begin{figure}
  \centering
  \epsfig{file=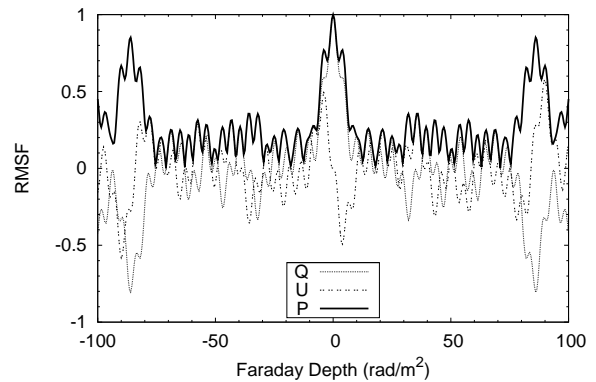,angle=0,width=0.45\textwidth}
  \caption{Rotation Measure Spread Function for a typical set of NVSS + 8 WSRT IF measurements.  The RMSF is the normalized (unitless), complex response to polarized emission in Faraday space for a given set of \lamsq \space sampling.}
  \label{fig:1060.rmsf.9samples}
\end{figure}

\emph{B. RM Synthesis/Clean.}  For each of the seven sources the FDF was constructed using the $q$,$u$ data, this time using the instrumental polarization corrected IF samples from the WSRT observations plus the NVSS data point.  Uniform weighting for all \lamsq~ samples was applied; we experimented with various weighting of the WSRT and NVSS samples used as input for RM Synthesis, but found negligible differences in the RM Clean solutions.
A representative RMSF is shown in Figure \ref{fig:1060.rmsf.9samples}, displaying the sidelobe structure due to the sparse \lamsq~ sampling.
The range of Faraday depth for the constructed FDF was this time limited to $\pm50$ \radmsq, reflecting the maximum RM due to the \lamsq~ separation of the IF averaged samples.
The full-channel FDFs were first searched for significant power beyond $\pm50$ \radmsq~ to ensure that this range of Faraday depths was large enough.
Our custom version of RM Clean (\citealt{brentjens05}, \citealt{heald09}) was used to deconvolve the complex RMSF from the FDF, drastically reducing sidelobes and producing a more lucid representation of the Faraday structure.
We used a gain factor of 0.1 and stopping criteria of either 200 iterations or a peak to RMS ratio of 1.5 in the residuals of $\tilde{p}$.
These convergence criteria were found to strike the optimal balance between minimizing the residuals and limiting spurious clean components.

\begin{figure}
  \centering
  \epsfig{file=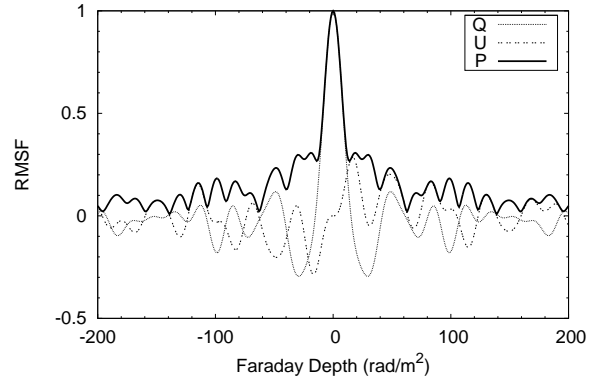,angle=0,width=0.45\textwidth}
  \caption{Rotation Measure Spread Function for a typical set of channels in the WSRT 350 MHz band.  The RMSF is the normalized (unitless), complex response to polarized emission in Faraday space for a given set of \lamsq~ sampling.  Roughly 400 channels were used to construct this RMSF.}
  \label{fig:1000.rmsf}
\end{figure}

The location (RM) and amplitude ($\tilde{p}_0$) of the two most dominant features in each cleaned FDF were extracted by Gaussian fitting to the cleaned Faraday spectrum.
To reduce polarization bias, which would enhance the amplitude of $\tilde{p}_0$ solutions, we subtracted the mean of the residuals in $\tilde{p}$ before performing the Gaussian fitting.
As determined in some of our experiments, and also noted by \cite{frick10}, the method of RM Synthesis/Clean has difficulty reproducing the correct phase information in the presence of multiple RM components.
For this reason, we have neglected $\chi$ in the solutions by fitting to $\tilde{p}$ only.  To exclude possible residual instrumental RM, which manifest near $\pm42$ \radmsq~ due to the 17 MHz modulation investigated by \cite{popping08}, we searched for components in the range $|$RM$|$ $<$ 40 \radmsq.
The RM Synthesis/Clean solutions for each source are summarized in Tables \ref{tab:1037tab} - \ref{tab:3063tab}.

\begin{figure}
  \centering
  \epsfig{file=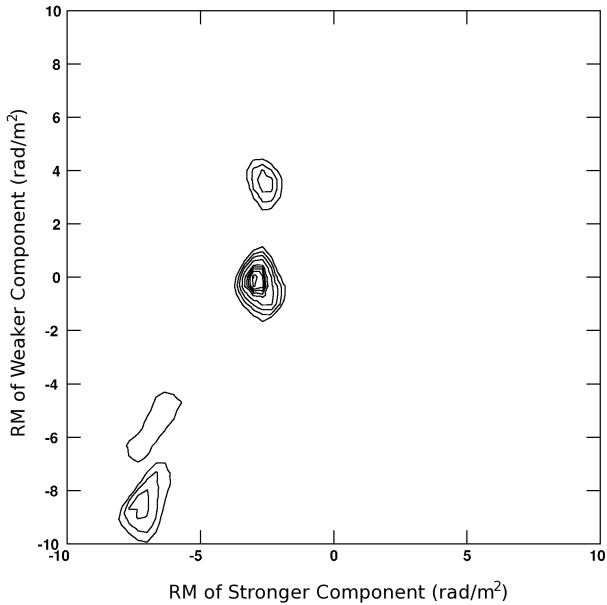,angle=0,width=0.45\textwidth}
  \caption{A cut through the $\chi^2$ surface for our $q$, $u$ vs. \lamsq \space model grid search for 3C33S.  The contours show the deepest minima in the surface, with the best fit RMs near -3 rad/m$^2$ (stronger component) and 0 rad/m$^2$ (weaker component).  Contour levels are at the probabilities of $10^{-6}$, $10^{-5}$, $10^{-4}$, $10^{-3}$, $10^{-2}$, $2.5\times10^{-2}$, $5\times10^{-6}$, $10^{-1}$.}
  \label{fig:RMgrid}
\end{figure}

\emph{C. 2 component models for $q$(\lamsq), $u$(\lamsq).}  This fit involves six parameters, with the amplitude of the fractional polarization, $p_0$, intrinsic polarization angle, $\chi_0$, and RM to be determined for each of two components. Because we expected (and sometimes found) multiple minima in $\chi^2$ in this six dimensional space, we minimized $\chi^2$ through a direct search of parameter space.  The explored ranges were tailored somewhat to the individual sources, but typical values were polarized fraction (0, 0.1), RM (-25, 25) \radmsq, and $\chi_0$ (0, 180) degrees. The values presented in Tables \ref{tab:1037tab} - \ref{tab:3063tab} represent the minimum of $\chi^2$ over these ranges.  Note that there are no n$\pi$ ambiguities when fits are done in $q$,$u$ space.  A slice through this $\chi^2$ surface for the two RM parameters for 3C33S is shown as an example in Figure \ref{fig:RMgrid}.  Each value in this space represents the minimum value of $\chi^2$ for fixed values of the two RMs, with all other parameters allowed to float.

The errors in RM were calculated by normalizing $\chi^2_{min}$, defining $\tilde{\chi}^2_{min}\equiv~ dof$. We then found the range of each of the two RMs for which the value of the normalized $\tilde{\chi}^2 \leq (dof + 1)$, allowing the other five parameters to float.  In a number of cases, there were additional minima within the $\tilde{\chi}^2 < (dof + 1)$ range, so no errors are quoted and these RM values are shown in brackets.  This procedure, of determining errors by adding 1 to the $\chi^2_{min}$ has a long history in the astrophysical literature (e.g., \citealt{avni76}, \citealt{wall96}), but has very serious problems as discussed below.  The probability contour levels in Figure \ref{fig:RMgrid} were assigned using the $\chi^2$ distribution for one degree of freedom \citep{lampton76} as is appropriate when assigning errors to each individual RM, and allowing the other RM and all other parameters to float.

\emph{D. Foreground rotation and depolarizing screen models for $q$(\lamsq), $u$(\lamsq).}  We followed a procedure similar to that of the two component model, finding the minimum $\chi^2$ for the three parameter function
\begin{eqnarray}
\mathbf{p}(\lambda^2) & = & q(\lambda^2)+iu(\lambda^2) \nonumber \\
                      & = & p_0 \exp (-2\sigma_{RM}^2\lambda^4) \exp [2i(\chi_0+\lambda^2 \rm{RM})]
\end{eqnarray}
similar to that described by \cite{burn66}. Errors in RM were determined in the same way as the two component model.

We quote errors on RM using methods similar to those in the literature, so that our uncertainties can be compared to them.  However, it is very rare in the literature to find $\chi^2$ values quoted for the fits, and therefore difficult to evaluate whether the models used are appropriate or not.  As we will show below, it is possible to get quite robust $\chi(\lambda^2)\propto\lambda^2$ behavior given an apparent RM quite different from the actual RMs for two component models.  \textit{Therefore, RM determinations using only $\chi(\lambda^2)$ and ignoring the fractional polarization behavior can provide no guidance regarding the appropriateness of the fit.}

The $\chi^2_{\nu}$ values in Tables \ref{tab:1037tab} - \ref{tab:3063tab} are often much greater than unity, showing that these models are not an adequate representation of the data.  In that case, the meaning of our errors is unclear. Our normalization of $\tilde{\chi}^2_{min}\equiv~ dof$ produces much more conservative errors than simply adding 1 to $\chi^2_{min}$.  However, as pointed out by \cite{lampton76}, this method produces the equivalent of a ratio of variances distribution, which has a very different probability distribution than $\chi^2$ itself.  In particular, they say \emph{``We stress again that if ($\chi^2_{min} >> dof$), \emph{no} formalism which uses distributions describing random fluctuations can provide the proper error estimator."}  Given this, our errors must be accepted only in the sense of providing comparisons to the literature, and our recommendations for future work are described in Section \ref{sec:recommendations}.

\subsubsection{Model Results and Comparisons}
\label{sec:modelResults}
\begin{figure}
 \centering
  \begin{tabular}{|c|}
  \hline
 \epsfig{file=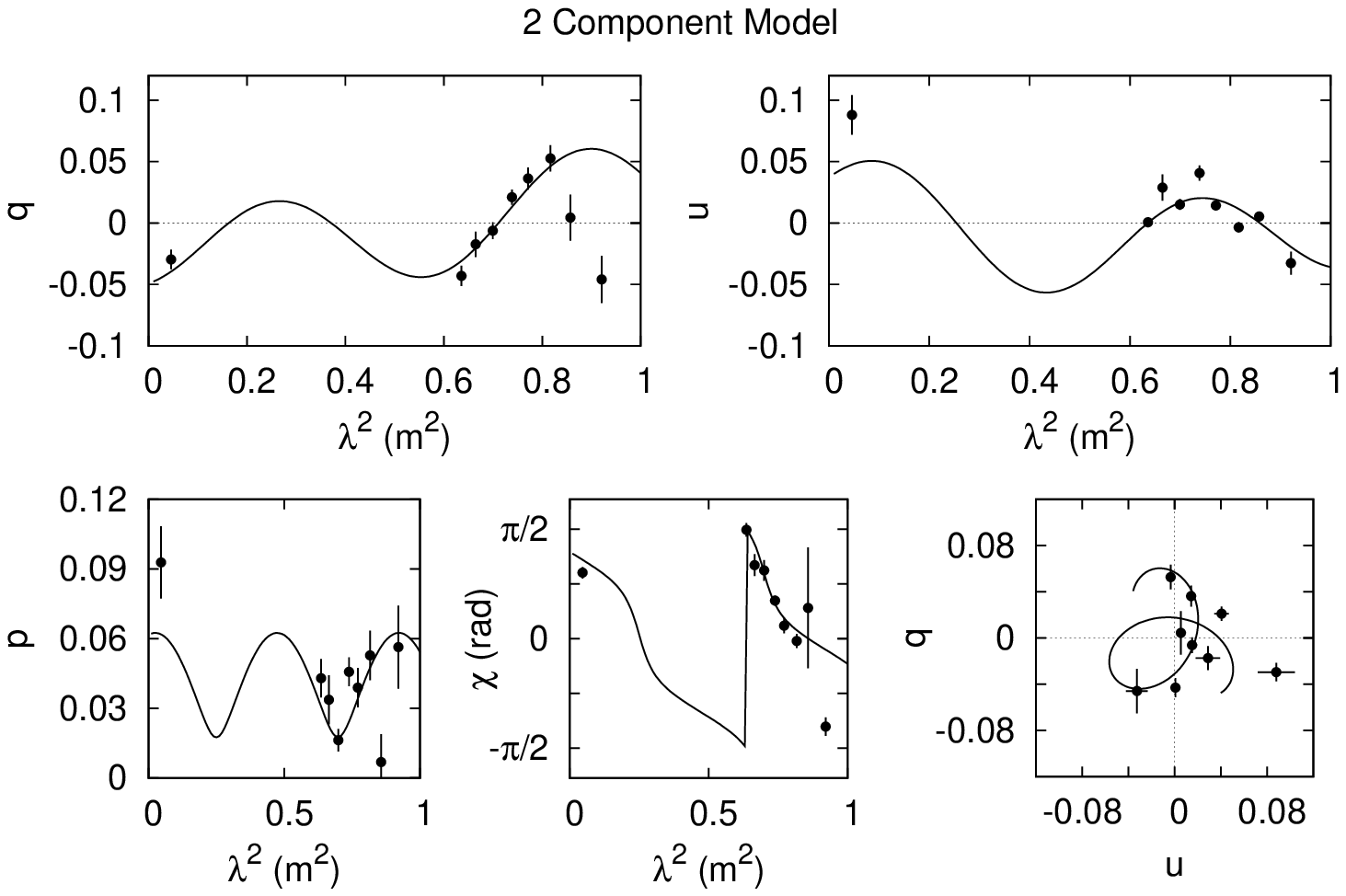,angle=0,width=0.45\textwidth} \\
  \hline
 \epsfig{file=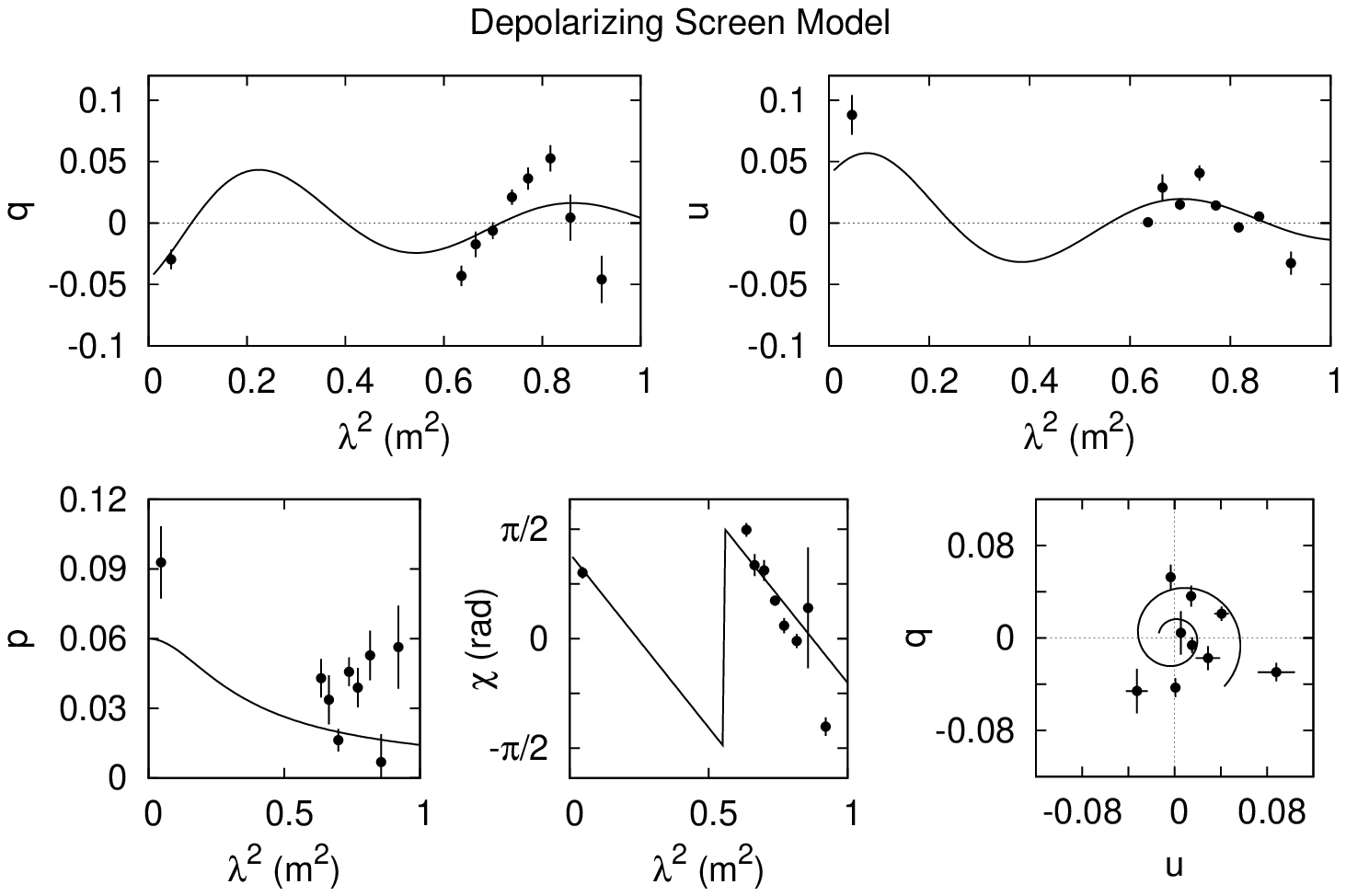,angle=0,width=0.45\textwidth} \\
  \hline
 \epsfig{file=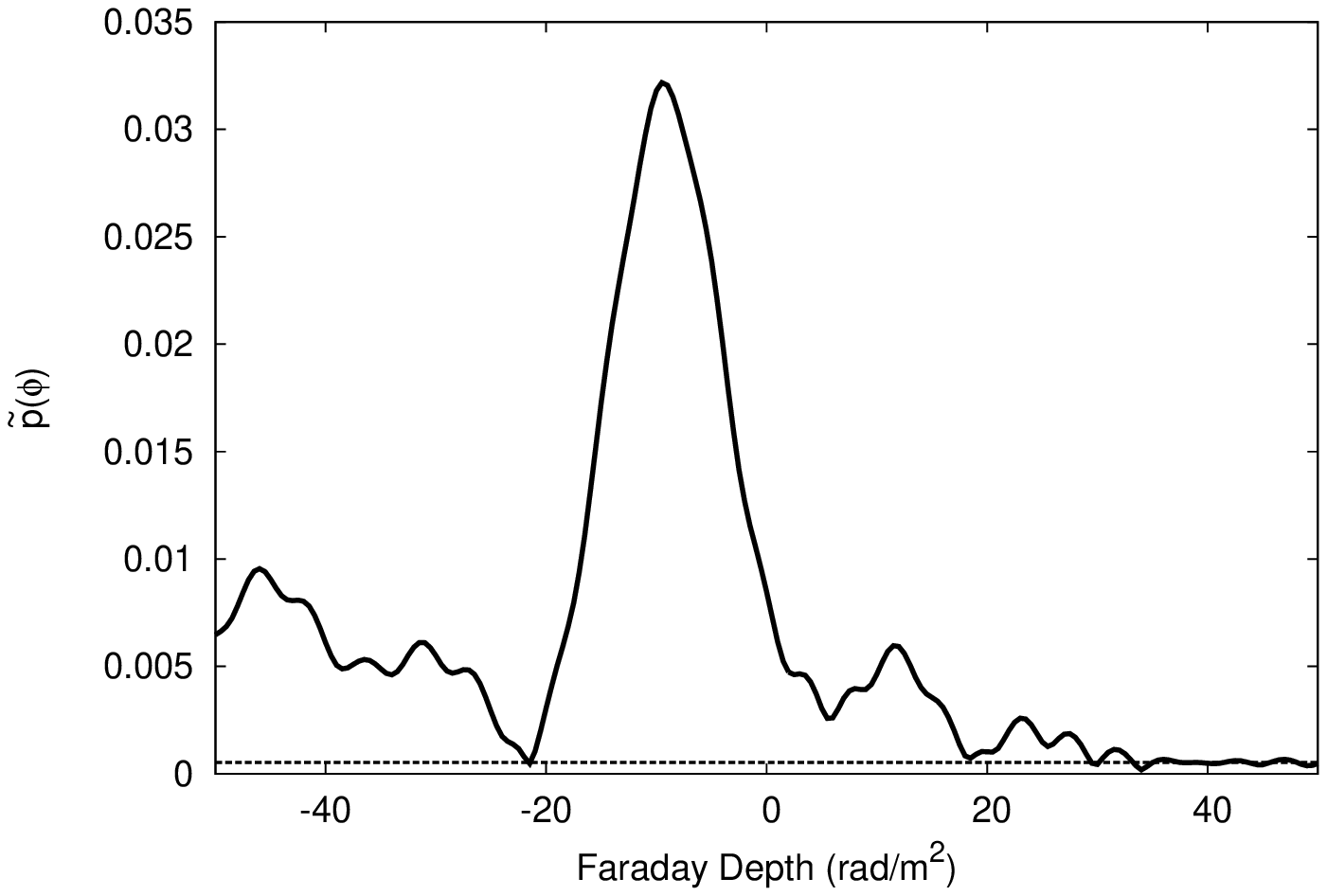,angle=0,width=0.45\textwidth} \\
  \hline
  \end{tabular}
 \caption{Polarization diagnostics for NVSS J010616+125116.  Model fits (lines) are plotted over the observed data (points).  Top panel: two component model. Middle panel: depolarizing screen. Bottom panel: magnitude of the cleaned fractional FDF (solid line) and rms of the residuals (horizontal dashed line).  The linear $\chi$(\lamsq) fit is omitted since it is nearly identical to the depolarizing screen.}
  \label{fig:1037.polDiags}
\end{figure}

\begin{figure}
 \centering
  \begin{tabular}{|c|}
  \hline
 \epsfig{file=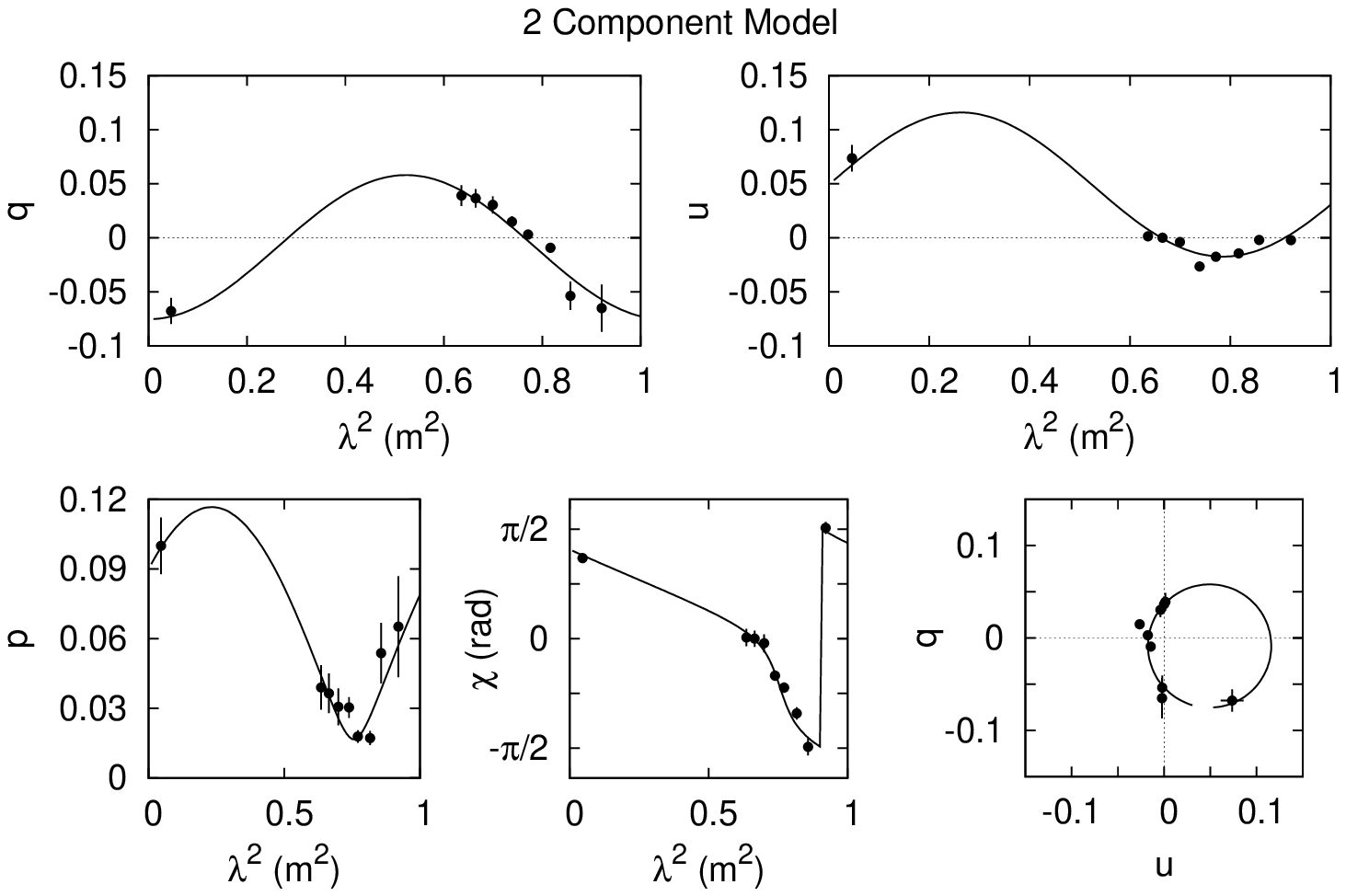,angle=0,width=0.45\textwidth} \\
  \hline
 \epsfig{file=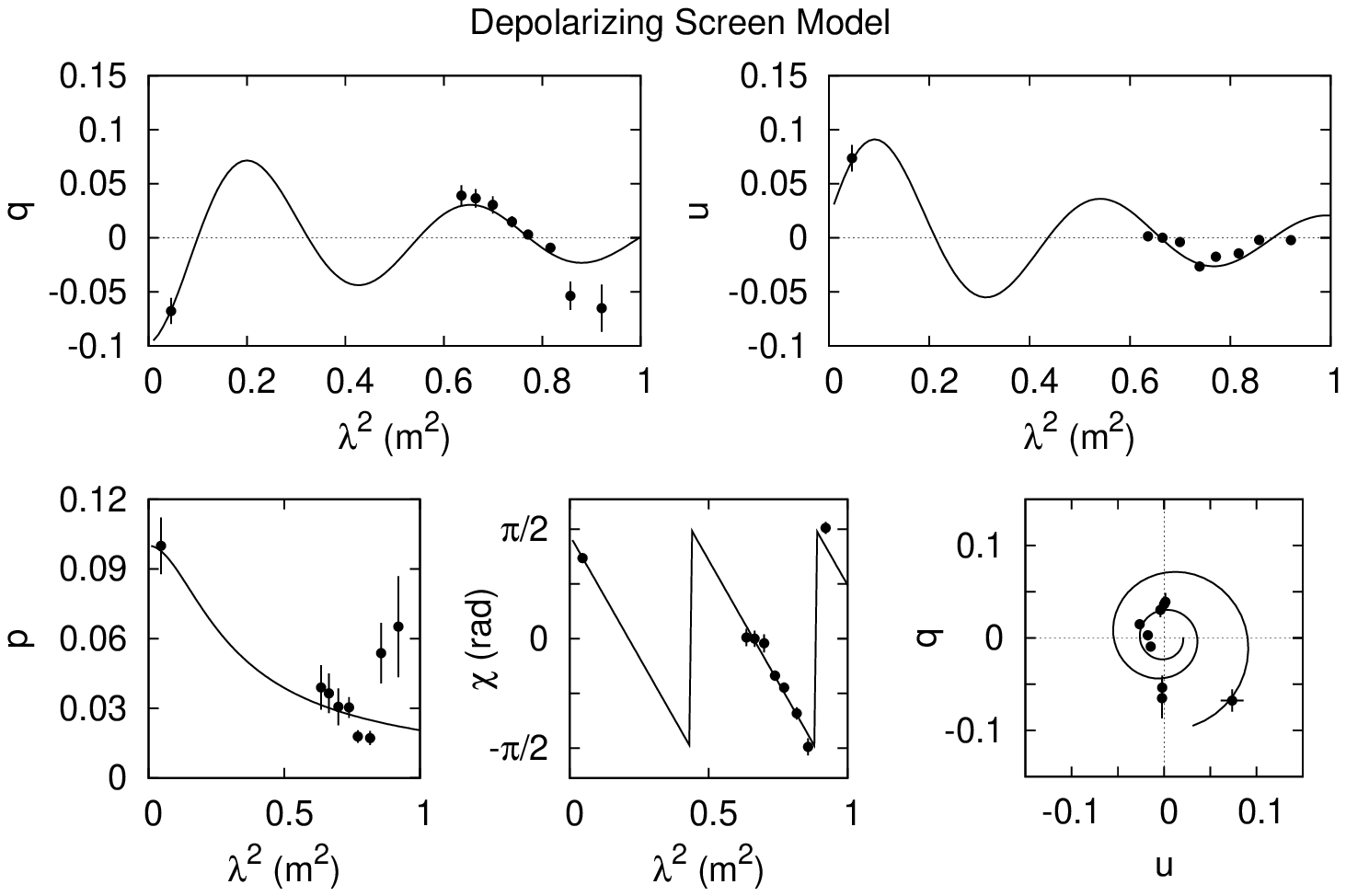,angle=0,width=0.45\textwidth} \\
  \hline
 \epsfig{file=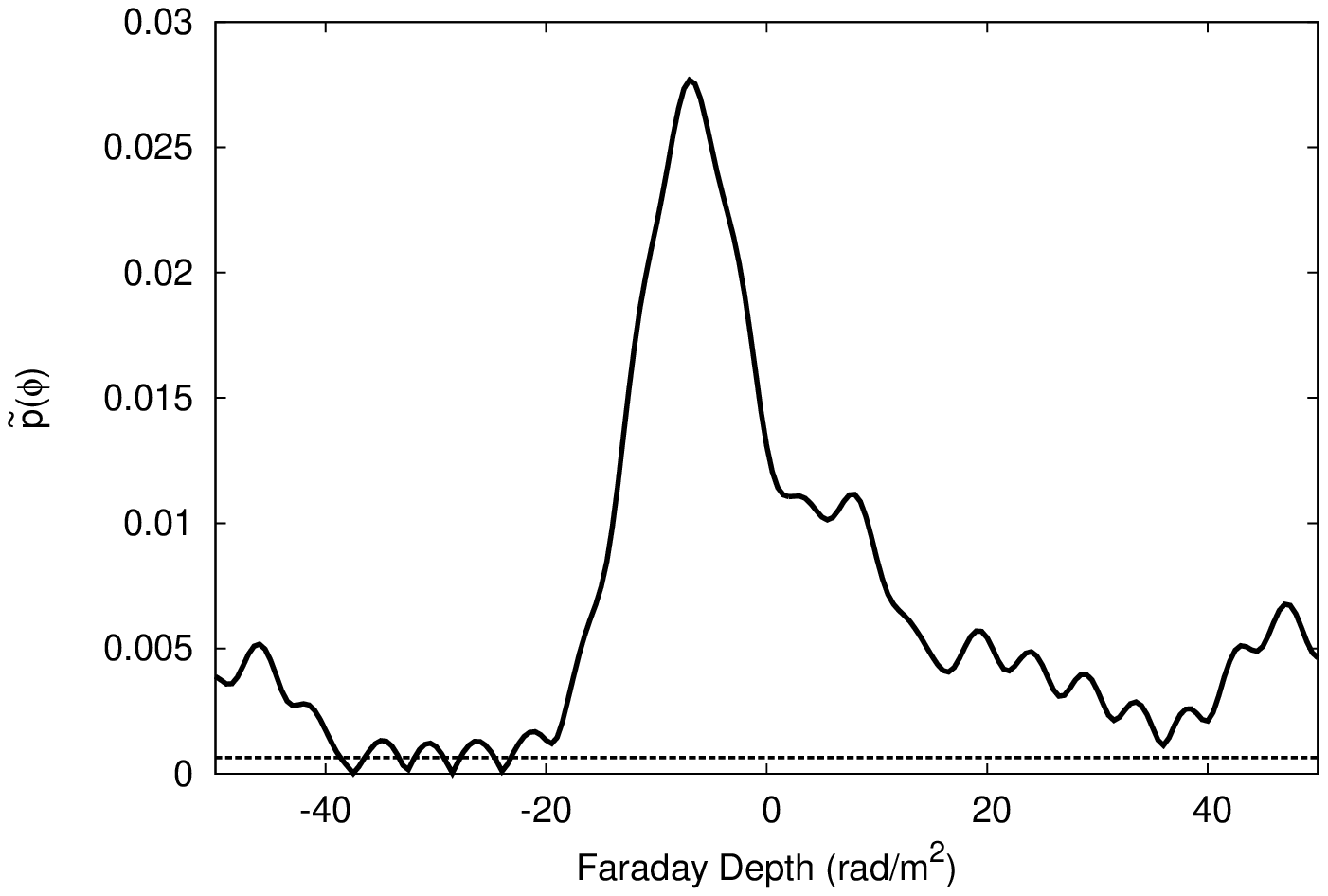,angle=0,width=0.45\textwidth} \\
  \hline
  \end{tabular}
 \caption{Polarization diagnostics for 3C33S.  Same layout as Figure \ref{fig:1037.polDiags}.}
 \label{fig:1060.polDiags}
\end{figure}

\begin{figure}
 \centering
  \begin{tabular}{|c|}
  \hline
 \epsfig{file=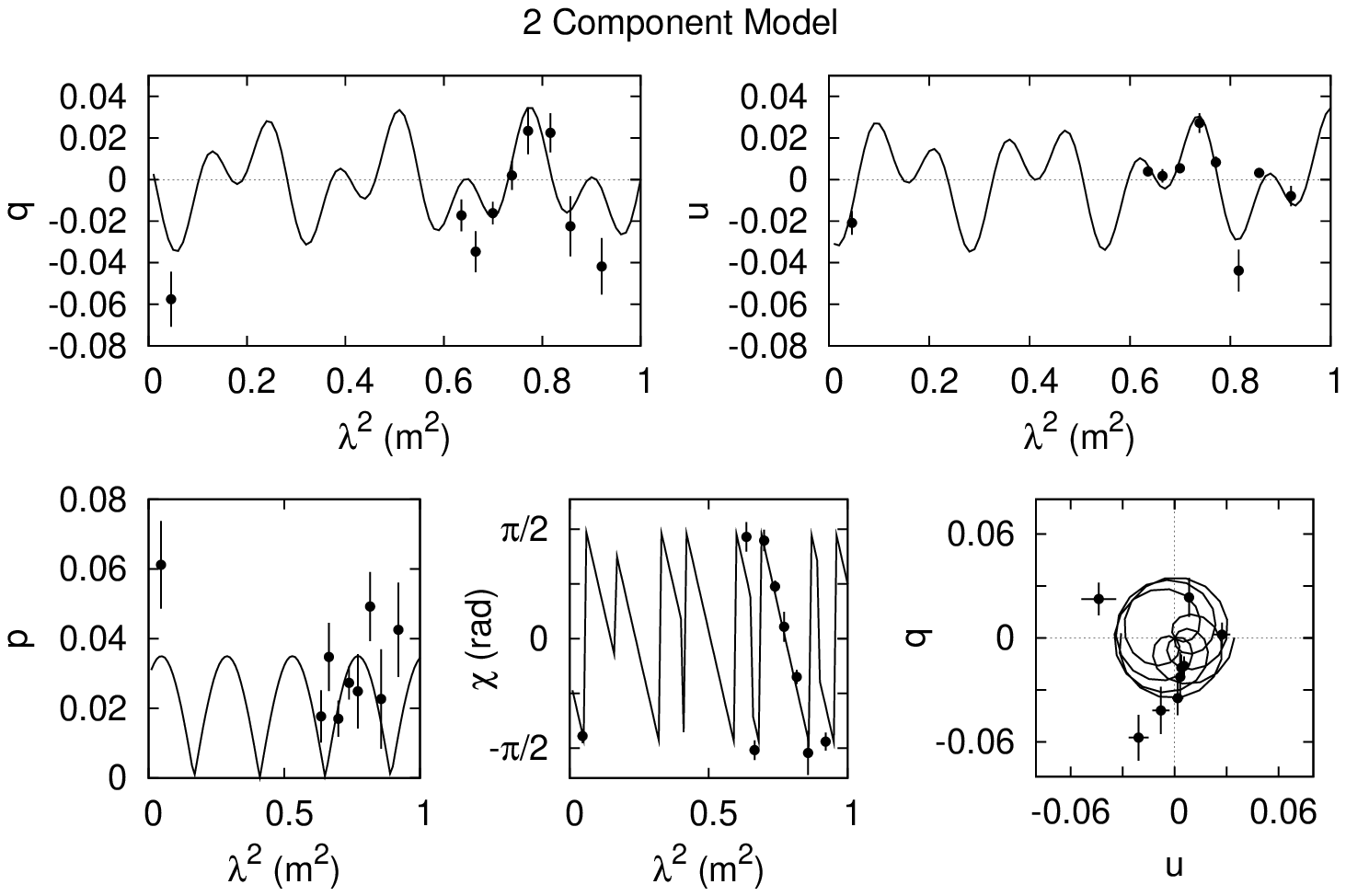,angle=0,width=0.45\textwidth} \\
  \hline
 \epsfig{file=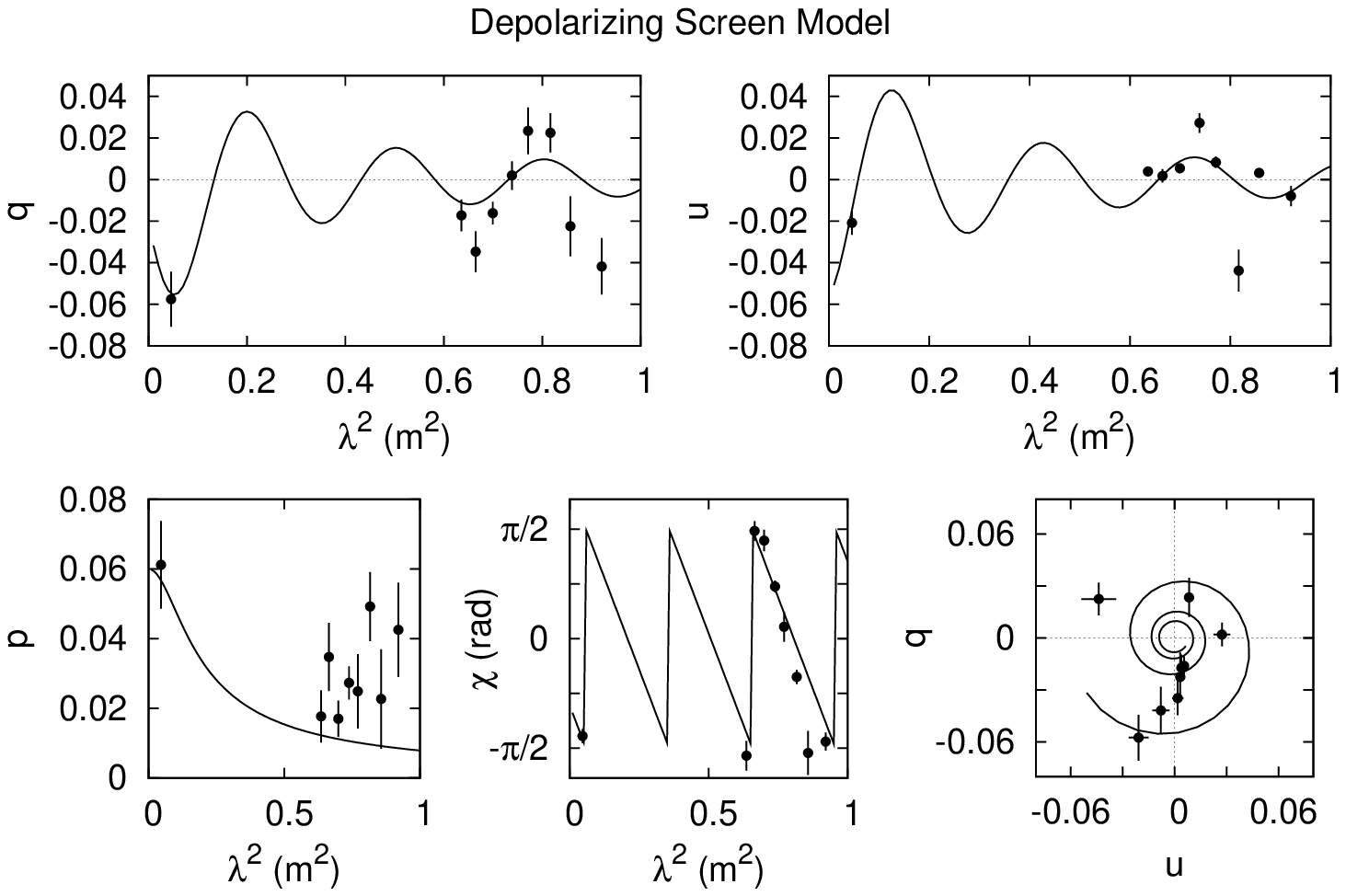,angle=0,width=0.45\textwidth} \\
  \hline
 \epsfig{file=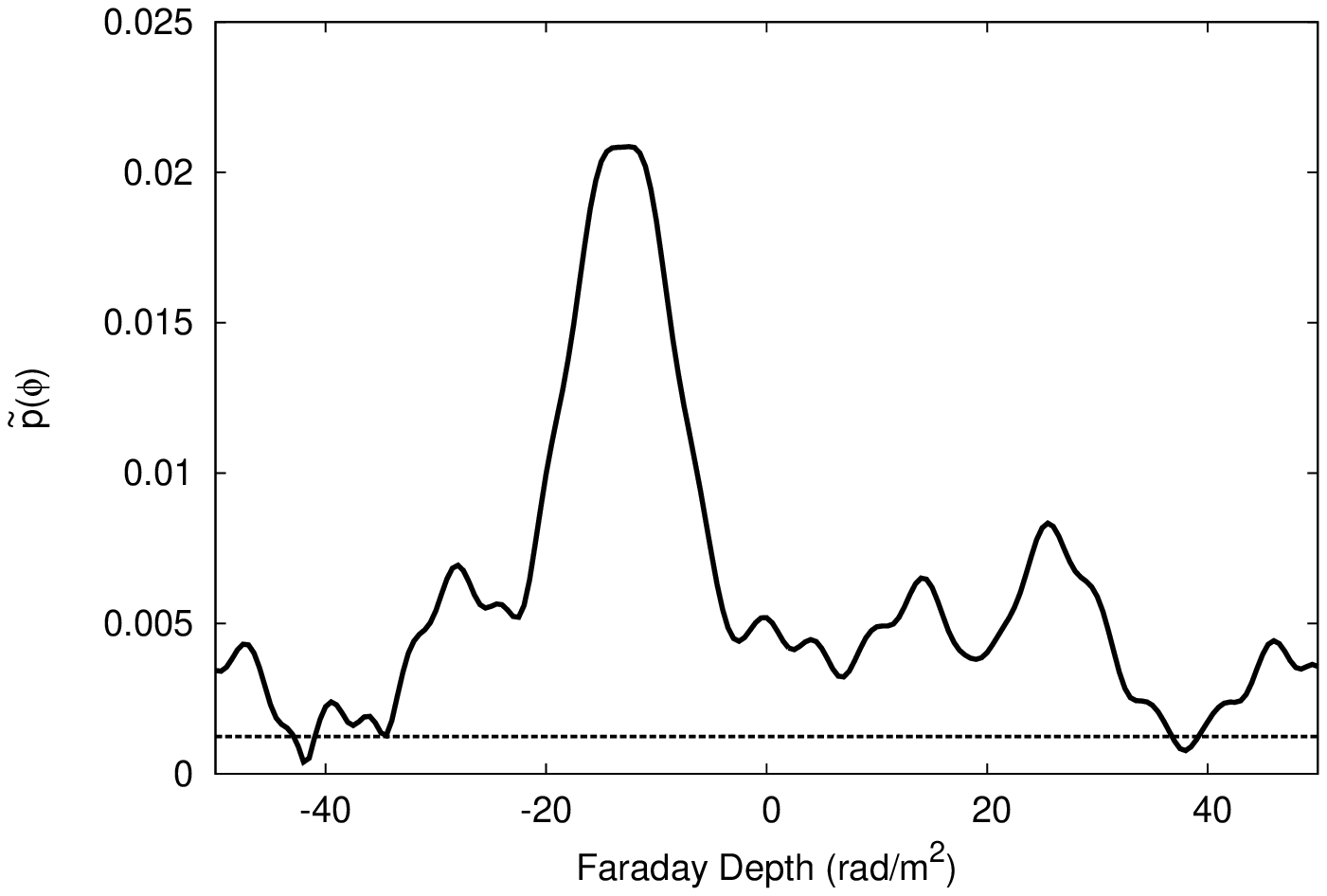,angle=0,width=0.45\textwidth} \\
  \hline
  \end{tabular}
 \caption{Polarization diagnostics for NVSS J011136+132437.  Same layout as Figure \ref{fig:1037.polDiags}.}
  \label{fig:1055.polDiags}
\end{figure}

\begin{figure}
 \centering
  \begin{tabular}{|c|}
  \hline
 \epsfig{file=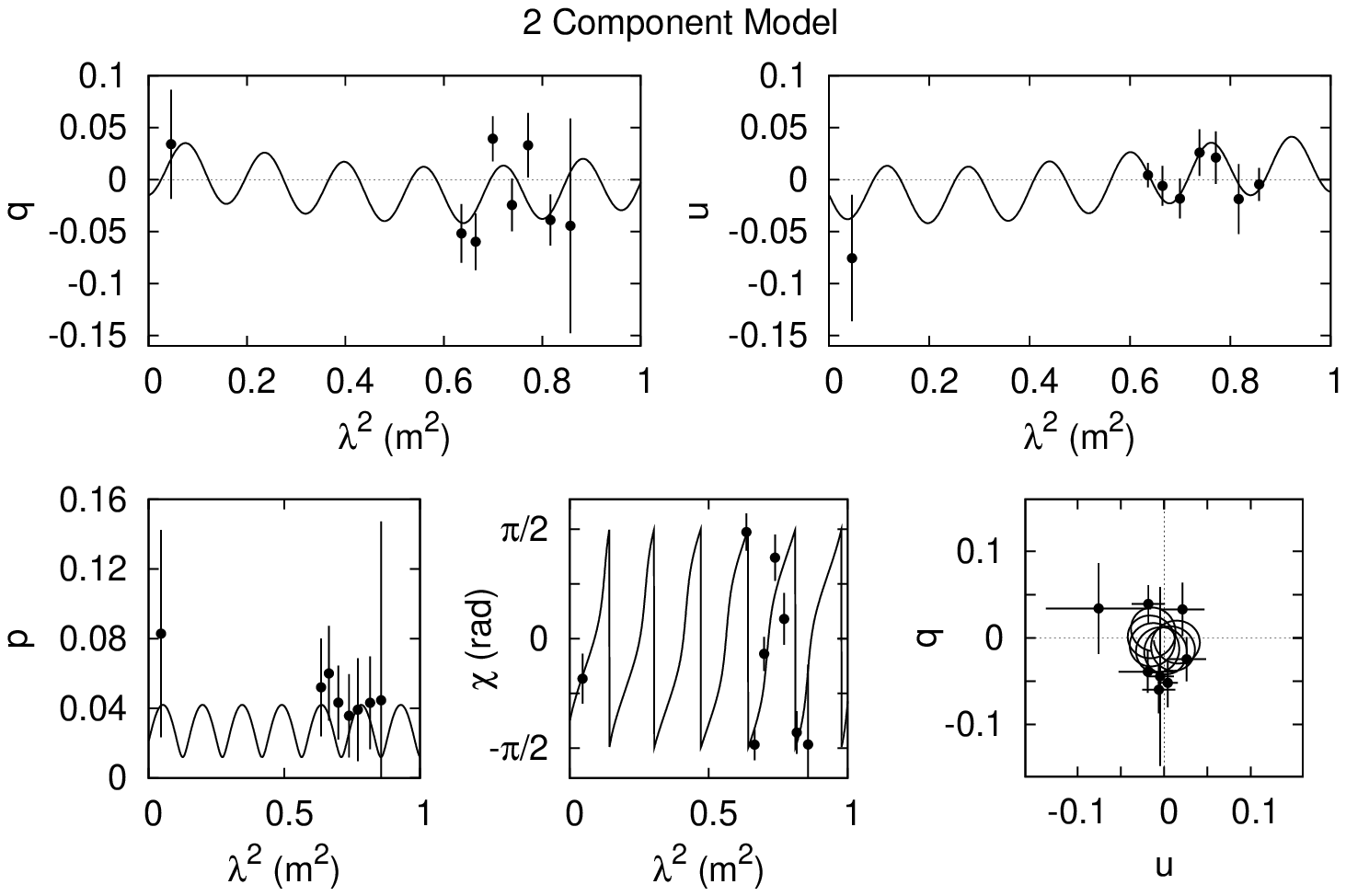,angle=0,width=0.45\textwidth} \\
  \hline
 \epsfig{file=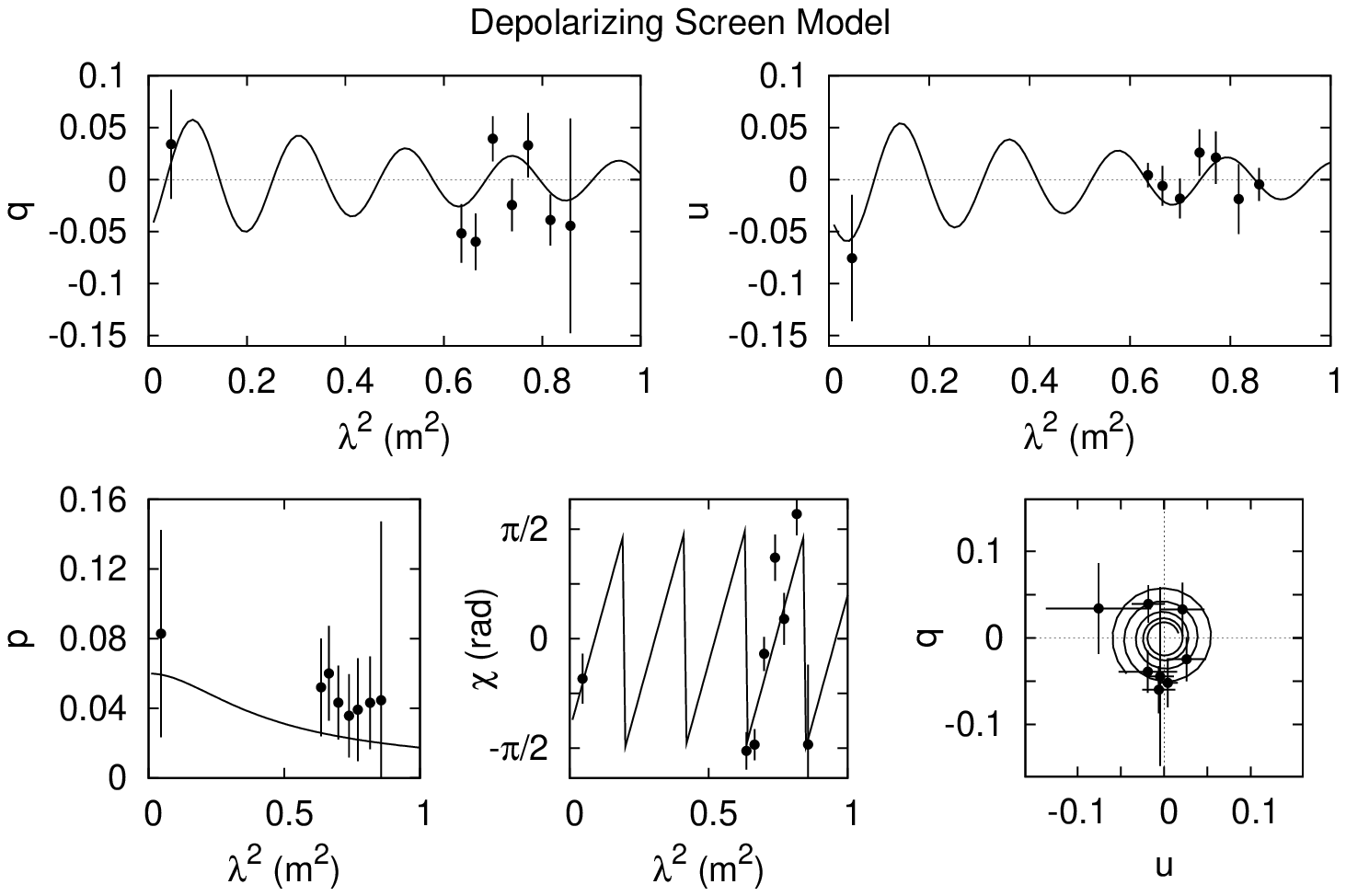,angle=0,width=0.45\textwidth} \\
  \hline
 \epsfig{file=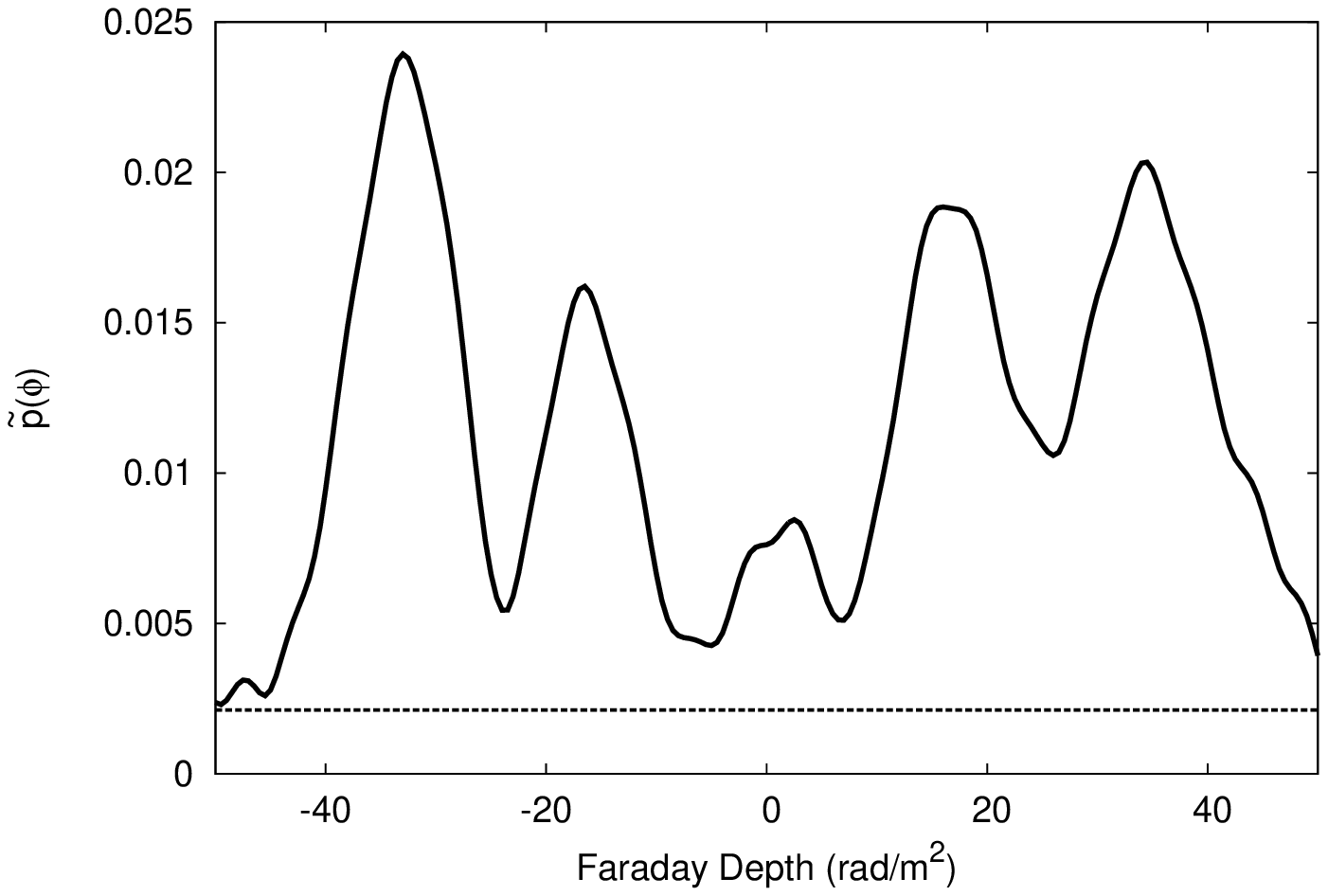,angle=0,width=0.45\textwidth} \\
  \hline
  \end{tabular}
 \caption{Polarization diagnostics for NVSS J011204+124118.  Same layout as Figure \ref{fig:1037.polDiags}.}
  \label{fig:1031.polDiags}
\end{figure}

\begin{figure}
 \centering
  \begin{tabular}{|c|}
  \hline
 \epsfig{file=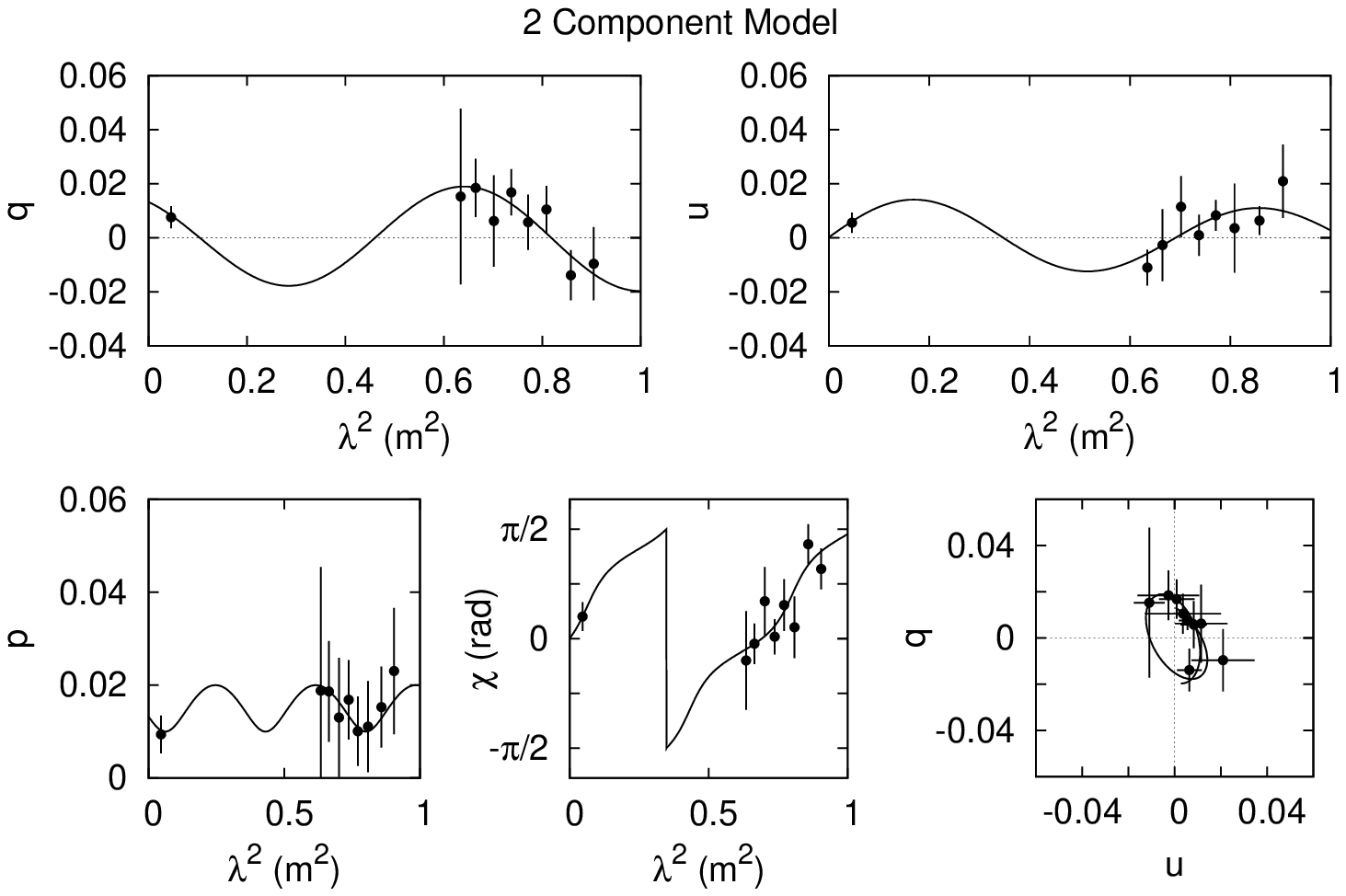,angle=0,width=0.45\textwidth} \\
  \hline
 \epsfig{file=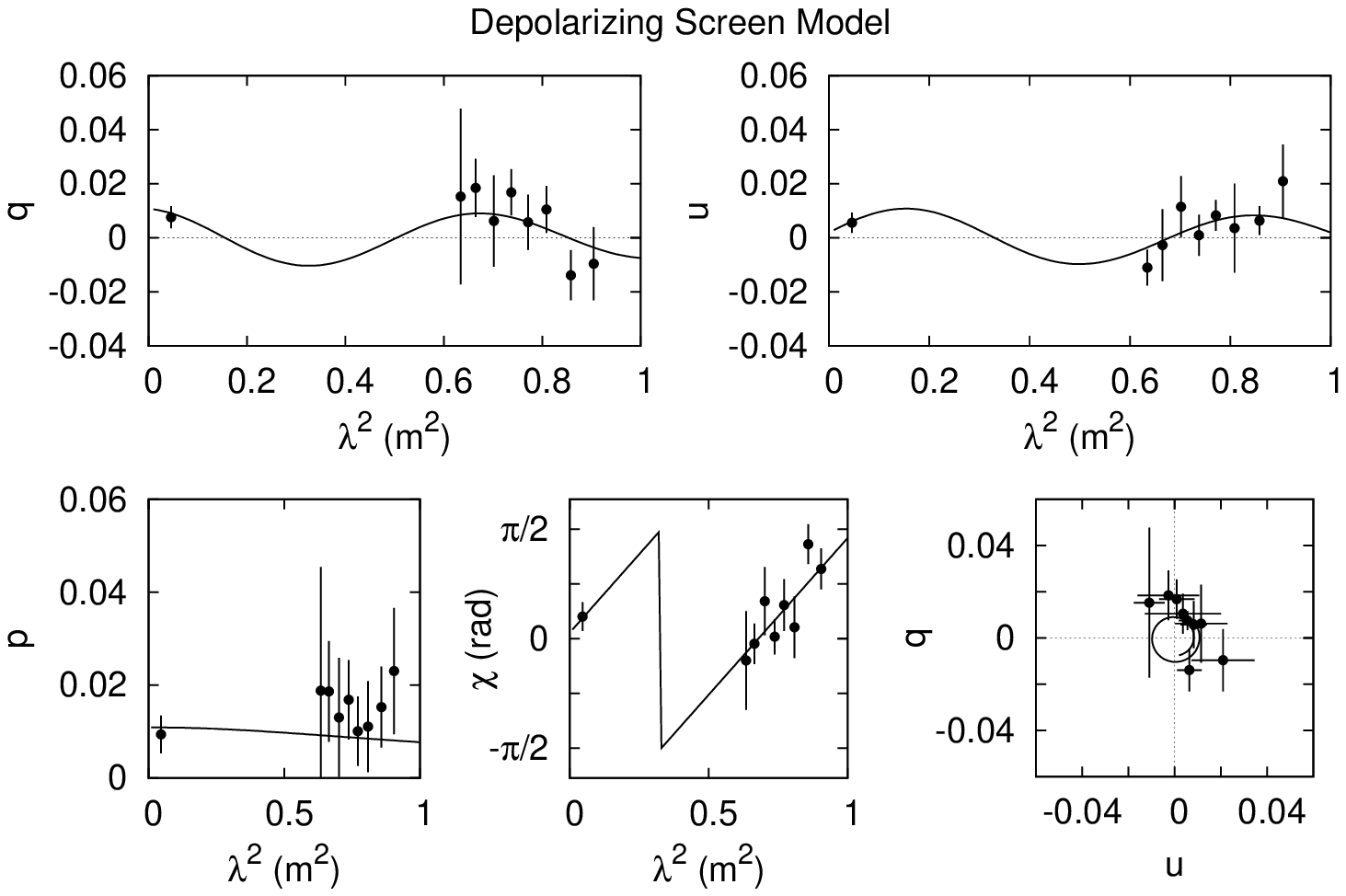,angle=0,width=0.45\textwidth} \\
  \hline
 \epsfig{file=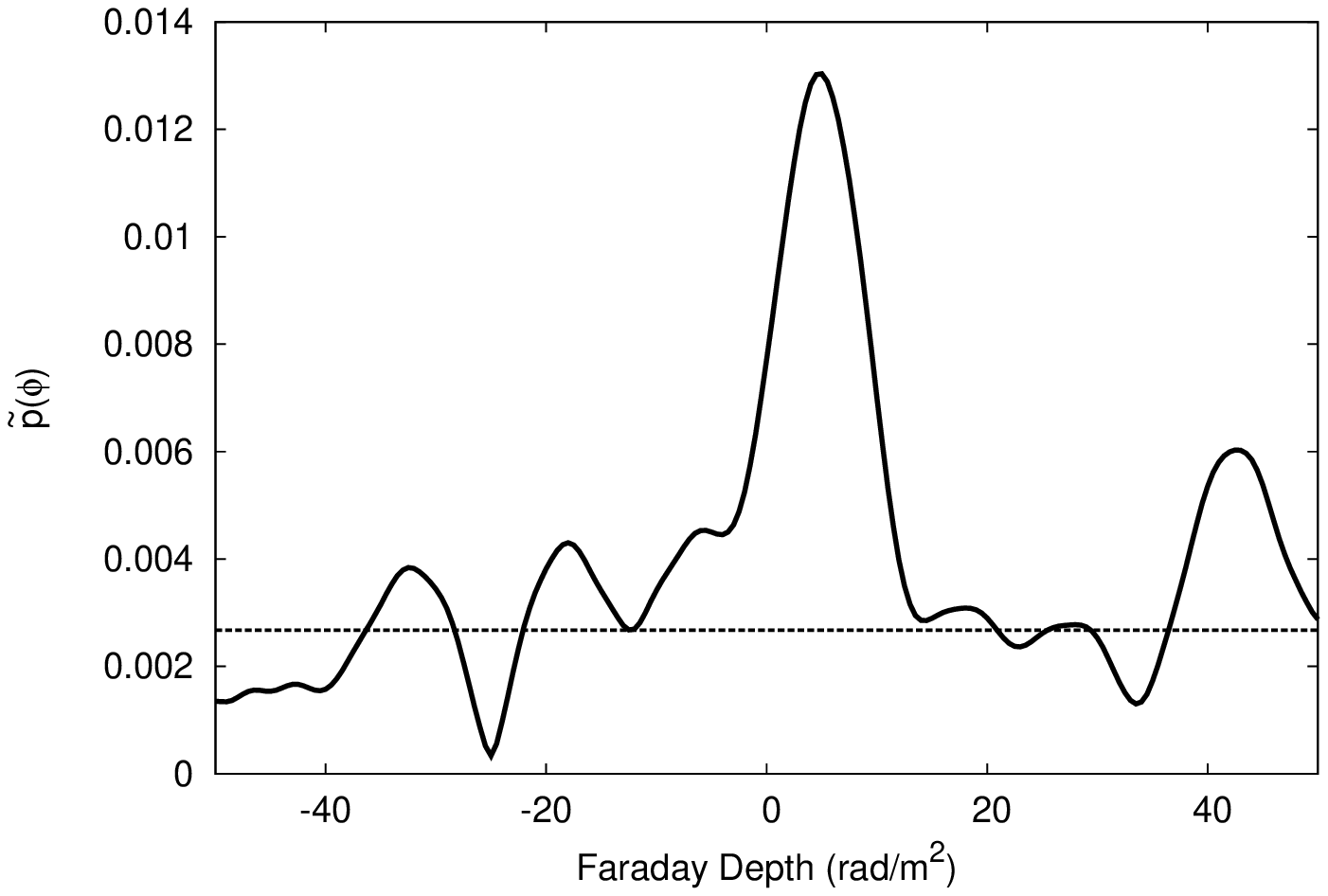,angle=0,width=0.45\textwidth} \\
  \hline
  \end{tabular}
 \caption{Polarization diagnostics for NVSS J125630+270108.  Same layout as Figure \ref{fig:1037.polDiags}.}
  \label{fig:4011.polDiags}
\end{figure}

\begin{figure}
 \centering
  \begin{tabular}{|c|}
  \hline
 \epsfig{file=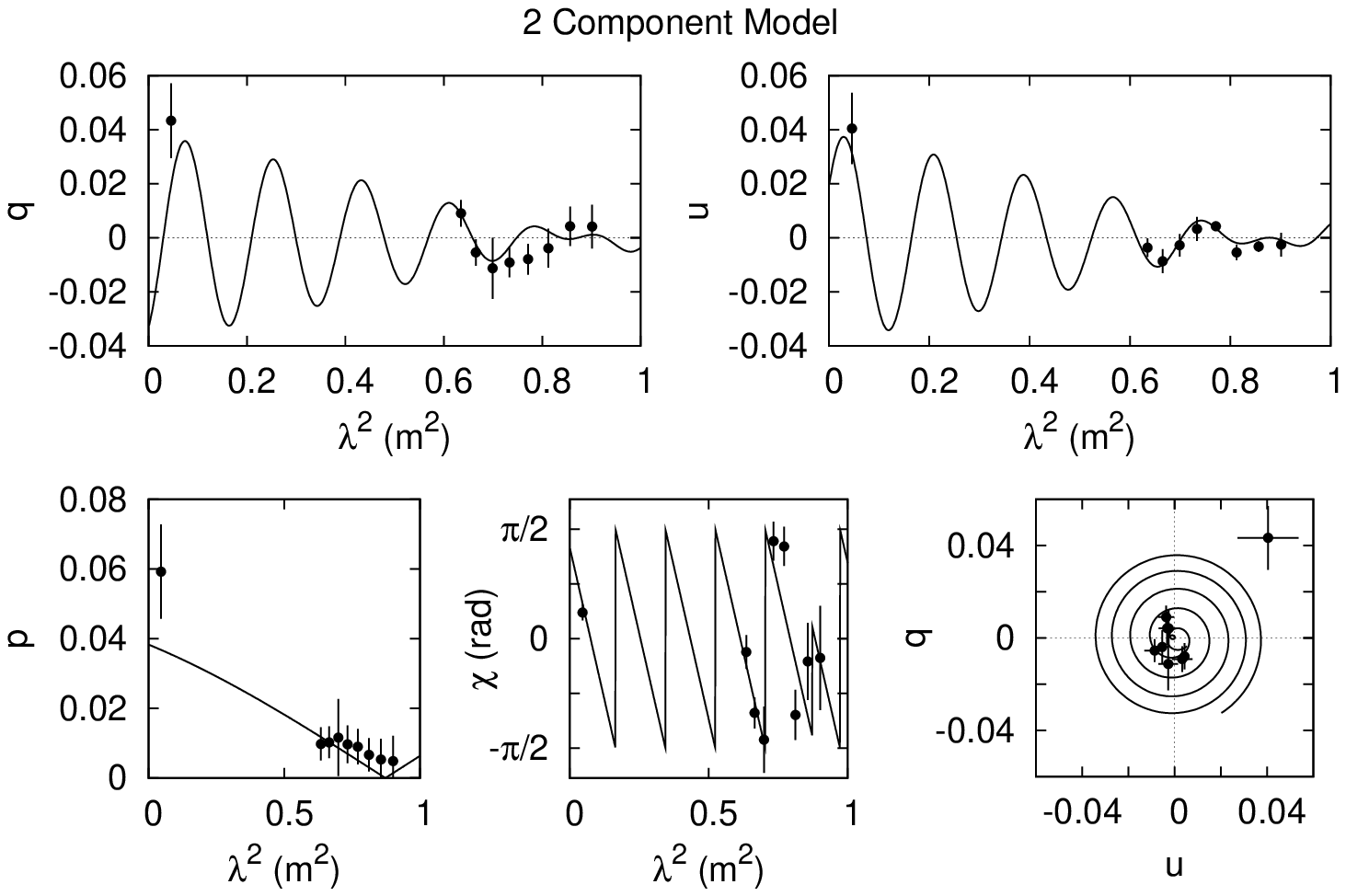,angle=0,width=0.45\textwidth} \\
  \hline
 \epsfig{file=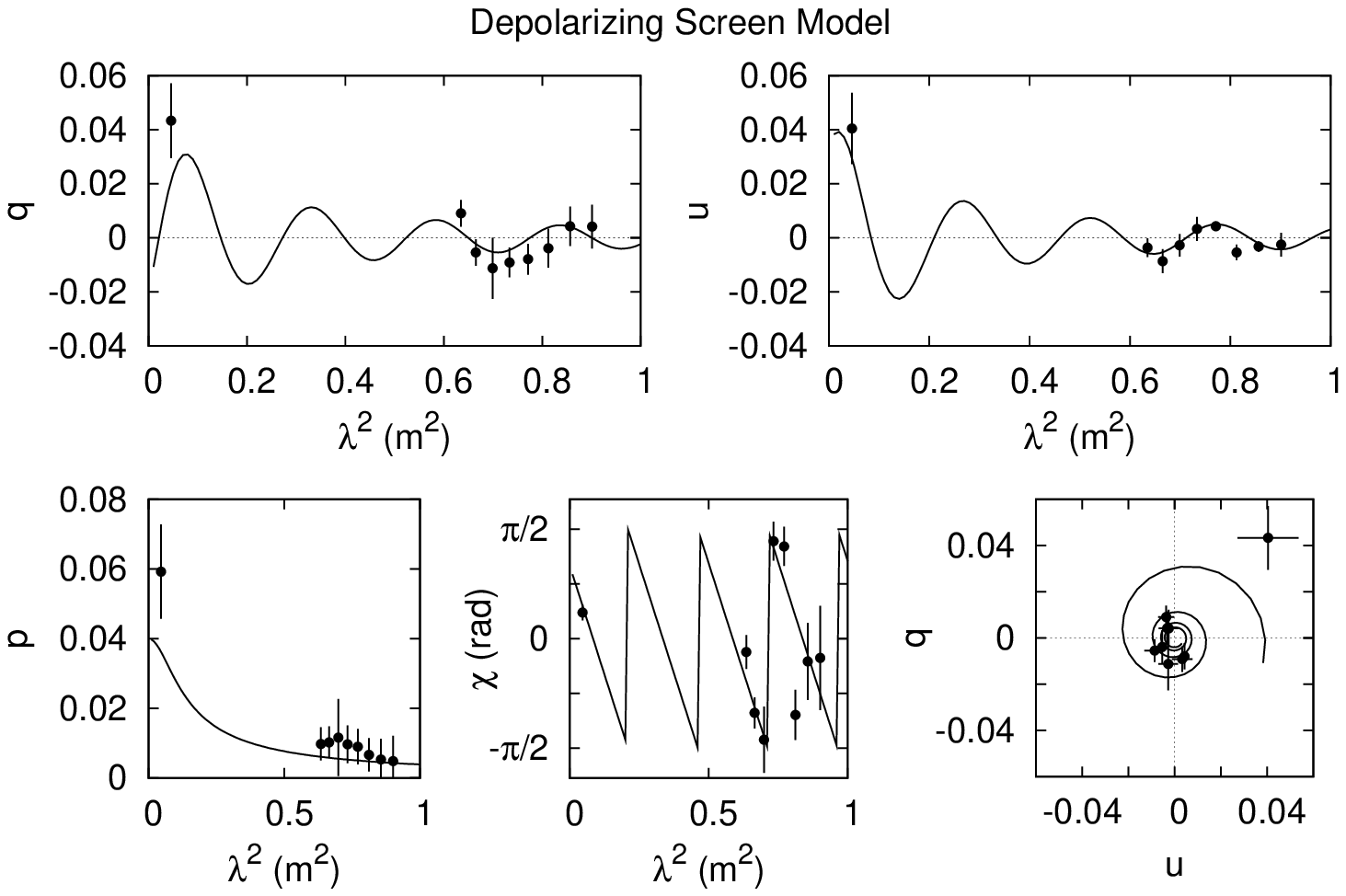,angle=0,width=0.45\textwidth} \\
  \hline
 \epsfig{file=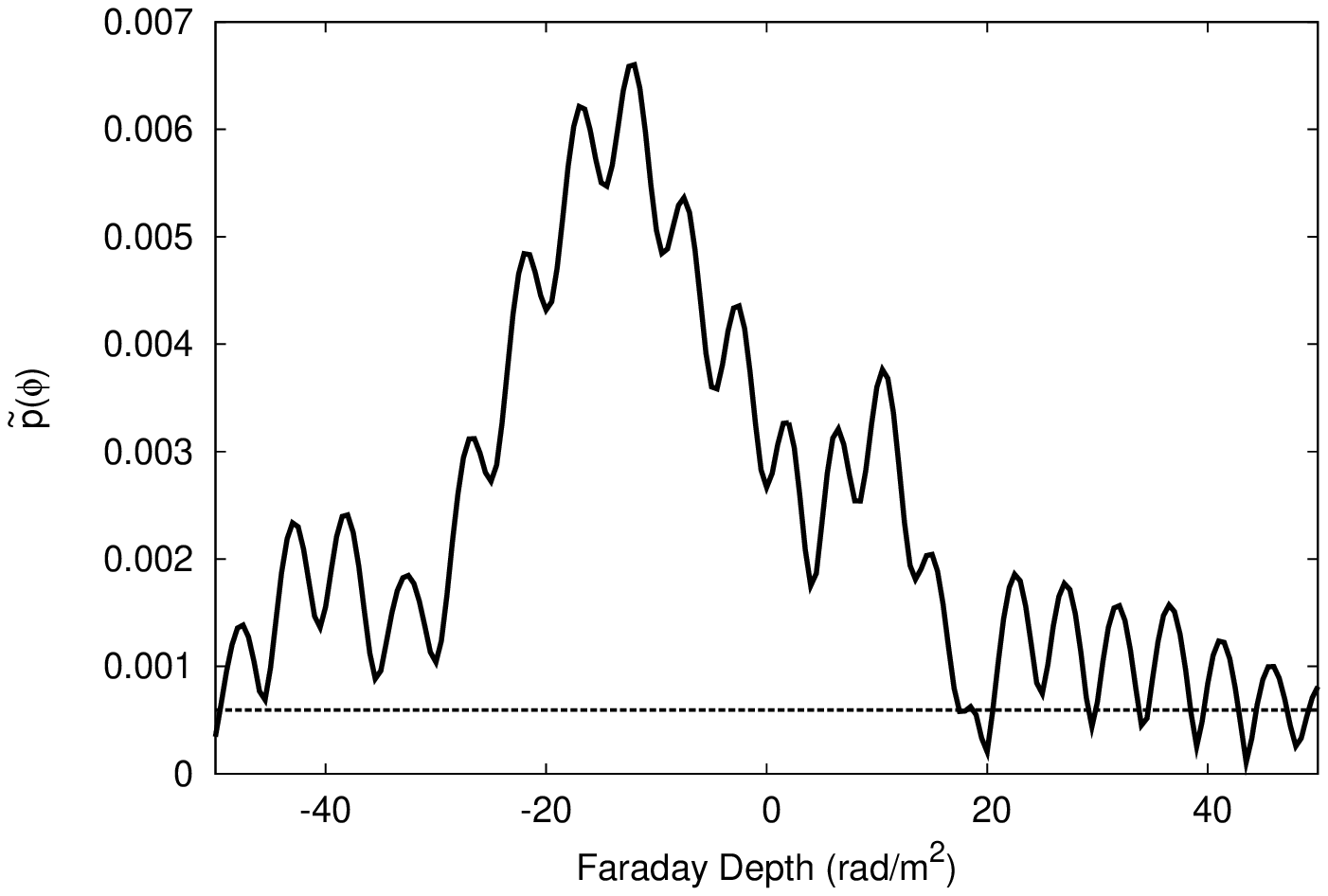,angle=0,width=0.45\textwidth} \\
  \hline
  \end{tabular}
 \caption{Polarization diagnostics for NVSS J162408+605400.  Same layout as Figure \ref{fig:1037.polDiags}.}
  \label{fig:3071.polDiags}
\end{figure}

\begin{figure}
 \centering
  \begin{tabular}{|c|}
  \hline
 \epsfig{file=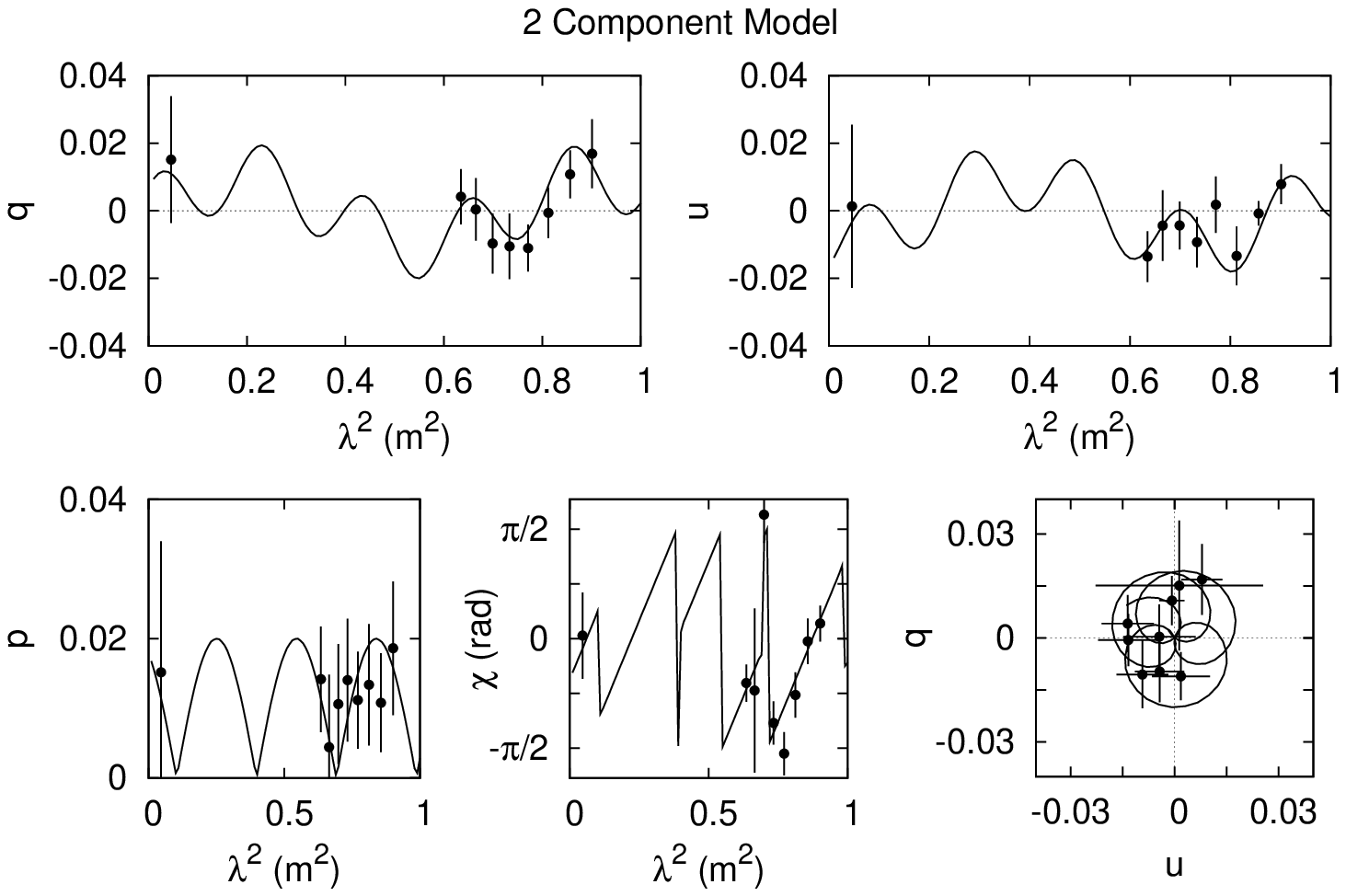,angle=0,width=0.45\textwidth} \\
  \hline
 \epsfig{file=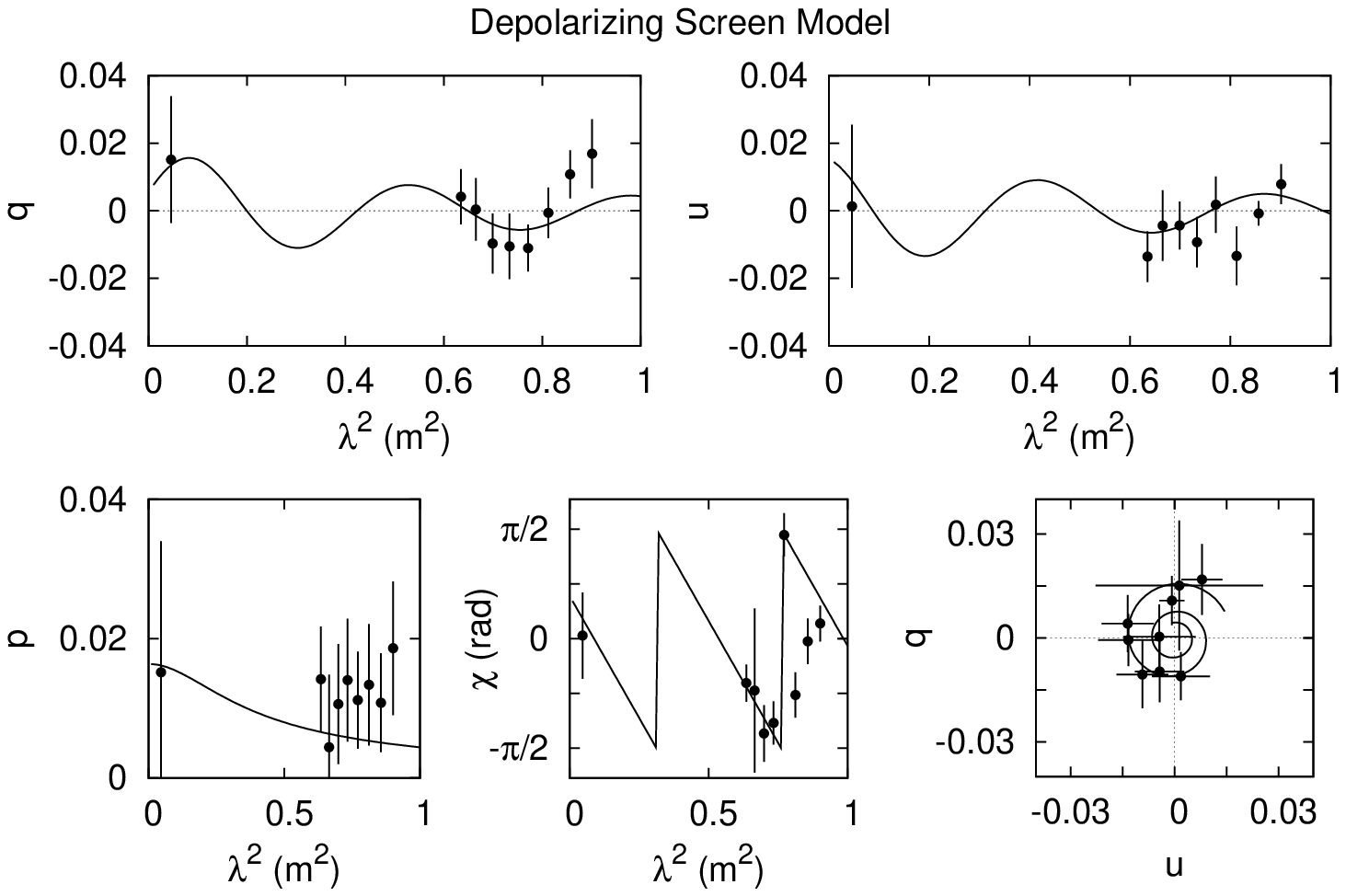,angle=0,width=0.45\textwidth} \\
  \hline
 \epsfig{file=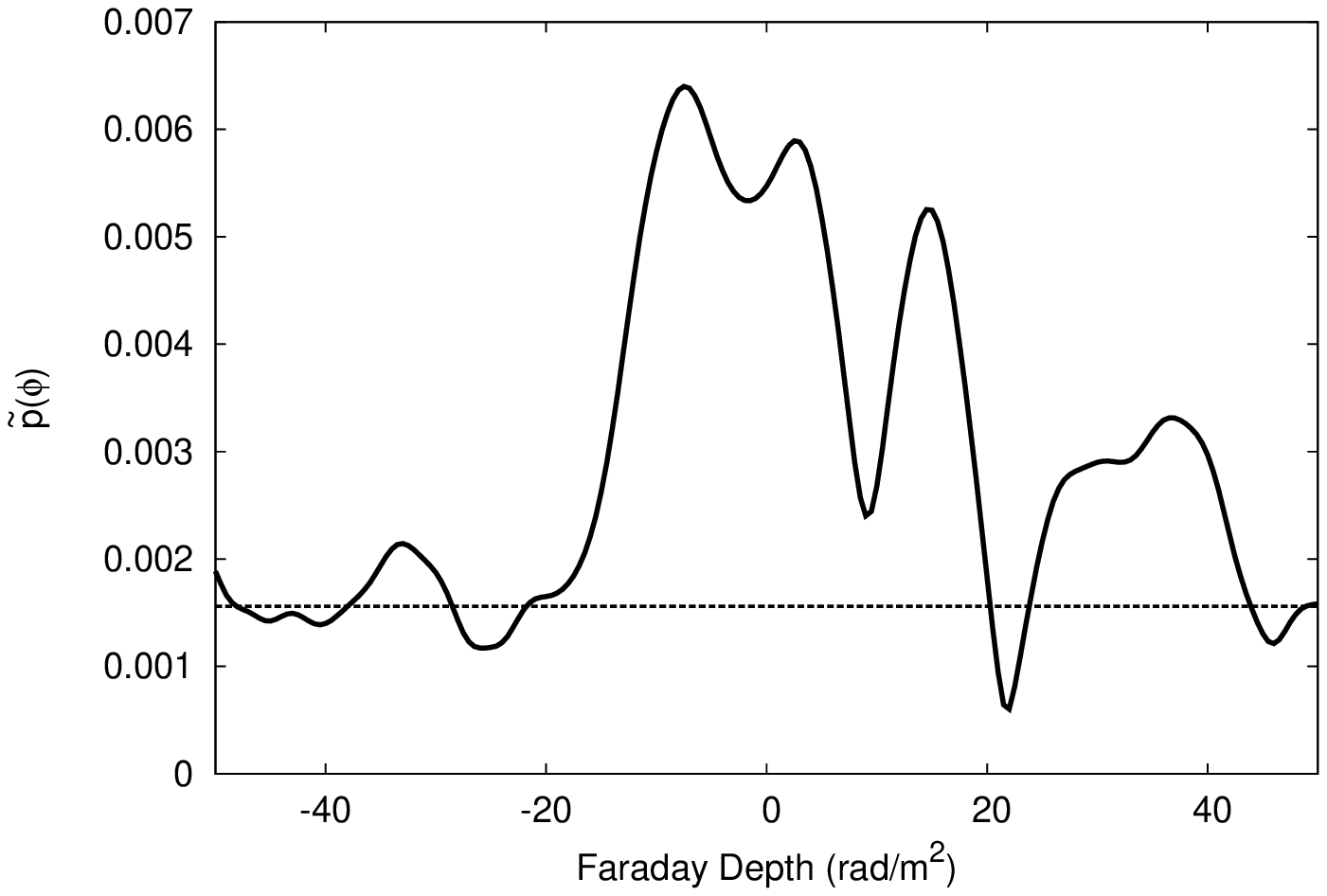,angle=0,width=0.45\textwidth} \\
  \hline
  \end{tabular}
 \caption{Polarization diagnostics for NVSS J162740+603900.  Same layout as Figure \ref{fig:1037.polDiags}.}
  \label{fig:3063.polDiags}
\end{figure}

The results of the various RM determinations for each source are shown in Tables \ref{tab:1037tab} - \ref{tab:3063tab} and in Figures \ref{fig:1037.polDiags} - \ref{fig:3063.polDiags}.
Oscillation visible in the restored $\tilde{p}(\phi)$ profile (e.g., Figure \ref{fig:3071.polDiags}) arises from the sinusoidal residuals in $\tilde{q}(\phi)$ and $\tilde{u}(\phi)$, which are added to the clean components, and provides a measure of the noise level in $\tilde{p}(\phi)$.  In our tests, more aggressive cleaning reduced the oscillation in $\tilde{p}$ by placing power in the clean components randomly across $\phi$ (thus reducing the residuals and producing nicer plots), but did not significantly change the amplitudes or Faraday depths of the major RM components as reported for each source.
If we look at the RM of the strongest component, we find that different models yield consistent results for some but not all sources.
We give a brief discussion of the modeling results for each source here, and a more extensive discussion of 3C33S in the following subsection.
One goal of this investigation is to explore the effect that ignoring the effects of depolarization, including the presence of multiple strong RM components, may have on the findings of the traditional linear $\chi$(\lamsq) method.
We compare our findings with those of \cite{taylor09} for sources where RMs were reported in their study.
Where appropriate, we use data from the VLA FIRST (Faint Images of the Radio Sky at Twenty centimeters; \citealt{becker95}) survey to supplement our analysis.

\begin{table*}
 \caption{Modeling results for NVSS J010616+125116}
 \centering
   \begin{tabular}{ccccccccc}
    \hline\hline
      Model & RM$_1$ (err) & $p_{01}$ & $\chi_{01}$ & RM$_2$ (err) & $p_{02}$ & $\chi_{02}$ & $\sigma_{RM}^2$ & $\chi^2_\nu$ \\
            & (\radmsq) & (\%) & ($^\circ$) & (\radmsq) & (\%) & ($^\circ$) & (\radmsq) & \\
    \hline
      Linear $\chi$(\lamsq) & -9.6 (0.12) & - & 80 & - & - & - & - & 1.2 \\
      Screen  & -9.5 (5) & 10.5 & 78 & - & - & - & 1.2 & 10.2 \\
      RM Synth/Clean & -9.1 (0.04) & 3.1 & - & 11.5 (0.2) & 0.5 & - & - & - \\
      Two Component$^1$ & [-5.0] & 4.0 & 75 & [2.0] & 2.2 & 65 & - & 8.8 \\
      \hline
   \end{tabular}

$^1$Brackets indicate multiple minima in $\chi^2$ surface -- no RM error reported.
\label{tab:1037tab}
\end{table*}

\begin{table*}
 \caption{Modeling results for 3C33S}
 \centering
   \begin{tabular}{ccccccccc}
    \hline\hline
      Model & RM$_1$ (err) & $p_{01}$ & $\chi_{01}$ & RM$_2$ (err) & $p_{02}$ & $\chi_{02}$ & $\sigma_{RM}^2$ & $\chi^2_\nu$ \\
            & (\radmsq)~ & (\%) & ($^\circ$) & (\radmsq)~ & (\%) & ($^\circ$) & (\radmsq)~ & \\
    \hline
      Linear $\chi$(\lamsq) & -6.8 (0.17) & - & 80 & - & - & - & - & 3.9 \\
      Screen  & -7.0 (0.15) & 8.5 & 86 & - & - & - & 1.0 & 6.0 \\
      RM Synth/Clean & -6.7 (0.06) & 2.6 & - & 8.0 (0.1) & 0.9 & - & - & - \\
      Two Component & -2.9 (0.1) & 6.7 & 85 & -0.05 (0.2) & 4.8 & 49 & - & 2.1 \\
      \hline
   \end{tabular}
\label{tab:3C33Stab}
\end{table*}

\begin{table*}
 \caption{Modeling results for NVSS J011136+132437}
 \centering
   \begin{tabular}{ccccccccc}
    \hline\hline
      Model & RM$_1$ (err) & $p_{01}$ & $\chi_{01}$ & RM$_2$ (err) & $p_{02}$ & $\chi_{02}$ & $\sigma_{RM}^2$ & $\chi^2_\nu$ \\
            & (\radmsq)~ & (\%) & ($^\circ$) & (\radmsq)~ & (\%) & ($^\circ$) & (\radmsq)~ & \\
    \hline
      Linear $\chi$(\lamsq) & -10.8 (0.31) & - & 130 & - & - & - & - & 6.5 \\
      Screen  & -10.75 (2.5) & 5.5 & 130 & - & - & - & 1.2 & 9.3 \\
      RM Synth/Clean & -13.1 (0.1) & 2.0 & - & 26.0 (0.3) & 0.6 & - & - & - \\
      Two Component & -11.2 (0.5) & 1.5 & 140 & -24.2 (0.6) & 1.5 & 175 & - & 7.9 \\
      \hline
   \end{tabular}
\label{tab:1055tab}
\end{table*}

\begin{table*}
 \caption{Modeling results for NVSS J011204+124118}
 \centering
   \begin{tabular}{ccccccccc}
    \hline\hline
      Model & RM$_1$ (err) & $p_{01}$ & $\chi_{01}$ & RM$_2$ (err) & $p_{02}$ & $\chi_{02}$ & $\sigma_{RM}^2$ & $\chi^2_\nu$ \\
            & (\radmsq)~ & (\%) & ($^\circ$) & (\radmsq)~ & (\%) & ($^\circ$) & (\radmsq)~ & \\
    \hline
      Linear $\chi$(\lamsq) & 20 (1.9) & - & 60 & - & - & - & - & 12.0 \\
      Screen  & 19.5 (0.5) & 7.5 & 94 & - & - & - & 1.0 & 1.9 \\
       RM Synth/Clean$^1$ & -32.8 (0.3) & 2.1 & - & 34.4 (0.5) & 1.7 & - & - & - \\
      Two Component$^2$ & [19.5] & 2.7 & 95 & [-2.0] & 1.5 & 160 & - & 2.12 \\
      \hline
   \end{tabular}

$^1$Four strong features exist in the cleaned FDF, including one near 17 \radmsq. \\
$^2$Brackets indicate multiple minima in $\chi^2$ surface -- no RM error reported.
\label{tab:1031tab}
\end{table*}

\begin{table*}
 \caption{Modeling results for NVSS J125630+270108}
 \centering
   \begin{tabular}{ccccccccc}
    \hline\hline
      Model & RM$_1$ (err) & $p_{01}$ & $\chi_{01}$ & RM$_2$ (err) & $p_{02}$ & $\chi_{02}$ & $\sigma_{RM}^2$ & $\chi^2_\nu$ \\
            & (\radmsq)~ & (\%) & ($^\circ$) & (\radmsq)~ & (\%) & ($^\circ$) & (\radmsq)~ & \\
    \hline
      Linear $\chi$(\lamsq) & 4.6 (0.31) & - & -175 & - & - & - & - & 0.9 \\
      Screen  & 4.5 (0.25) & 1.0 & 4 & - & - & - & 0.0 & 1.03 \\
      RM Synth/Clean & 4.8 (0.3) & 1.1 & - & -17.7 (1.3) & 0.2 & - & - & - \\
      Two Component & 4.5 (0.25) & 1.5 & 10 & -4.0 & 0.5 & 130 & - & 0.851 \\
      \hline
   \end{tabular}
\label{tab:4011tab}
\end{table*}

\begin{table*}
 \caption{Modeling results for NVSS J162408+605400}
 \centering
   \begin{tabular}{ccccccccc}
    \hline\hline
      Model & RM$_1$ (err) & $p_{01}$ & $\chi_{01}$ & RM$_2$ (err) & $p_{02}$ & $\chi_{02}$ & $\sigma_{RM}^2$ & $\chi^2_\nu$ \\
            & (\radmsq)~ & (\%) & ($^\circ$) & (\radmsq)~ & (\%) & ($^\circ$) & (\radmsq)~ & \\
    \hline
      Linear $\chi$(\lamsq) & -16.8 (0.4) & - & 65 & - & - & - & - & 3.2 \\
      Screen  & -17 (0.15) & 5.5 & 64 & - & - & - & 1.4 & 1.2 \\
      RM Synth/Clean & -14.6 (1.5) & 0.5 & - & 10.3 (0.4) & 0.2 & - & - & - \\
      Two Component$^1$ & [-17.0] & 2.5 & 95 & [-18.0] & 2.5 & 55 & - & 1.99 \\
      \hline
   \end{tabular}

$^1$Brackets indicate multiple minima in $\chi^2$ surface -- no RM error reported.
\label{tab:3071tab}
\end{table*}

\begin{table*}
 \caption{Modeling results for NVSS J162740+603900}
 \centering
   \begin{tabular}{ccccccccc}
    \hline\hline
      Model & RM$_1$ (err) & $p_{01}$ & $\chi_{01}$ & RM$_2$ (err) & $p_{02}$ & $\chi_{02}$ & $\sigma_{RM}^2$ & $\chi^2_\nu$ \\
            & (\radmsq)~ & (\%) & ($^\circ$) & (\radmsq)~ & (\%) & ($^\circ$) & (\radmsq)~ & \\
    \hline
      Linear $\chi$(\lamsq) & -7.8 (1.5) & - & 65 & - & - & - & - & 4.0 \\
      Screen$^1$  & [-7.0] & 2.0 & 34 & - & - & - & 1.0 & 1.8 \\
      RM Synth/Clean$^2$ & -6.4 (0.8) & 0.5 & - & 14.6 (0.5) & 0.4 & - & - & - \\
      Two Component & 4.3 (0.5) & 1.0 & 133 & 15.0 (0.5) & 1.0 & 160 & - & 1.58 \\
      \hline
   \end{tabular}

$^1$Brackets indicate multiple minima in $\chi^2$ surface -- no RM error reported. \\
$^2$Three strong RM components are present in FDF, including one near 4 \radmsq.
\label{tab:3063tab}
\end{table*}

\emph{NVSS J010616+125116.}  This source is resolved as a double source (separation $\sim$60$''$) in the original NVSS image, but appears as a single source when convolved to the WSRT field resolution (325$''$x70$''$).  We adopt the name of the brighter NVSS source (peak $I_{1.4}$=102 mJy/beam); the secondary source is NVSS J010615+124210 (peak $I_{1.4}$=72 mJy/beam); the two sources have similar $p_{1.4}$.
The dominant RM is found near -9 \radmsq~ for the linear $\chi(\lambda^2)$, depolarizing screen, and RM Synth/Clean methods.
The two component model, however, finds the dominant RM component near -5 \radmsq.  It is possible that the relatively strong secondary RM component found near +2 \radmsq~ in the two component fit has drawn the other solutions away from the true intrinsic Faraday structure.  The presence of multiple minima in the $\chi^2$ surface, however, casts uncertainty on the two component result.
For comparison, \cite{taylor09} determined the RM of NVSS J010616+125116 to be $-16.8 \pm 14.7$ \radmsq; no RM was reported for NVSS J010615+124210.

\emph{3C33S.}  This source is also known as NVSS J010850+131831.  The dominant RM found by the linear $\chi(\lambda^2)$, depolarizing screen, and RM Synth/Clean methods are near -7 \radmsq.
This is in disagreement with the two component modeling, which finds no significant component near -7 \radmsq; rather, the dominant component is found near -3 \radmsq~ with a relatively strong second component near 0 \radmsq.  For comparison, \cite{taylor09} determined the RM to be $3.4 \pm 1.9$ \radmsq.  In addition, \cite{law2011} performed RM Synthesis on 3C33S using two bands, each 100 MHz wide, centered at 1.43 and 2.01 GHz with the ATA.  After cleaning they found a single RM at $-12.3 \pm 0.4$ \radmsq.  That they found a single RM is not unexpected, considering the FWHM of their RMSF of 141 \radmsq, but the RM value found would not fit our 350 MHz observations.  Given the high signal to noise in our $Q$, $U$ data, this is the strongest case yet for interference between two strong RM components causing other methods to misinterpret the true Faraday structure.  In the next section, we will use idealized models to demonstrate how this comes about.

\emph{NVSS J011136+132437.}  The dominant RM component is found near -11 \radmsq~ for the linear $\chi(\lambda^2)$ and depolarizing screen models.
The two component method finds equal amplitudes for both RM components, with one near -11 \radmsq~ and the other near -24 \radmsq.  It doesn't appear that a secondary component has affected the outcome of the single component methods.
RM Synth/Clean nearly agrees, finding the dominant RM component at -13 \radmsq.
The secondary RM component found by RM Synth/Clean and the two component model are in disagreement, however, in both location and relative amplitude.
All methods have a high $\chi^2_{\nu}$, suggesting that no solution is to be trusted. 
For comparison, \cite{taylor09} determined the RM to be $-13.8 \pm 3.3$ \radmsq, in agreement with our findings.  

\emph{NVSS J011204+124118.}  The dominant RM is found near +20 \radmsq~ for the linear $\chi(\lambda^2)$, depolarizing screen, and two component models.
The two component fit finds a secondary component with $p_{02} / p_{01} > 0.5$ near -2 \radmsq, but it doesn't appear to have affected the outcome of the single RM methods.
RM Clean finds the dominant RM component near -33 \radmsq, but three other peaks of significant amplitude are found in the Faraday spectrum, including relatively strong components near +34 and +17 \radmsq.
\cite{taylor09} do not report a RM for this source.

\emph{NVSS J125630+270108.}  All four methods find the dominant RM component to lie near +4.5 \radmsq.  Secondary components for RM Synthesis and the two component method are of relatively weak amplitude, and likely do not contribute significantly to the solutions found by the single RM methods.
Due to the lack of depolarization from 1.4 GHz to 350 MHz, it is not surprising that the traditional linear fit to $\chi$(\lamsq) is in agreement with the other methods.
\cite{taylor09} do not report a RM for this particular source (unresolved in both NVSS and FIRST), but using their data we determined the weighted mean RM of the 17 sources within 2\textdegree~ (with an entry in \citealt{taylor09}) to be $\approx$$2.5 \pm 1$ \radmsq.  This suggests that a Galactic foreground (rotating) screen is the single dominant component of Faraday structure for this source, a situation which is unique in our modeling results.

\emph{NVSS J162408+605400.}  The dominant RM component is found to lie near -17 \radmsq~ for the linear $\chi(\lambda^2)$ and depolarizing screen models, while RM Synth/Clean finds the dominant component near -15 \radmsq.  The two component model shows two equal amplitude RM components near -17 \radmsq~ and -18 \radmsq.  The presence of multiple minima in the $\chi^2$ surface casts uncertainty on the two component solution.
\cite{taylor09} do not report a RM for this source.
The oscillation visible in the cleaned FDF shown in Figure \ref{fig:3071.polDiags} is due to the low amplitude of the clean components relative to the amplitude of the residuals in RM Clean.  As mentioned in the beginning of this section, cleaning further would reduce the level of apparent oscillation in the cleaned FDF, but would not appreciably change the locations or amplitudes of the fitted RM components.

\emph{NVSS J162740+603900.}  The dominant RM component is found by the linear $\chi(\lambda^2)$, depolarizing screen, and RM Synth/Clean methods to be near -7 \radmsq.
The cleaned FDF displays three strong RM features, near -7, +4, and +15 \radmsq, but it is likely that the components at -7 and +4 \radmsq~ are blended, contributing power to each other and increasing their peak amplitudes.
These results contrast with the two component method, which finds two dominant RM components of equal amplitude near +4.5 and +15 \radmsq.
Again, it seems likely that two RM components are interfering in a way which confounds the other methods.
\cite{taylor09} do not report a RM for this source.

\begin{figure*}
 \centering
  \begin{tabular}{|c|c|}
  \hline
 \epsfig{file=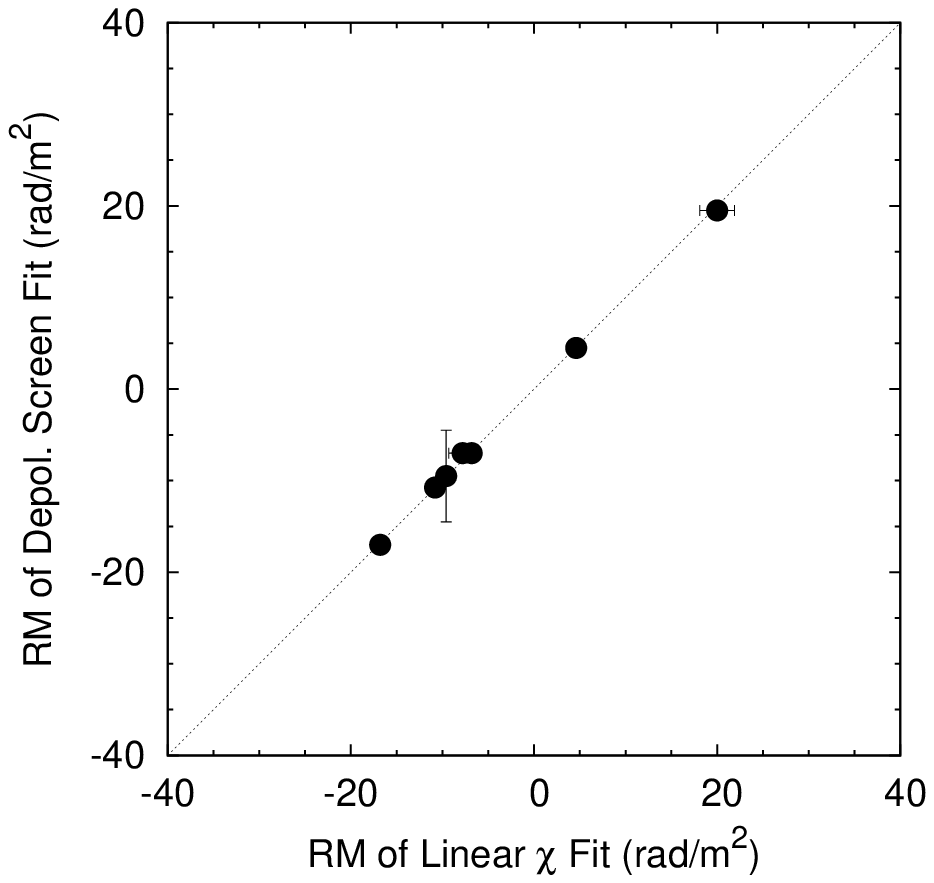, width=.45\textwidth} &
 \epsfig{file=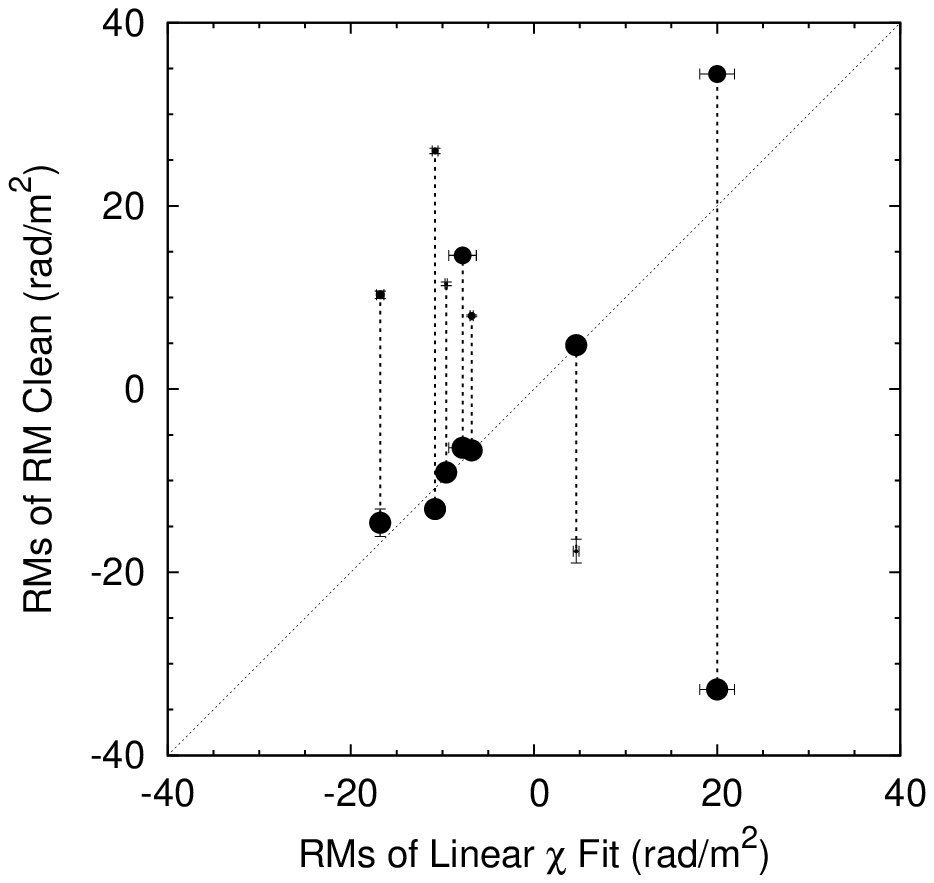, width=.45\textwidth} \\
  \hline
 \epsfig{file=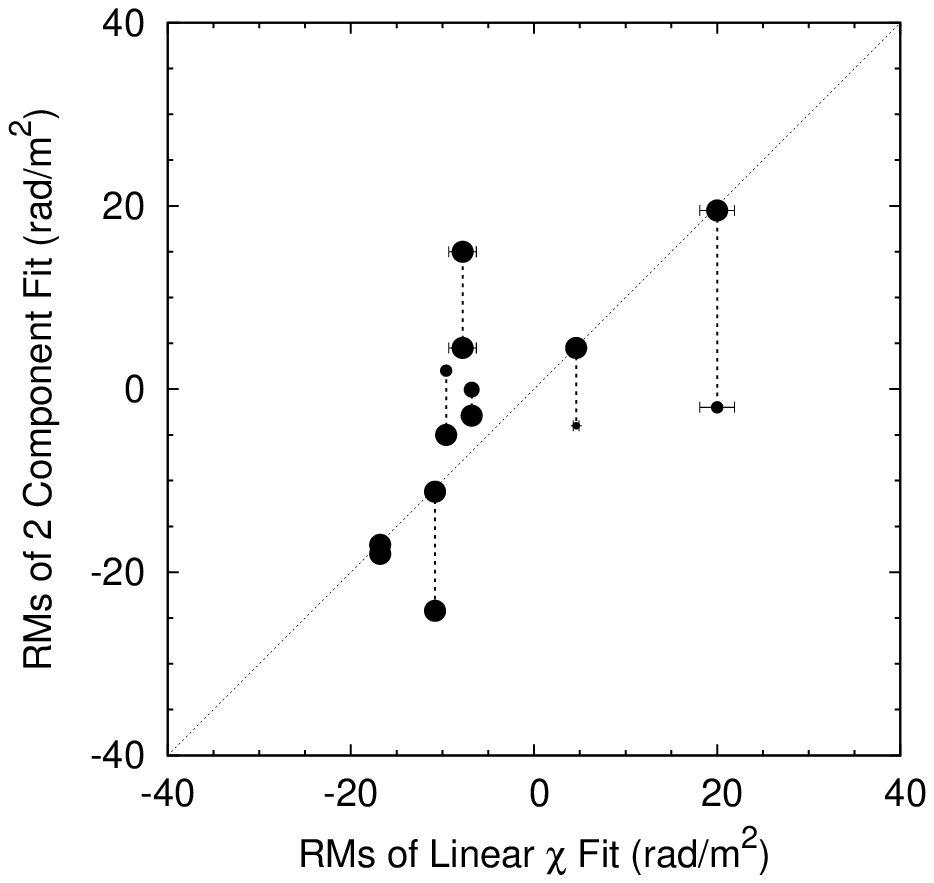, width=.45\textwidth} &
 \epsfig{file=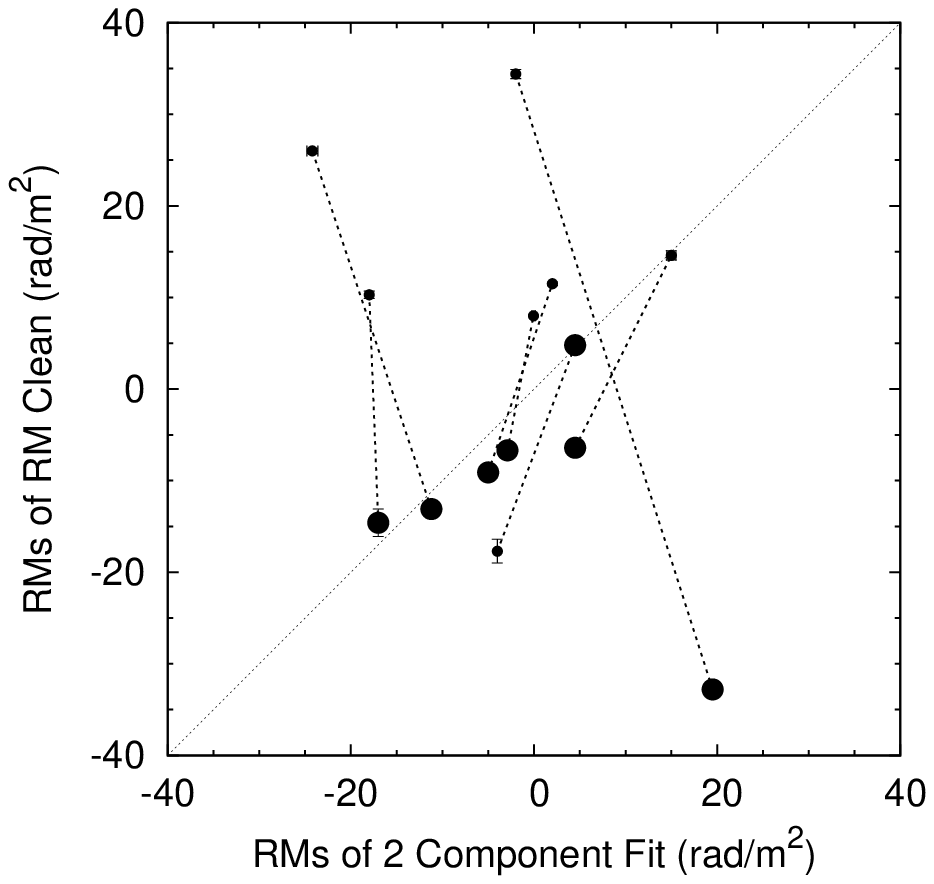, width=.45\textwidth} \\
  \hline
  \end{tabular}
\caption{Comparison of the methods for RM determination for the seven modeled sources.  \textbf{Upper left:} depolarizing screen vs. $\chi$(\lamsq) fit.  \textbf{Upper right:} RM Synth/Clean vs. $\chi$(\lamsq) fit.  \textbf{Lower left:} two component fit vs. $\chi$(\lamsq) fit.  The two strongest RMs are plotted for RM Synthesis/Clean and two-component model fitting, connected by thick dashed lines for each source.  Pointsize for the primary component is fixed, while the pointsize of the secondary component (relative to that of the primary) is proportional to the ratio of amplitudes for the RM components (i.e. $p_2/p_1$), as listed in Tables \ref{tab:1037tab}-\ref{tab:3063tab}.  \textbf{Lower right:} RM Synth/Clean vs. two component fit.  Large and small points show RM$_1$ and RM$_2$, respectively, as listed in  Tables \ref{tab:1037tab}-\ref{tab:3063tab}. Errors from the fitting techniques are plotted, but are smaller than the pointsize for most sources.}
\label{fig:RMcomp}
\end{figure*}

A comparison between the different methods of determining RMs for each source are shown in Figure \ref{fig:RMcomp}.  As expected, the linear $\chi(\lambda^2)$ and depolarizing screen fits agree well for the dominant RM value since $n\pi$ angle shifts were inserted into the data for the $\chi$(\lamsq) fits to best match the depolarizing screen models.  The RM of the dominant component found by RM Synth/Clean agrees fairly well with the linear $\chi(\lambda^2)$ fit method for six of the seven sources, although only three agree within the formal errors.  The dominant RM found by the two component model fit, however, finds agreement with the linear $\chi$(\lamsq) fit method in only four of the seven sources analyzed.
It is apparent that in six of the seven sources the traditional linear fit to $\chi$(\lamsq) is incapable of providing a description of the source's true Faraday structure, instead providing what may be referred to as a ``characteristic'' RM.  This is due to the fact that fitting to $\chi$(\lamsq) does not consider the behavior of $p$(\lamsq), which is variable in most of our sources when the measurements across a large range of \lamsq~ are considered.  One must consider depolarization models, such as the depolarizing screen or interference between multiple RM components, if the true Faraday structure is to be described.

We note that the $\chi^2_{\nu}$ values for these fits are quite high in many cases, suggesting more complicated models would be needed to properly fit the data.  Some of the data appear anomalous when the apparent behavior of neighboring points is taken into account.  These data could be contaminated by residual instrumental problems; we have attempted to incorporate these effects into our errors.  Given infinite resources, our instrumental errors would approach zero; polarization calibration is notoriously difficult at low frequencies and we must therefore proceed with our best effort, given the current technological limitations.  By removing ``anomalous'' data we would be biasing the solutions toward simple Faraday structures in the model fitting and degrading the ability of RM Synthesis to resolve multiple components closely spaced in Faraday depth.  In addition, we found that some of the discrepancies between the fits were due to flaws in the techniques themselves, which we discuss in Section \ref{sec:3C33S} using the case of 3C33S.

\subsection {3C33 South}
\label{sec:3C33S}
3C33S is the southern lobe of the $z = 0.059$ radio galaxy 3C33 near the pointing center of the Aries-Pisces field.
At an off-axis radius of $\sim$650$''$, its mean $p$ of $\sim$4\% in the 350 MHz band is well above the mean WSRT instrumental contribution of $\lesssim$0.5\% at that radius.
In the NVSS image, convolved to the common beamsize used for our WSRT images, 3C33S displays a fractional polarization of 10\%.
Prior studies at $\lambda6$ cm and $\lambda2$ cm by \cite{rudnick81}, which resolve the structure of the lobe, find the fractional polarization ranging from $\sim$15\% at the radio peak to more than $\sim$60\% in the lower surface brightness regions.
Previous studies have quoted an integrated RM of -12 rad/m$^2$ for 3C33 (e.g., \citealt{berge67}, \citealt{simard81}), although this RM determination may be contaminated by the northern lobe.
\cite{rudnick88} finds the RM to be $\approx$-7 \radmsq~ in the southern hotspot region using unpublished 20 cm and 6 cm data.

Table \ref{tab:3C33Stab} and Figure \ref{fig:1060.polDiags} summarize the modeling results for 3C33S.
The model fit for the depolarizing screen also yielded RM $=-7\pm0.15$ \radmsq, as did the $\chi$(\lamsq) fit alone (RM $=-6.8\pm0.17$ \radmsq).  Similarly, RM Synthesis/Clean found a dominant component with $p=2.5\%$ at -6.7 \radmsq, with a weaker $p=0.6\%$ component at +8 \radmsq.  This result is robust for various weighting of the NVSS sample with respect to the WSRT samples before computing the FDF.  While increasing the weighting of the NVSS sample can have a large effect on the sidelobe level and structure in the RMSF and, hence, the constructed FDF, RM Clean yields a similar solution each time:  a dominant peak near -7~\radmsq~ and a secondary peak near +8~\radmsq.  Thus, there appears to be good agreement between the literature, depolarizing screen, $\chi$(\lamsq) and RM Synthesis/Clean results that the dominant RM component in 3C33S is at -7~\radmsq.  Weaker RM components in the FDF, such as the $p=0.4\%$ one at +23 \radmsq, are increasingly unreliable (see Figure \ref{fig:1060.polDiags}).

However, our two component fit to these same data give quite different values, -3 and 0 \radmsq.  Which of these determinations is correct?  While direct comparison of the $\chi_{\nu}^2$ values for the best model fits is inappropriate, it is clear from Figure \ref{fig:1060.polDiags} that the two component model provides a better fit to the data than the depolarizing screen model.  It does a much better job of explaining the two longest wavelength observations (particularly in $q$), where $p$ rises from a minimum near $\lambda^2\sim0.8$ m$^2$.  The observation of a minimum in $p$ is compelling evidence against a simple depolarizing screen.  In any case, a slight change in the errors assigned to the original data points could change the relative goodness of fit for these two alternatives. Polarization data at $\lambda=9$ cm \citep{rudnick83} agrees in $p$ with our two component and depolarizing screen models, but have been excluded from the fits since they were integrated over both lobes and therefore not reliable for these purposes.

We note that such a discrepancy can be quite important depending on the scientific issues under investigation.  First, these two models (a single component at -7 \radmsq~ or two components, at -3 and 0 \radmsq) represent quite different physical structures in the source.  For example, the magnetic field in 3C33S very closely tracks the bow-shocked shape leading edge \citep{rudnick88}, and a small toroidal sheath could give rise to two dominant RM components.  Alternatively, the surrounding medium might have a depolarizing screen with very fine scale structure ($\ll$1$''$, $\sim$1 kpc) that is independent of the geometry of the source.  If we were not interested in the Faraday structure, we could simply look at the weighted mean of the two component fit, which yields -1.7 \radmsq.  However, the difference between this value and the -7 \radmsq~ from other models represents  a factor of greater than 4 in any derived densities or magnetic field strengths.  If similar discrepancies are found at shorter wavelengths, e.g., 1~GHz, then the physical parameters involved would be $\sim$10$\times$ larger.

The discrepancy between the -7 \radmsq~ and weighted -1.7 \radmsq~ fits is not due to the inaccuracies of the measurements, as determined from the formal errors.  The error in our $\chi$(\lamsq) fit is small (RM $=-6.8\pm0.17$) and the errors in the two component fit are even smaller (RM $=-2.9\pm0.1$). Thus, using our linear $\chi$(\lamsq) fit we would have ruled out the two component weighted mean with high confidence. Similarly, our RM Synth/Clean results would have ruled out the two component fit.  It might be further argued that we shouldn't have expected to distinguish between values of -7 and -1.7 using this method, since the FWHM of the RMSF's main lobe is $\sim$12 \radmsq.  This argument ignores the standard practice of quoting uncertainties in the location of a peak at a value of $\sim$FWHM/(2$\times$signal:noise).  In the case of 3C33S, the error in the dominant RM peak (from Gaussian fitting to the cleaned FDF) is 0.06 \radmsq.  Again, we would have ruled out the (unresolved) combination of peaks near -3, 0 \radmsq~  with high confidence.

Because the discrepancies between the results of various models, in particular RM Synthesis/Clean, were much larger than our calculated errors, we carried out a series of experiments with infinite signal to noise models using two RM components.

\section{Experiments with two RM components}
\label{sec:RMexperiments}

\subsection{RM Synthesis}
Although many different polarization diagnostic experiments could (and should) be done, we focused on two-component models for several reasons.  First, a two-component model produced a good fit to the 3C33S data.  Second, two Faraday components might be a reasonable expectation for double radio galaxies that are unresolved.  In addition, when angular resolution becomes sufficient to resolve depolarizing Faraday screens, there will always be places where the beam overlaps two neighboring structures, producing two Faraday components.  Finally, recent work by \cite{law2011}, where RM Synth/Clean was performed on 37 polarized radio sources using the Allen Telescope Array (ATA), showed that two or more components were detected with high confidence in $\sim$25\% of their sources.  We now discuss a few simple experiments to demonstrate some of the potential pitfalls when RM Synthesis/Clean is employed.

\begin{figure}
  \centering
  \epsfig{file=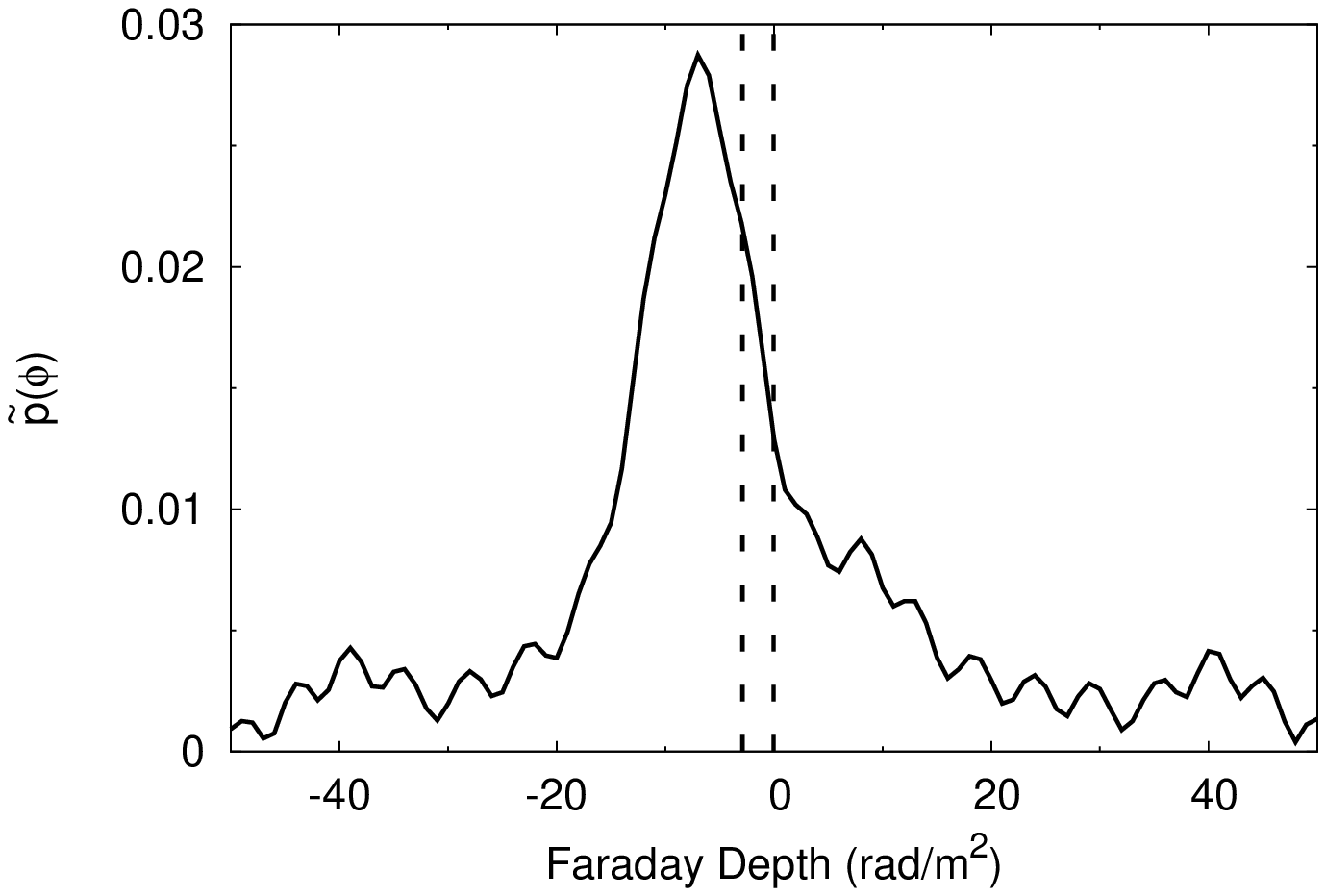,angle=0,width=0.45\textwidth}
  \caption{Cleaned FDF for the best fit two-component model of 3C33S.  The two input RM components are at -2.9 and -0.05 \radmsq~ as discussed in the text, but the dominant peak in the Faraday spectrum is near -7 \radmsq~ with a secondary feature near +8 \radmsq.  Vertical dashed lines show the location of the two input RMs, -2.9 and -0.05 \radmsq.}
  \label{fig:3c33.bestModel.rmclean.9}
\end{figure}

Our first experiment was to adopt a model fixed to the best two-component fit to the 3C33S data, with components at -3 and 0 \radmsq.  Synthetic $q$,$u$ spectra were constructed for the same NVSS + 8-WSRT \lamsq~ locations as in our previously discussed observations.  The results of RM Synth/Clean are shown in Figure \ref{fig:3c33.bestModel.rmclean.9}.  It bears a remarkable resemblance to the observed FDF for 3C33S, displaying a single dominant peak near RM $\sim-7$ \radmsq~ and a low amplitude secondary feature near RM $\sim+8$ \radmsq, even though the input RMs were at -3 and 0 \radmsq.  Thus, the FDF, with or without cleaning, produces RM power at what we can now state is the wrong value, since we know the input model parameters.  This is true whether you examine the clean components at high RM resolution or their convolved version which reflects more closely the limitations in resolving  multiple RM components.  In the convolved case, one would expect the FDF to still reflect the weighted mean of the input components; it does not.

\begin{figure*}
 \centering
  \begin{tabular}{|c|c|}
  \hline
 \epsfig{file=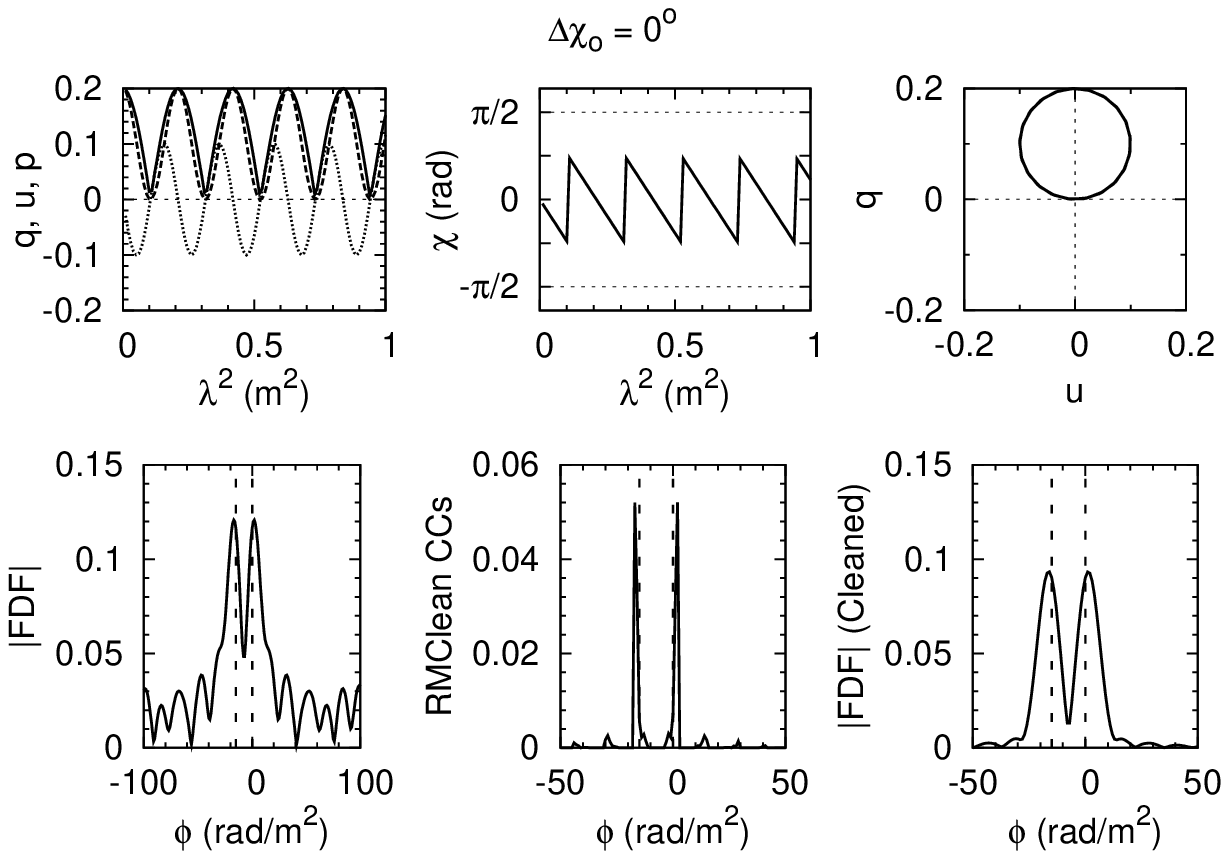,width=0.45\textwidth} &
 \epsfig{file=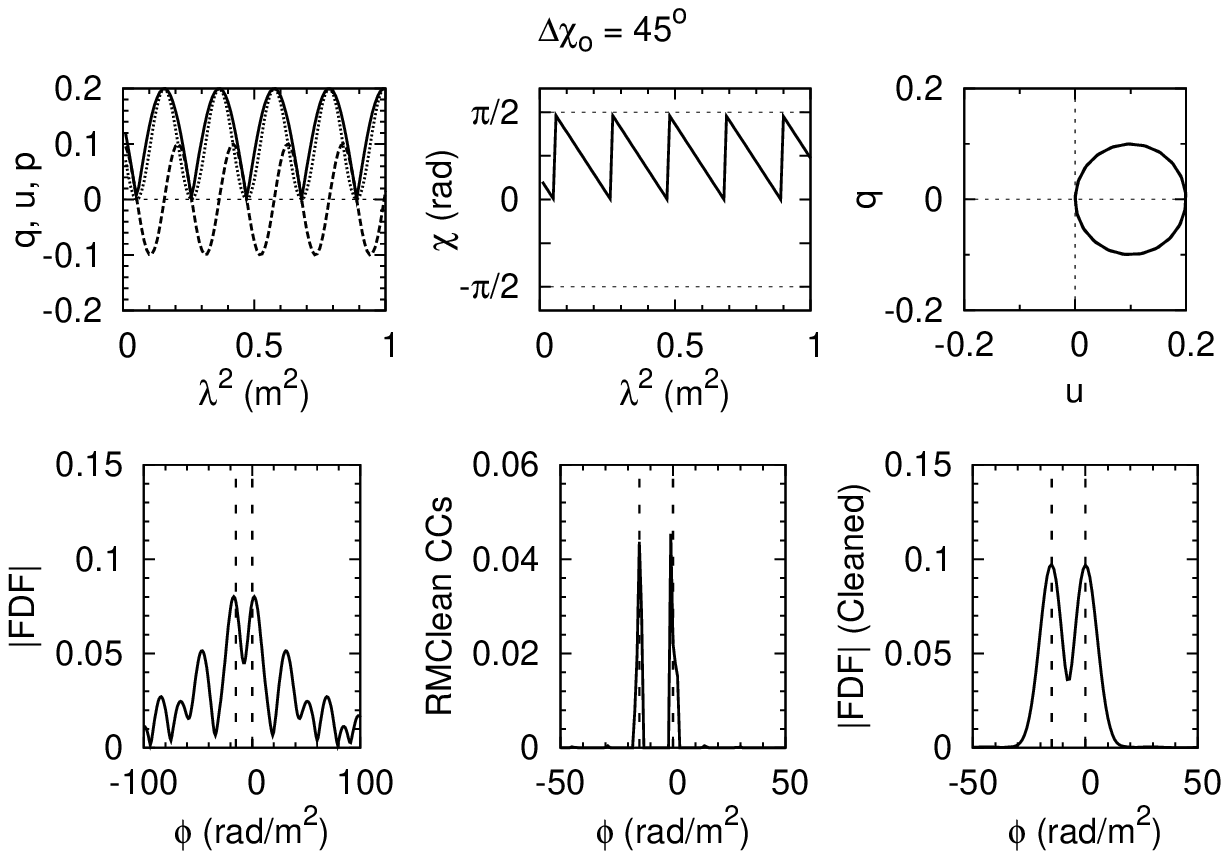,width=0.45\textwidth} \\
  \hline
 \epsfig{file=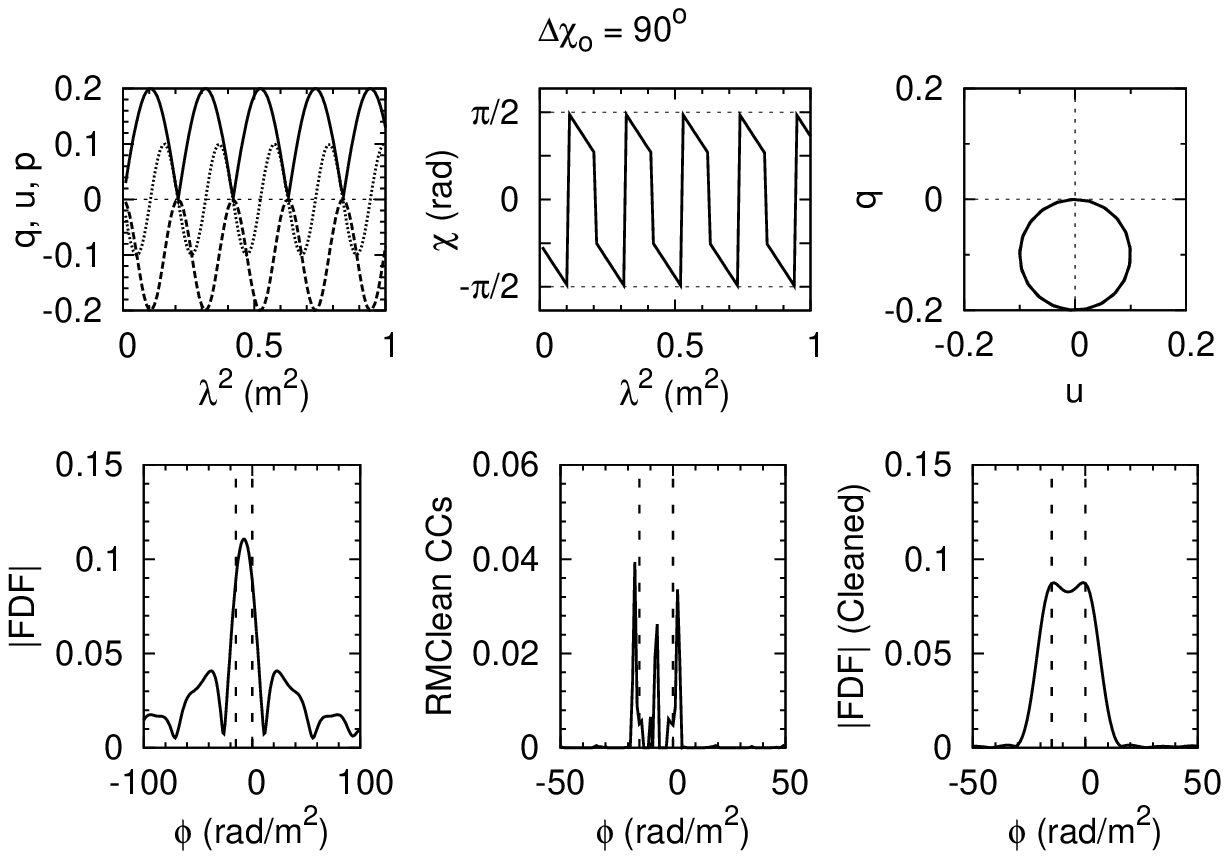,width=0.45\textwidth} &
 \epsfig{file=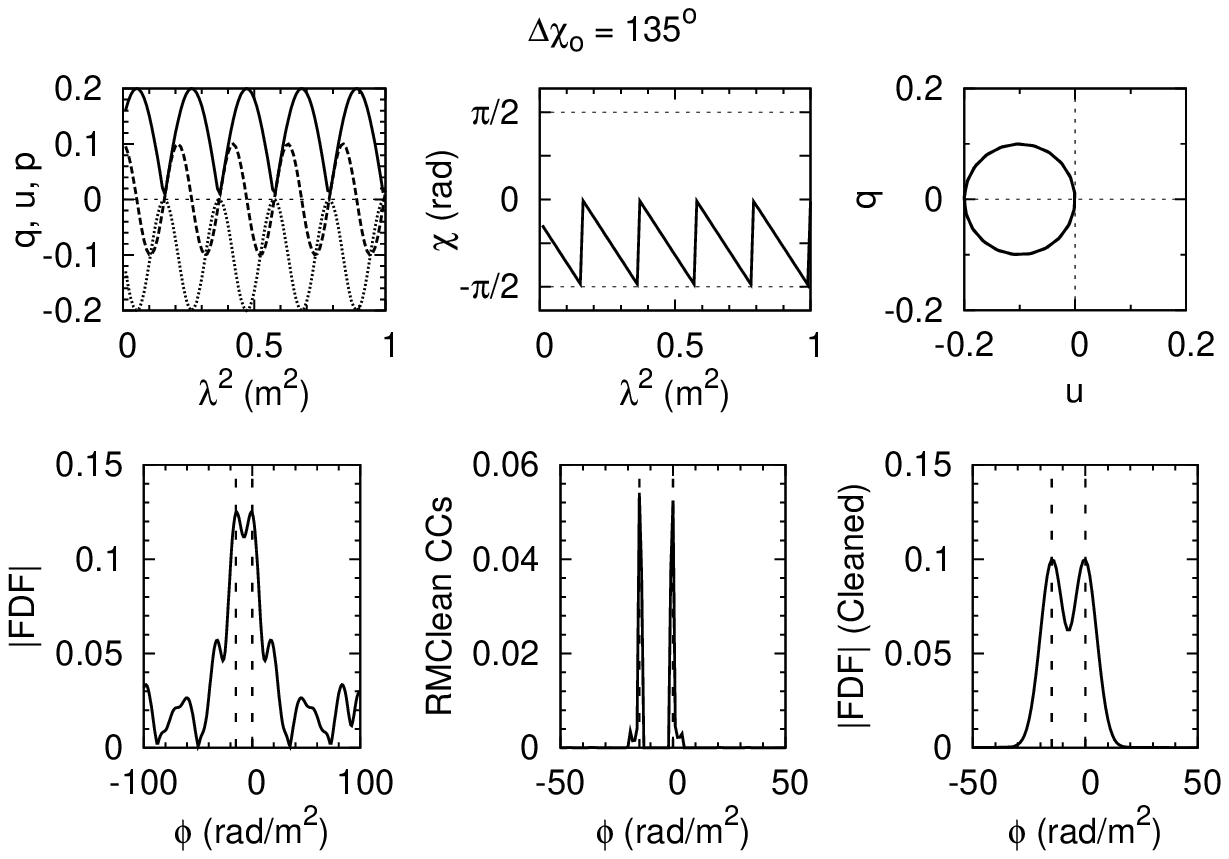,width=0.45\textwidth} \\
  \hline
  \end{tabular}
\caption{Illustration of the effect of relative phase between two RM components upon the results of RM Synthesis/Clean for various $\Delta \chi_0$ configurations.  Although the two components are separated by more than the FWHM (12 \radmsq) of the RMSF, RM Synthesis/Clean, using the same channels as in Figure \ref{fig:1000.rmsf}, fails to properly reproduce the solution for certain relative phases.  The model RMs are at -15, 0 \radmsq, shown by vertical dashed lines in the FDFs.  Plotted in each panel are:  \textbf{top left:} Fractional polarization, $q$ (dashed), $u$ (dotted), $p$ (solid); \textbf{top center:} polarization angle (radians); \textbf{top right:} $q$ vs. $u$; \textbf{bottom left:} Dirty FDF; \textbf{bottom center:} RM Clean clean components; \textbf{bottom right:} Cleaned FDF.}
\label{fig:rm-15-0}
\end{figure*}

Another case, demonstrating the impact of the relative phase of the two polarized components, involves using two components of equal amplitude with RMs of -15 and 0 \radmsq.  These are separated by more than the FWHM of the RMSF, 12 \radmsq, constructed from $\sim$400 channels in the WSRT 350 MHz band and shown in Figure \ref{fig:1000.rmsf}.  Nominally, then, they should appear well-separated in the FDF.
Figure \ref{fig:rm-15-0} shows the results of using four different values for the difference in $\chi_0$ for the two components.
In three cases, RM Synthesis/Clean successfully resolved the two components.  In the fourth case, with a difference in $\chi_0$ of 90$^{\circ}$, the raw FDF was dominated by a single peak near the mean RM of -7.5 \radmsq, along with considerable sidelobe power.  Cleaning produced an apparent triple component structure, with power at RMs of -17, +2 and -7.5 \radmsq, instead of the input values of -15 and 0 \radmsq.

We have also asked colleagues to run these and other models through their own RM Synthesis and cleaning programs, to verify that coding problems were not at fault. \textit{The above problems with RM Synthesis and Clean are robust to its exact implementation.}  They occur when the separation of the RMs is on the order of the FWHM of the RMSF, $2\sqrt{3}$/\Deltalamsq.  Under these conditions, the number of cycles of $Q$ and $U$ within the bandwidth differ by one or less for the two components.  RM synthesis is therefore not able to reliably resolve them into separate Fourier components.  \emph{However, the two components do not simply blend in this case, as two nearby sources would blend in total intensity.  Instead, they interfere to create complicated structures in $Q$(\lamsq), $U$(\lamsq) (i.e. $P$(\lamsq) and $\chi$(\lamsq)) which causes RM Synthesis to put power at values other than the input RM.}  In some cases, RM Clean is able to recover from this interference; in other cases it is not.

Some of the shortcomings of RM Synthesis arise not from a fault in the technique, but rather a limit of our measurement abilities.  One can use radio aperture synthesis as an analogy from which to draw insight; limitations in baseline sampling for aperture synthesis are in some ways analogous to limitations in \lamsq~ sampling for RM Synthesis.
However, the RM interference that we have illustrated here is considerably more complicated.  These experiments reflect the interference between two RM components and are reminiscent of other types of interference that are better understood.  Polarization canals \citep{shukurov2003}, e.g., do not represent actual dips in polarization, but simply the interference, in one beam, between two components separated by 90\textdegree~ in polarization angle at some observed wavelength.  Similarly, rotation measure involves the trend of $\chi$(\lamsq) over a range of wavelengths, and the mapping between multiple RM components and $\chi$(\lamsq) is not yet fully understood.  This illustrates the need for sufficiently broad \lamsq~ coverage in polarization observations when performing RM experiments, where detection of potential maxima or minima in $p$(\lamsq) is also critical to help diagnose the Faraday structure.  These methods are also subject to a variety of degeneracies, some of which we illustrate in the following section.

\subsection{Pseudo-\lamsq~ behavior}
\label{sec:pseudolamsq}
Another insidious quality of two component models is that they commonly produce 
\begin{equation}\label{rm.eqn}
RM(\lambda^2) \equiv d\chi(\lambda^2)/d\lambda^2~\approx~constant 
\end{equation}
over substantial ranges in \lamsq~ space.  Although it may be obvious that sparsely sampled data (especially using only two or three \lamsq~ data points) could lead to mistakes, it is assumed that continuous sampling over a significant range of wavelengths (e.g., (\lamsqmax-\lamsqmin)/\lamsq~ $>0.25$) can verify whether RM(\lamsq) $\approx$ constant.  This is not always true, as we now illustrate.

\begin{figure}
  \centering
  \epsfig{file=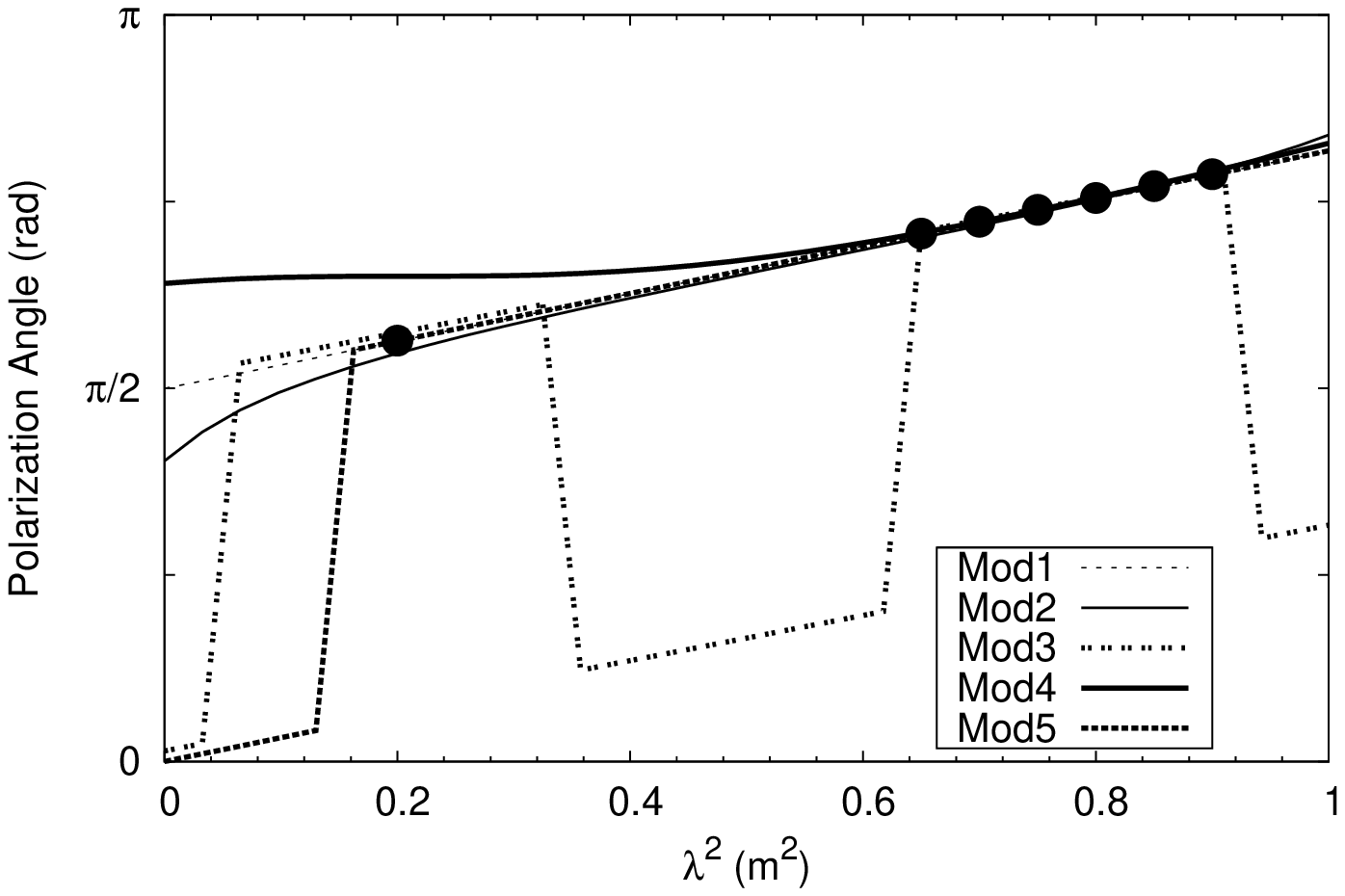,angle=0,width=0.45\textwidth} \\
  \epsfig{file=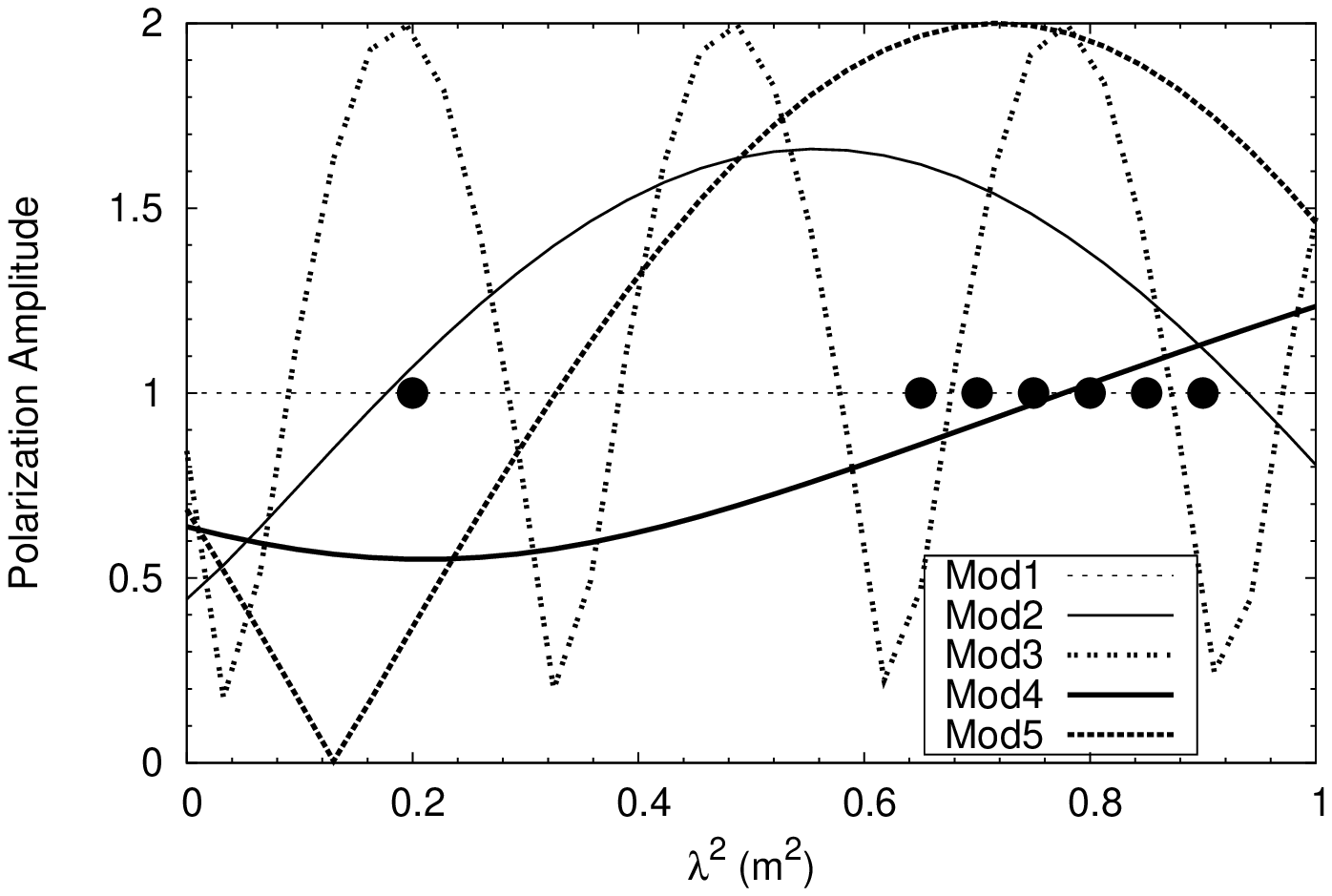,angle=0,width=0.45\textwidth}
  \caption{Various models illustrating the importance of considering polarization amplitude as well as angle in the long wavelength regime.  A linear fit to $\chi$(\lamsq) yields the same ``characteristic'' RM in each case, but inspection of the amplitude behavior reveals the complicated nature of the various Faraday structures listed in Table \ref{tab:pseudolong}.  Top: Polarization angle vs. \lamsq.  Bottom: Polarization amplitude vs. \lamsq.}
  \label{fig:LongFpol}
\end{figure}

Figure \ref{fig:LongFpol} shows five different models, all of which produce excellent RM(\lamsq) $\approx$ constant over the WSRT 350~MHz band, which covers $\sim$35\% in \lamsq~ space.  In addition, three of these models would also yield the same excellent \lamsq~ behavior with an additional point at 1.4~GHz (e.g., NVSS).  The model parameters are listed in Table \ref{tab:pseudolong}.  If one were examining the behavior of $\chi$(\lamsq)~ alone, as is done in most of the existing literature, there are a wide variety of two component models which easily fit the data but have very different values of RM than the one observed.
The key to ruling out such two component models, and thus to have a reliable determination of RM, lies in their $p$(\lamsq) behavior, which is quite different for each model.  A better way to avoid these mistakes is to simply fit the function $\mathbf{p} = pe^{2i\chi} = q + iu$ to the $q$(\lamsq) and $u$(\lamsq) data, and determine whether a satisfactory fit has been achieved.

\begin{table}
	\caption{Model Parameters for Long Wavelength Pseudo-\lamsq~ Experiment$^1$}
		\begin{tabular}{ ccccccc }
		\hline\hline
		Model & RM$_1$      & RM$_2$     & $p_1$ & $p_2$ & $\chi_1$ & $\chi_2$ \\
		  ID  &(rad/m$^2$) & (rad/m$^2$) &  (\%) & (\%)  & ($^{\circ}$) & ($^{\circ}$) \\
		\hline
		Mod1 & 1 & - & 1 & 0 & 90 & -  \\
		Mod2 & -0.5 & 2 & 0.66 & 1& -43 & -123 \\
		Mod3 & -4.4 & 6.3 & 1 & 1 & -30 & -145 \\
 		Mod4 & 1 & 2 & 1.1 & .55 & -75 & 3 \\
		Mod5 & -97 & 99 & 1 & 1 & -35 & 35 \\
		\hline
		\end{tabular}

  $^1$See Figure \ref{fig:LongFpol}.
	\label{tab:pseudolong}
\end{table}

\begin{figure}
  \centering
  \epsfig{file=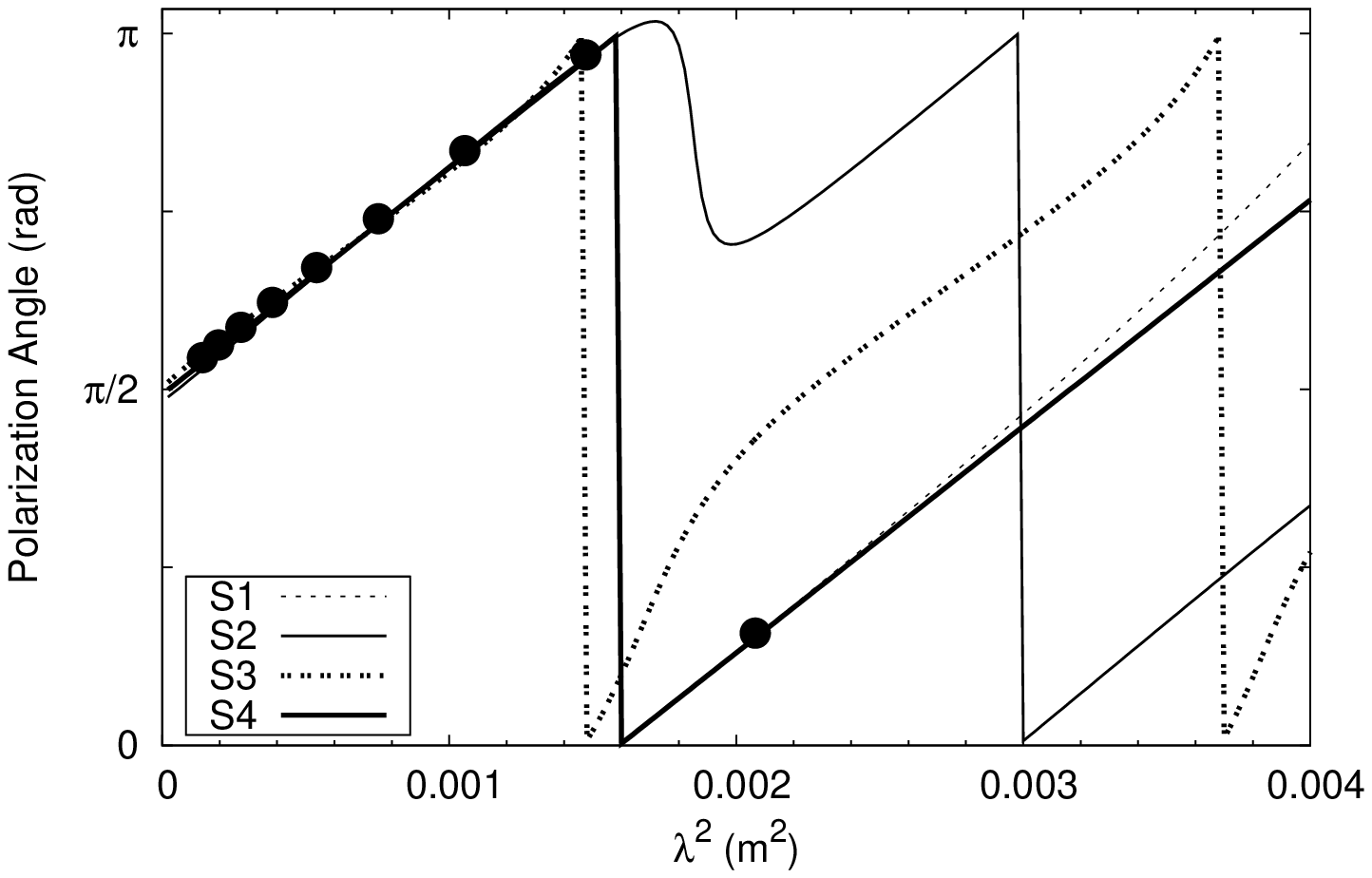,angle=0,width=0.45\textwidth} \\
  \epsfig{file=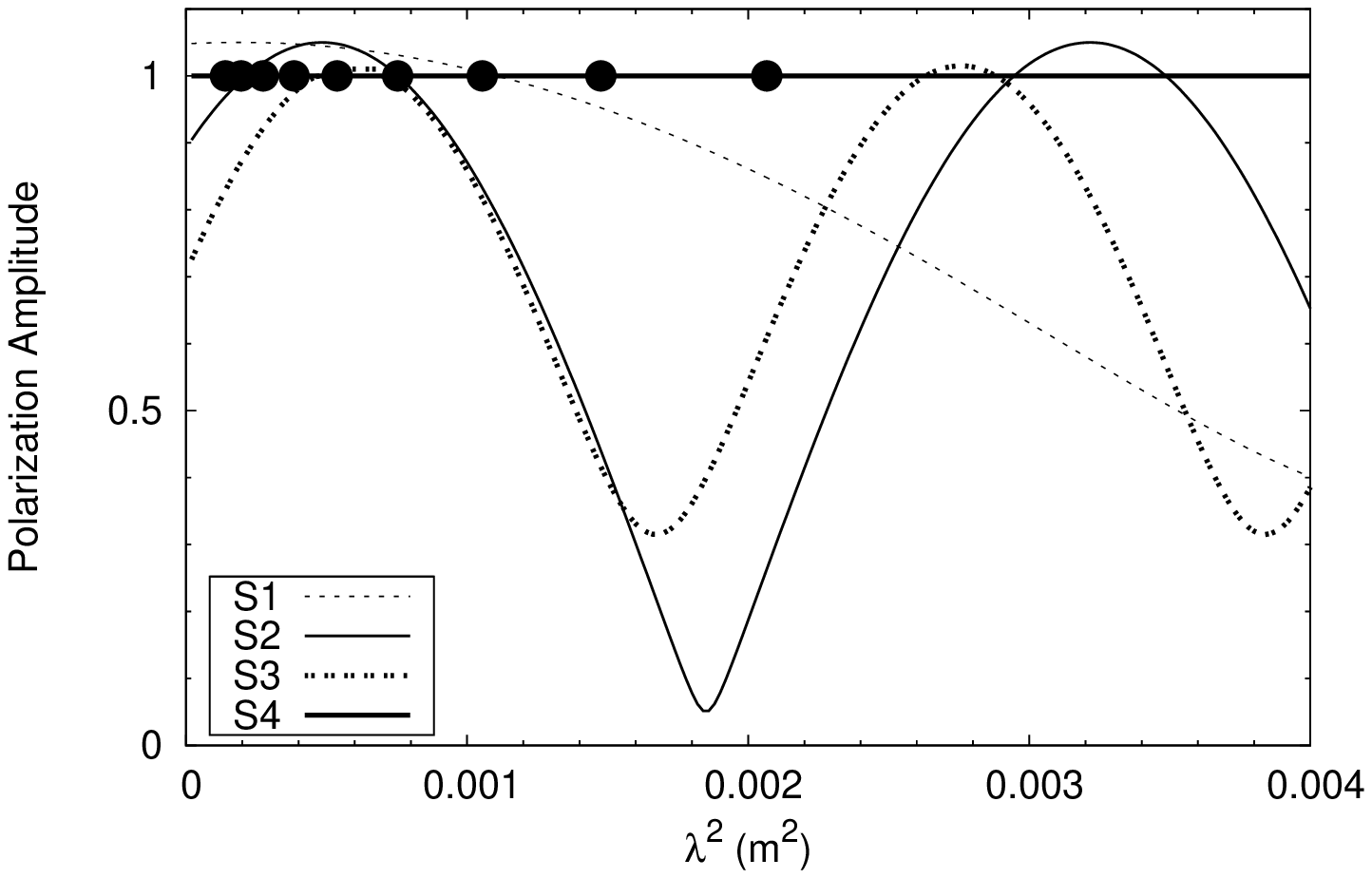,angle=0,width=0.45\textwidth} \\
  \epsfig{file=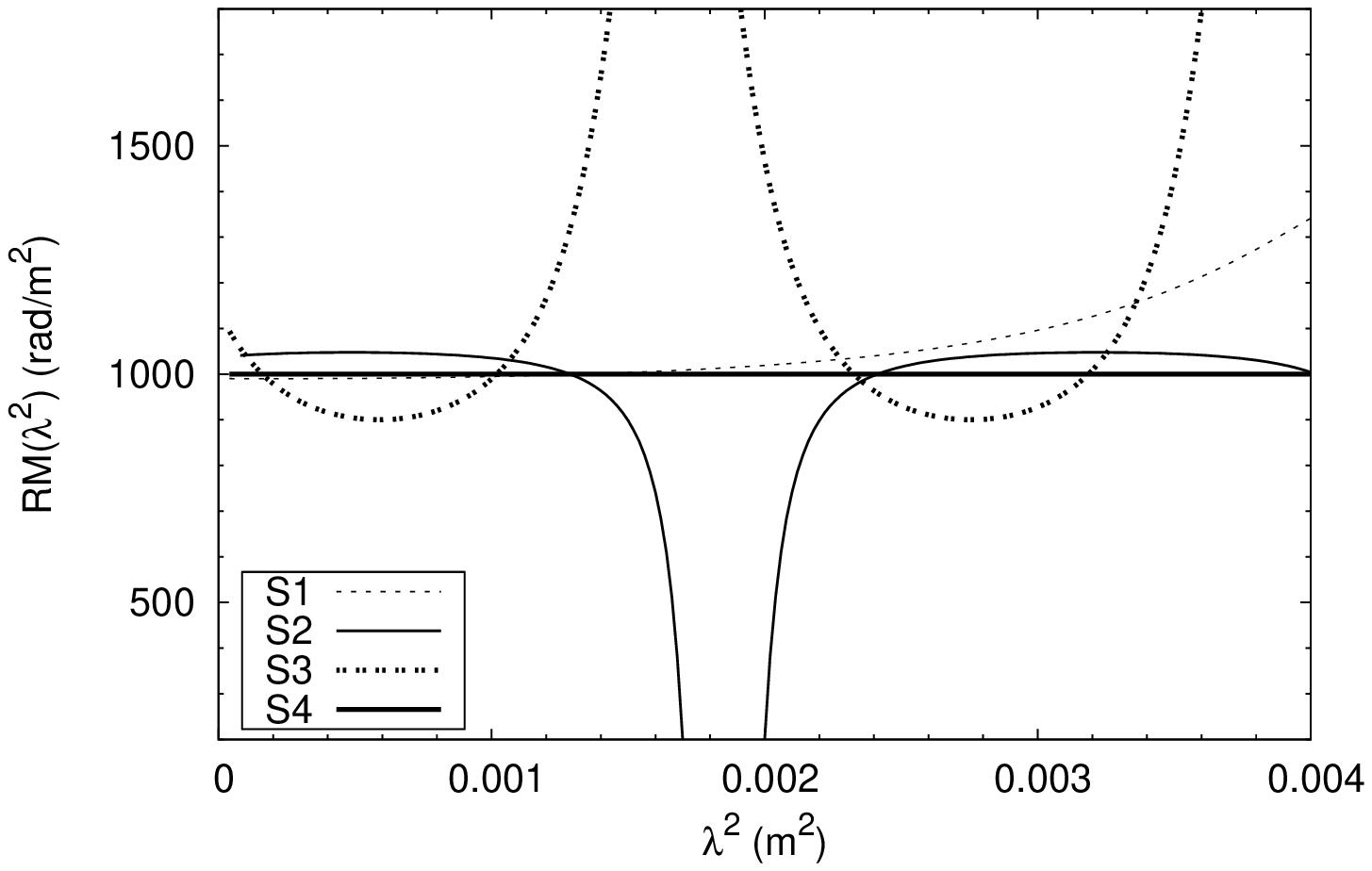,angle=0,width=0.45\textwidth}
  \caption{Various models illustrating the importance of considering polarization amplitude as well as angle in the short wavelength regime.  A linear fit to $\chi$(\lamsq) yields the same ``characteristic'' RM in each case, but inspection of the amplitude behavior reveals the complicated nature of the various Faraday structures listed in Table \ref{tab:pseudoshort}.  Top: Polarization angle vs. \lamsq.  Center: Polarization amplitude vs. \lamsq.  Bottom: RM(\lamsq).}
  \label{fig:ShortFpol}
\end{figure}

It is tempting to assume that there is a ``short wavelength'' limit where these problems can be safely ignored.  We now show that is not true.  First, we define a ``short wavelength'' set of observations as one in which there is reasonable sampling in \lamsq~ space and \lamsqmin $\ll$ \lamsqmax.  Thus, one can verify whether RM(\lamsq)~$\approx$~constant down to effectively zero wavelength. The models described in Table \ref{tab:pseudoshort} and shown in Figure \ref{fig:ShortFpol} demonstrate that this does not exclude two component models with RMs very different than the ones measured by fitting RM(\lamsq) = constant.  In the examples shown, the $\chi$(\lamsq) data alone follow very closely a constant RM = 1000 \radmsq.  However, they actually contain components that range from -50 to 1650 \radmsq.  Again, the key is to examine the $p$(\lamsq) behavior, as seen in Figure \ref{fig:ShortFpol}, or better, as noted before, to fit a model directly to the $q$(\lamsq) and $u$(\lamsq) data.

\begin{table}
	\caption{Model Parameters for Short Wavelength Pseudo-\lamsq~ Experiment$^1$}
		\begin{tabular}{ ccccccc }
		\hline\hline
		Model & RM$_1$      & RM$_2$     & $p_1$ & $p_2$ & $\chi_1$ & $\chi_2$ \\
		  ID  &(rad/m$^2$) & (rad/m$^2$) &  (\%) & (\%)  & ($^{\circ}$) & ($^{\circ}$) \\
		\hline
      S1 & 1110 & 750 & 0.7   & 0.35 & -92   & 92  \\
      S2 & 1650 & 500 & 0.5   & 0.55 & -110  & 102 \\
      S3 & 1400 & -50 & 0.665 & 0.35 & 76    & 125 \\
      S4 & 1000 & -   & 1     & 0    & 88.85 & -   \\
		\hline
		\end{tabular}

  $^1$See Figure \ref{fig:ShortFpol}.
	\label{tab:pseudoshort}
\end{table}

We also show in Figure \ref{fig:ShortFpol} the full RM(\lamsq) for these models.  The wide variations in this number show that when there are two interfering components, measuring the RM using data at closely spaced wavelengths, or only using very sparse sampling, can render the observed RM virtually meaningless.

\subsection{Recommendations regarding RM measurements}
\label{sec:recommendations}
There is no simple prescription for producing reliable rotation measures because it depends on the specific scientific goals.  We begin the discussion of those issues below, but here we simply offer some general guidelines to inform the future practice of Faraday structure determinations:

\begin{itemize}
  \item {Fitting of models to $q$(\lamsq) and $u$(\lamsq) (or equivalently, $p$(\lamsq) and $\chi$(\lamsq)) is the only reliable way to determine the underlying Faraday structure.  In particular, results derived from $\chi$(\lamsq) alone or RM Synthesis alone are subject to large ambiguities.}

  \item{RM Synthesis/Clean, as it is currently implemented, can serve as a first order indicator of the location of power in Faraday depth space, and guide more detailed modeling.}

  \item{Plots of $q$ vs. $u$ provide another useful diagnostic of the appropriateness of any models.}

  \item{Results for RM determinations should always specify not only the formal errors, but also the $\chi^2$ or RMS residuals of the fits.  This, along with documenting the coverage in \lamsq~ space, will allow for an analysis of what ambiguities are permitted by the data.}

  \item{The allowable space for ambiguities can be significantly reduced by broadening the \lamsq~ coverage, increasing the sampling, and ensuring that regions of \lamsq~ space are observed where RM(\lamsq) $\neq$ constant and $p$(\lamsq) $\ne$ constant.}

  \item{Scientifically useful results are possible in the presence of ambiguities if the underlying assumptions are both documented and valid, as discussed in Section \ref{sec:discussion}.}

	\item{Alternative methods of parameter determination, such as Maximum Likelihood, should be considered in the presence of low signal to noise.  In this case, least squares fitting may yield a low $\chi^2_{\nu}$ statistic, but may not necessarily yield the appropriate solution.  For example, \cite{guidetti08} use four frequency samples in a linear fit to $\chi$(\lamsq) to determine RMs for a number of cluster sources.  In the limit of infinite signal to noise, it doesn't matter how closely spaced the points are; with four points and two parameters, i.e. \textit{dof}=2, $\chi^2_{\nu} \approx 1$ would truly signify a good fit.  However, we note that for each source two of their samples are at nearly the same frequency, and these measurements agree within errors.  This essentially guarantees a value for $\chi^2_{\nu}$ of order unity, perhaps giving false confidence in the appropriateness of the model.  Thus, if minimization of $\chi^2_{\nu}$ is to be used, we caution that the effective degrees of freedom should first be carefully considered.}
\end{itemize}

\section{Discussion}
\label{sec:discussion}

\subsection{Depolarization}
The median depolarization ratio between 350 MHz and 1.4 GHz for our seven modeled sources is $p_{350}/p_{1.4}\sim$ 0.3.  This is the same as the median of the upper limits for our sample as a whole.  If this depolarization is due to a random foreground screen (and not to the interference between two components), then this corresponds to a Burn law $\sigma_{RM} \sim$ 1 \radmsq~, where the depolarization is $\exp(-2\sigma_{RM}^2\lambda^4)$.  It is likely that the overall sample is even more depolarized, since we observed the upper limits to drop as the polarization fraction at 1.4 GHz increased (see Figure \ref{fig:depolUL}).  This result has two implications, one for observations at low frequencies and one concerning the environment of radio galaxies.

Assuming a Faraday dispersion as above, we can estimate, e.g., the depolarization that would be observed by LOFAR\footnote{http://www.lofar.org} which has a high frequency band covering 120-240 MHz, and a low frequency band covering 30-80 MHz.  If the Burn law were to remain roughly accurate for integrated polarizations, then the depolarization would peak at 0.008 at the high end of the high frequency band, and drop by many orders of magnitude at low frequencies, essentially making polarizations undetectable.  However, as pointed out by \cite{tribble91}, the falloff from a Gaussian depolarizing screen is likely to be considerably slower, dominated by the small patches around extrema in RM, where the RM gradient is near zero.  If we start with a characteristic integrated polarization of $\sim$3\% at 1.4 GHz and extrapolate with only a \lamsq~ dependence from our depolarization results at 350 MHz, then we would expect fractional polarizations of 0.1\% - 0.5\% in LOFAR's high band, and 0.01\% - 0.05\% in the low band.  These are not likely to be detectable.  It is not clear, at present, whether even well-resolved extragalactic sources will have small enough Faraday dispersions to be observed in polarization at these low frequencies.

Using the more physical units introduced by \cite{garrington91}, our observed characteristic lower limit to the Faraday dispersion is $\sim$1.5 cm$^{-3}$~$\mu$G pc. For the purposes of calculating some very rough estimates of what these limits mean for field strengths around radio galaxies, we assume that the depolarization occurs in a foreground screen completely unaffected by the radio galaxy.  Assume that we need $\sim$10 independent patches across a 100 kpc source in order to depolarize it, and a fiducial electron density of $n_e$=10$^{-3}$ cm$^{-3}$. 
The resulting magnetic field is then B/$\mu$G $\geq$ 0.1($n_e$/10$^{-3}$)($r$/10)(cm$^{3}$ kpc$^{-1}$), where $r$ is a typical scale size of magnetic field fluctuations and we have ignored the $\sqrt{N}$ averaging along each line of sight for this order of magnitude calculation.  Fields of this strength are less than those found in clusters of galaxies, but greater than expected in the more filamentary WHIM outside of clusters \citep{ryu08}, especially if one factors in the much lower densities in those regions.

Thus, radio galaxies appear to be associated with thermal, magnetized plasmas with much higher values of $n_e B$ than expected for filamentary regions, but similar to those found in clusters.  This could result because of the bias for radio galaxies to be found in high density regions \citep{dezotti10}.  Alternatively, effects very local to the parent galaxy, such as emission-line regions (e.g., \citealt{pedelty89}) could be responsible for the ubiquitous depolarization.  This leaves very few radio galaxies available to probe cosmological filaments, except, perhaps some Mpc-scale sources \citep{saripalli09}.  As we seek to understand the causes of the $\sim$1.5 cm$^{-3}$~$\mu$G pc limits, however, it will also be important to readdress the questions of internal depolarization, e.g., due to a mixing layer \citep{bicknell90} between the radio source (with possibly much higher fields) and its low density environment.

\subsection{Science Implications of RM Ambiguity}
We have shown that there is considerable ambiguity (sometimes $>$100\%) in the determinations of RMs using the methods universally used in the literature, and even in the more recent RM Synthesis technique.  We now briefly examine the implications this has for different types of scientific investigations.

\subsubsection{Galactic Foreground}
The use of polarized extragalactic sources to characterize the magnetic field structure of our Galaxy has a long history (e.g., \citealt{simard80}, \citealt{brown01}) plus a major recent advancement \citep{taylor09}.  Our investigations do not reveal any (signed) bias in RM determinations, therefore, we would expect that the average RM of a group of extragalactic sources in some area of the sky should be a fair measure of the true value.  However, the structure function of galactic fluctuations will have contributions from the RM ambiguities discussed here, as well as attempts to measure the intrinsic differences in RM between sources, especially on the smallest scales.

\subsubsection{Fluctuations through Galaxy Clusters}
The situation with respect to cluster measurements is much more complicated.  The clusters are expected to have fields that are tangled on scales substantially smaller than the cluster.  Therefore the mean RM of a distant extragalactic source seen through the cluster should be zero, but the scatter in such RMs should be larger than for background sources not seen through clusters (e.g., \citealt{kim91}, \citealt{clarke01}, \citealt{bonafede10}).  In this case, the quantity being measured is the RM scatter, which will be increased because of the ambiguities discussed in this paper.  In the ideal world, this scatter should be no different for sources seen through clusters (``the sample'') than for sources not seen through clusters (``the control''), so again the measurements should be unbiased.

However, unless the sample and control have exactly the same properties, both intrinsically and in terms of the observations leading to their RMs, it is impossible to know how the RM ambiguities would affect their comparison.  For example, if RM determinations include a short wavelength point for some sources, as opposed to others, a different range of possible underlying RMs will be present for the two cases.  Or, if the sources in the sample or control are statistically different physically (e.g., FRI vs. FRII sources), then the ambiguities can have different effects and contaminate the test.  All of these problems are present in the well-cited studies by \cite{kim90}, \cite{kim91}, and \cite{clarke01}, as discussed by \cite{rudnick04}. Similar contamination can be present if RMs from one experiment are compared to RMs from another, with different wavelength coverage, different editing for non-\lamsq~ behavior, etc. (e.g., \citealt{johnston04}).  Since our modelling shows that RMs can be affected by factors of order unity, it is not possible to assess the reliability of these cluster background experiments.
Two types of studies are required to address this issue.  First, the prevalence of multiple RM components within observing beams must be estimated; \cite{pizzo11} found multiple components at $\sim$15$''$ resolution in all three radio galaxies near the center of Abell 2255.  Second, statistical predictors are needed to quantify the likely errors in RM for a given distribution of multiple components.

\subsubsection{Faraday structure of radio galaxies}
Increasingly detailed studies of the Faraday structure of individual radio galaxies are now becoming available (e.g., \citealt{laing87}, \citealt{zavala02}, \citealt{laing06}, \citealt{osullivan08}, \citealt{govoni10}).  In many cases, the rotation measures are assumed to be entirely in the unperturbed foreground, and thus a fair measure of the magnetic field structure of the environment, usually a cluster of galaxies.  However, as suggested by \cite{bicknell90} and \cite{rudnick03}, and now demonstrated convincingly by \cite{laing08}, the radio source itself may change the observed RM structure.  This issue aside, the question remains how the newly described RM ambiguities could affect these measurements.

The studies of individual radio galaxies involve higher order characterizations of the RM distribution, such as the structure function, so they are much more sensitive to possible ambiguities.  In addition to increasing the overall scatter in RMs, contributions from ambiguities are likely to change as a function of scale.  If we assume an unperturbed foreground screen, then when the observing beam is much smaller than the smallest angular scale of RM variations, a single component dominates and the RM determination can be free of ambiguities.  \cite{feain2009} took advantage of this situation in a Faraday structure study of the radio lobes of Centaurus A, using background sources.

In the limit where the observing beam is much larger than the characteristic scale of variations, then we approach the Burn limit of a depolarizing screen, and the effect of ambiguities is minimized.  However, as pointed out by \cite{tribble91}, the situation is typically much more complicated, and the observed polarized emission will be dominated by regions where the angular RM distribution is at an extremum, with only small gradients.  The emission is then a complicated function of the beam size and the angular structure of the magnetic field fluctuations.  Detailed modeling is required in such cases, and it is not certain whether a clear diagnosis of the Faraday structure is possible, in practice.  In particular, we may not be able to distinguish between the physically distinct cases of fully external screens, thin mixing layers of relativistic and thermal emission or fully mixed plasmas (e.g., \citealt{cioffi80}).

The intermediate situation, where two or three different RM components dominate within an individual observing beam, is the most sensitive to the ambiguities discussed in this paper.  The resulting complex interference patterns in \lamsq~ space can give rise to erroneous RMs, and will increase the observed scatter preferentially on these angular scales.  This situation will necessarily arise whenever the minimum angular scale of RM fluctuations is being approached.  The only effective way to deal with this will be using Monte Carlo or numerical simulations (e.g., \citealt{guidetti08}, \citealt{guidetti10}), where we expect there to be differences in shape between the input structure function and the observed structure function on scales of the order of the beamsize.

\section{Conclusions}
\label{sec:conclusions}

We have presented our polarization analysis of compact radio sources observed with the WSRT at 350 MHz.  Using the observations of 585 sources in six fields, we computed a simple analytic model of the off-axis instrumental polarization (which can rise to several percent in $q$ at the primary beam radius).  After correction of the observations using this model, only a small fraction of the sources were determined to have significant polarization at this frequency.  By supplementing our observations with data from the NVSS, we have assessed the depolarization of our sample, finding the median depolarization ratio from 1.4 GHz for the strongest sources to be $p_{350}/p_{1.4}<0.2$.

We modeled the Faraday structure of seven sources using various methods, including the traditional linear fit to $\chi$(\lamsq), as well as $q$, $u$ vs. \lamsq~ fitting to two simple depolarization models -- a foreground screen and two interfering RM components.  In addition, we applied the novel Rotation Measure Synthesis and RM Clean techniques.  A comparison of the RMs determined by various methods has shown agreement in many sources, and yet failure to reproduce the $q$, $u$ observations casts doubt upon the validity of those solutions.  In only one of the seven sources modeled, where depolarization from 1.4 GHz was not present, did the linear $\chi$(\lamsq) fit offer a solution that sufficiently reproduced the $q$, $u$ observations.  Of the remaining six sources, RM Synthesis/Clean suggested multiple significant ($p_{02}/p_{01}\geq0.5$) RM components in three sources, while the two-component model found a significant secondary RM component in all six.  Thus, a ``characteristic'' RM may be said to exist for any source, but the true Faraday structure may not always be adequately described by this alone.  This point is well demonstrated by our detailed analysis of the southern lobe of the radio galaxy 3C33.  Previous studies, as well as our own linear $\chi$(\lamsq) fit and RM Synthesis/Clean analyses have found a single, dominant RM of -7 \radmsq, in sharp contrast to the $q$, $u$ observations which strongly suggest two significant RM components near -3 and 0 \radmsq.

To further explore the possible shortcomings of the linear $\chi$(\lamsq) fit and RM Synthesis/Clean methods, we have performed a few simple experiments.  By constructing synthetic $q$, $u$ spectra using the best fit two-component model for 3C33S, we find that RM Synthesis may place power at incorrect Faraday depths when multiple, closely spaced RM components interfere.  In this case, both RM Synthesis and the linear $\chi$(\lamsq) fit find a consistent solution, but one that does not agree with the known model inputs.  The vulnerability of RM Synthesis is further demonstrated by a second experiment, which illustrates the role of phase in RM ambiguity.  In this experiment, we show that two RM components, separated by more than the FWHM of the RMSF, may still yield an incorrect solution under RM Synthesis depending upon the relative phase (i.e. intrinsic polarization angle) between the two components.  A third experiment shows the dangers of a common assumption, that RM determinations made at high frequencies are sufficient.  We show that the \lamsq~ coverage must be broadened as much as possible to explore the true depolarization behavior of the source.

With modeling of our WSRT observations and experiments on synthetic observations, we have touched upon some of the ambiguities that exist in rotation measure determinations.  We caution that care must be taken when designing RM experiments and choosing one or more methods of analysis, stressing the importance of considering both degree and angle of polarization (or equivalently, $q$ and $u$) over as wide a range of \lamsq~ space as possible to see a more global picture of the polarization behavior and produce a more accurate description of the Faraday structure.

\acknowledgements
We gratefully acknowledge assistance and comments from the anonymous referee, Rainer Beck, Michiel Brentjens, Peter Frick, Bryan Gaensler, Leonia Kogan, Frazer Owen, Dominick Schnitzeler, Dmitry Sokoloff, and Rodion Stepanov.
Partial support for this work at the University of Minnesota was provided by the U.S. National Science Foundation grant AST 0908668.
The Westerbork Synthesis Radio Telescope is operated by the ASTRON (Netherlands Institute for Radio Astronomy) with support from the Netherlands Foundation for Scientific Research (NWO).
We acknowledge the use of NASA's \textit{SkyView} facility (http://skyview.gsfc.nasa.gov) located at NASA Goddard Space Flight Center.
This research has made use of the NASA/IPAC Extragalactic Database (NED) which is operated by the Jet Propulsion Laboratory, California Institute of Technology, under contract with the National Aeronautics and Space Administration.
This research has made use of the VizieR catalogue access tool, CDS, Strasbourg, France.

\end{document}